\documentclass[12pt]{article}
\usepackage{amsmath,amssymb,amsfonts,amsthm,mathtools}
\usepackage{booktabs,array,graphicx,float}
\usepackage{enumitem}
\usepackage{titlesec}
\usepackage{geometry,setspace,natbib,hyperref,xcolor}
\hypersetup{colorlinks=true,citecolor=blue,linkcolor=blue,urlcolor=blue,
 pdftitle={Probability-Feasible Historical Transport},
 pdfauthor={Ulrich Hounyo}}

\newtheorem{assumption}{Assumption}
\newtheorem{theorem}{Theorem}
\newtheorem{proposition}{Proposition}
\newtheorem{corollary}{Corollary}
\newtheorem{lemma}{Lemma}

\newcommand{\E}{\mathbb E}

\newcommand{\Pcal}{\mathcal P}
\newcommand{\Fcal}{\mathcal F}

\newcommand{\Pp}{\mathbb P}
\newcommand{\1}{\mathbf 1}
\newcommand{\cF}{\mathcal F}
\newcommand{\cQ}{\mathcal Q}
\newcommand{\norm}[1]{\left\lVert #1\right\rVert}
\newcommand{\TV}{\mathrm{TV}}
\newcommand{\co}{\operatorname{co}}
\newcommand{\PP}{\mathsf P}
\newcommand{\Dcal}{\mathbb D}
\newcommand{\argmin}{\operatorname*{arg\,min}}

\title{\textbf{Probability-Feasible Historical Transport}\\[0.6em]
\large\normalfont Sharp Bounds and Falsification Tests for Distributional Pre-Trend Designs}
\author{Ulrich Hounyo\thanks{Department of Economics, University at Albany, State University of New York. Email: \texttt{khounyo@albany.edu}.}}
\date{\today}

\begin{document}
\maketitle

\begin{abstract}
\noindent Distributional difference-in-differences studies often assume that the post-policy gap between treated and comparison groups is a mixture of pre-policy gaps. This historical-transport assumption can fail in ways that one-statistic checks miss. When policy events are pooled, the average counterfactual can be a valid distribution although no event's counterfactual is, and pooled bounds can exclude the true average effect. Mixtures of past gaps can assign negative probability to outcomes, and cutoff-by-cutoff bounds can be incompatible. We characterize sharp bounds and give breakdown radii, a grid-free confidence region, and tests with and without an experiment. The tests reject transport in the National Supported Work data and expose impossible state counterfactuals behind a pooled minimum-wage bound. For the 1993 Earned Income Tax Credit expansion they pass through 1995, when bounds put single mothers' employment gain relative to childless women at 5 to 7 points; the gain stays positive after dropping early welfare-waiver states.
\end{abstract}


\newpage
\section{Introduction}\label{sec:intro}

Difference-in-differences studies increasingly ask how a policy changes a whole outcome distribution: the share of workers paid below a wage, the probability of having any earnings, a quantile. When parallel trends is in doubt, the pre-policy record is the natural place to look for discipline. A recent proposal, developed for average effects by \citet{BanKedagniChenDenteh2026}, replaces parallel trends with a weaker requirement: the treated--comparison gap after the policy must resemble the gaps observed before it, in the sense of being a mixture of them. This paper asks what that requirement means, and what it can deliver, when the object of interest is a distribution.

The question is not a technicality. Write $\mu^0_{g,t}$ for the distribution of the untreated outcome in group $g\in\{0,1\}$ at date $t$, where $g=1$ is the treated group and $t=1$ is the first post-policy date, and let $S$ be a set of pre-policy dates. The distributional version of the requirement is
\begin{equation}\label{eq:intro-hull}
\mu^0_{1,1}-\mu^0_{0,1}=\sum_{s\in S}\lambda_s\bigl(\mu^0_{1,s}-\mu^0_{0,s}\bigr)
\quad\text{for some weights }\lambda_s\ge0\text{ with }\textstyle\sum_s\lambda_s=1 .
\end{equation}
Two things can go wrong when an analyst uses \eqref{eq:intro-hull}. The comparison group's current distribution plus a mixture of past differences is a signed measure, so it can assign negative probability to some outcomes, and such a ``counterfactual'' describes no population. And an analyst who bounds the effect at several wage cutoffs, optimizing the weights separately for each cutoff, reports endpoints that may come from incompatible mixtures; each endpoint is sharp alone, yet no single counterfactual distribution produces them together. Neither problem is visible one scalar at a time, and both matter for economic conclusions. The bunching estimator of \citet{CengizEtAl2019}, for instance, nets the jobs missing below a new minimum wage against the excess jobs just above it; a bound on that difference is meaningful only if both counts come from one counterfactual wage distribution. A third problem arises when, as is now standard, several policy events are pooled: an average of event-level counterfactuals can look like a distribution even when none of them is one, and the pooled bound can then miss the average effect it is read as measuring. All three appear in the data we study.

\paragraph{What the paper does.} The sharp identified set imposes \eqref{eq:intro-hull} with common weights for all outcome events and intersects the result with the set of probability distributions (Theorem~\ref{thm:sharp-id}); on a grid of $K$ cells, the bound on any linear contrast is a linear program in the $|S|$ historical weights with $K$ constraints. The paper is about what this restriction implies for the way distributional pre-trend designs are used:
\begin{itemize}[leftmargin=1.5em,itemsep=1pt,topsep=3pt]
\item \emph{Pooled events can hide impossible counterfactuals.} Stacking events and checking feasibility only for the average is the practice most likely to mislead. When one historical weight is shared across events, the averaged counterfactual can be a valid distribution even though, for every admissible weight, some event's counterfactual is not (Theorem~\ref{thm:stacking}). The stacked bound then describes a composite that no event produces, not an average of event effects. This has consequences beyond interpretation: when events draw on their histories with different weights, the stacked bound excludes the true average effect in 29 percent of our simulated designs. We give an exact test with one linear program per event and cell, the sharp sets for three models of pooling (common weights, event-specific weights and a pooled restriction), and a family that interpolates between them by bounding how far event weights may differ. The same structure governs covariates: transport within covariate cells requires coherence cell by cell, which unconditional checks do not see. In the minimum-wage data, the stacked counterfactual is a distribution while each state's is not, under common and under state-specific weights.
\item \emph{When coherence changes a conclusion.} The probability restriction narrows the bound on a policy contrast if and only if every historical mixture that attains the unrestricted bound is infeasible (Proposition~\ref{prop:functional-gain}); it can reverse a sign. When no feasible mixture exists, a single nonnegative outcome score has negative expectation under every historical mixture and refutes the model as a whole (Theorem~\ref{thm:common-witness}). Bounding cutoffs separately and projecting candidates onto the simplex each change the identified set.
\item \emph{Grids.} For a contrast defined on both grids, a coarser grid can only widen the bound, but it can conceal infeasibility entirely. A confidence region computed by one linear program over all intervals removes the choice of grid (Propositions~\ref{prop:partition-refinement}--\ref{prop:grid-choice}).
\item \emph{Breakdown and inference.} We give an exact formula, a single linear program in which both terms share one weight vector, for the radius at which the sign of an effect on an outcome share is no longer determined, with a confidence interval (Proposition~\ref{prop:breakdown}). It differs from the distance needed to make the model feasible: in the NSW data a repair of under one percent of the probability mass restores feasibility, while the sign of the effect survives departures of nearly forty percent in total variation. Measured instead by how far earnings must move (Wasserstein distance), the sign is lost at an average shift of about \$1,050, well below the \$2,630 departure the experiment reveals; the choice of distance matters, and the programs accommodate either. Because the error of a mixture never exceeds the largest error at a historical date, simultaneous bands for finitely many historical candidates yield confidence sets, a coherence test and, when an experiment identifies the missing distribution, a test of transport itself with a lower confidence bound on its error (Theorem~\ref{thm:coherent-inference}, Proposition~\ref{prop:benchmark-test}). With independent individual histories and Dvoretzky--Kiefer--Wolfowitz radii these are valid in finite samples; bootstrap and studentized versions are sharper and valid as samples grow. For researchers without an experiment, the studentized coherence test is the workhorse: in simulations it is conservative under the null and has power where unstudentized bands have none. It is a test of moment inequalities with a nuisance weight in a simplex, and generic moment-inequality procedures apply to the same restrictions. With data sampled in clusters, such as states, the same results hold on a fixed grid with bands calibrated over clusters, in finite samples or uniformly as the number of clusters grows; a one-sided calibration is valid and more powerful, and the exact local limit of the coherence statistic, the value of a small linear program in the binding cells, gives its power function and explains why critical values from all vertex cells are conservative.
\end{itemize}

\paragraph{What the applications show.} The three applications have different purposes. The first asks whether historical transport recovers a known truth. In the National Supported Work (NSW) demonstration, randomization reveals the 1978 earnings distribution that participants would have had without the program. We construct historical-transport bounds from the 1974--75 earnings of participants and of two survey comparison samples and compare them with the experimental control distribution. Transport fails by a wide margin. Measured by the largest vertical gap between cumulative distribution functions (CDFs), the 95\% lower confidence bound on the distance between the truth and the nearest transported distribution is 0.20 for the Current Population Survey (CPS) comparison sample and 0.23 for the Panel Study of Income Dynamics (PSID) sample, and the implied employment effect (0.38 to 0.49) is three to four times the experimental one (0.11). The failure has an economic cause that is visible before the program: eligibility required recent joblessness, so the pre-program nonemployment gaps between participants and survey respondents (49 to 62 points) far exceed the post-program gap revealed by the experiment (22 to 24 points). Historical gaps do not bracket the post-policy gap when selection works through the pre-period outcome itself. Without using the experiment, a studentized coherence test already rejects transport from PSID ($p=0.013$). With CPS, where the coherence test does not reject, only the experiment reveals the failure, which shows the limit of what pre-program data can detect. When half of the experimental controls replace the survey sample, so that transport holds by design, the benchmark test rejects in 1 of 200 random splits and the coherence test in none.

The second application is a setting in which the assumption deserves a hearing: the 1993 expansion of the Earned Income Tax Credit (EITC), which raised the credit for families with children. Treated and comparison groups (single women with and without children) differ by a stable characteristic, not by a recent outcome. We fixed the design in writing before estimation. The diagnostics pass: the coherence test does not reject, a placebo benchmark at 1993 does not reject ($p=0.96$), and the comparison group's earnings distribution stays within its historical range through 1995. By 1995 the bounds place the employment gain of single mothers relative to childless women at 4.7 to 7.0 points, an effect of the credit and of the state welfare waivers of those years; by 1996 at 11.5 to 13.9 points, but in 1996 the comparison group's distribution leaves its historical range, and the protocol flags that year. Sampling uncertainty is large relative to these bounds: the 95 percent outer region excludes zero in 1996 but not in 1995, whether women or whole states are resampled. Dropping the states with welfare waivers approved by 1995 leaves every diagnostic passing, including in 1996, and moves the 1996 bound to 8.4 to 11.8 points, a range that coherence narrows; dropping every state with an approval through 1996 widens the bounds, and the 1995 bound then includes zero. Common weights also deliver what separate bounds cannot: in 1995, every counterfactual with an employment gain above 6.2 points implies a fall in the share of mothers with earnings in the credit's phase-out range.

The third application shows the pooling problem in a standard design, and how much coherence can add on a realistic grid. In the 118-bin state--quarter wage distributions of \citet{CengizEtAl2019}, all four historical candidates for five 2014 minimum-wage increases put negative mass on some wage bin, and imposing probability feasibility halves the width of the bound on the change in the population share employed below \$9.25, to $[-0.52,-0.32]$ percentage points. But although the stacked counterfactual is a distribution, no state's counterfactual is, whether the historical weights are common to all states or chosen separately for each, so the bound is not an average of state effects. The audit adds further reasons not to read it causally: the 2014 labor market lies outside its historical range, dropping a historical quarter in which two of the increases had already been signed or approved makes the coherent set empty, and weighting events by population moves the bound to include zero.

Two lessons follow. Probability coherence is necessary for any distributional counterfactual built from history, and imposing it can sharpen bounds substantially; ignoring it produces endpoints that no population attains. But coherence is a check, not a validation. Whether history can discipline a counterfactual depends on whether the post-policy environment lies within the historical one, and that can and should be audited before transport is used. Settings in which selection reflects stable differences between groups, rather than a transitory pre-program state, are where the assumption deserves a hearing, and the EITC application shows that the diagnostics then pass and the bounds can be informative.

\paragraph{Related literature.} The paper extends the historical-hull idea of \citet{BanKedagniChenDenteh2026} from average effects to distributions. Imposing their scalar restriction with one weight for every bounded outcome score is equivalent to \eqref{eq:intro-hull} (Proposition~\ref{S-prop:all-scores}); with weights that may differ by score the equivalence fails. Our model is thus the distributional closure of theirs, and for a single prespecified score their bound is already sharp; the new questions are feasibility, joint attainability, aggregation across events, breakdown and testing. Other distributional difference-in-differences models impose different restrictions: rank invariance over time in changes-in-changes \citep{AtheyImbens2006}, copula stability with panel data \citep{CallawayLi2019}, transformation invariance \citep{RothSantAnna2023}, distribution regression \citep{FernandezValEtAl2024}, and distributional synthetic controls \citep{Gunsilius2023}. Section~\ref{sec:content} compares their economic content with ours and says when each is the better choice. Bounded deviations from parallel trends \citep{ManskiPepper2018,RambachanRoth2023} restrict scalar trends. Coherence is a system of moment inequalities indexed by a weight in a simplex, so the tests connect to inference for moment inequalities \citep{ChernozhukovHongTamer2007,AndrewsSoares2010,RomanoShaikhWolf2014}. The pooling result parallels the staggered-adoption literature, which shows that two-way fixed-effects and pooled estimators need not average event-level effects when effects vary across events \citep{deChaisemartinDHaultfoeuille2020,GoodmanBacon2021,SunAbraham2021,CallawaySantAnna2021}; here the issue arises even for identification, before any estimation: a pooled distributional counterfactual can be a valid distribution when no event-level one is. Rearranging an estimated distribution function \citep{ChernozhukovFernandezValGalichon2010} enforces monotonicity without the common-weight constraint, and so answers a different question. The bounds are linear programs in the tradition of \citet{MogstadSantosTorgovitsky2018} and \citet{KaidoMolinariStoye2019}, and the relaxation is a breakdown analysis in the sense of \citet{MastenPoirier2020}. The NSW exercise follows \citet{LaLonde1986}, \citet{DehejiaWahba1999}, \citet{SmithTodd2005}, \citet{CalonicoSmith2017} and \citet{ImbensXu2025}; to our knowledge it is the first experimental evaluation of a distributional pre-trend method. The EITC application follows the comparison of single women with and without children in \citet{EissaLiebman1996} and \citet{MeyerRosenbaum2001}; \citet{Kleven2019} questions how much of the employment gain after 1993 the credit explains. The minimum-wage application builds on \citet{CengizEtAl2019}, \citet{DubeLesterReich2010} and \citet{CallawaySantAnna2021}.

Section~\ref{sec:model} presents the model and its economic content, Section~\ref{sec:matters} the results on when coherence matters, Section~\ref{sec:quantile} quantiles, Section~\ref{sec:breakdown} the relaxation, and Section~\ref{sec:inference} inference and simulations. Sections~\ref{sec:nsw}, \ref{sec:eitc} and~\ref{sec:mw} contain the applications. Proofs not in the text are in the Appendix.

\section{Model}\label{sec:model}

\subsection{Setup}

There are two groups, $G\in\{0,1\}$, and dates $t=-T_0,\ldots,0,1$; group $G=1$ is treated from $t=1$ on. Outcomes take values in a standard Borel space $(\mathcal Y,\mathcal B)$, which covers real outcomes, finite categories, and mixed outcomes such as earnings with a mass at zero. Units have potential outcomes $Y_t(0)$ and $Y_t(1)$ and observed outcome $Y_t$. Write
\[
\mu^d_{g,t}(A)=\Pp\{Y_t(d)\in A\mid G=g\},\qquad \mu_{g,t}(A)=\Pp\{Y_t\in A\mid G=g\},\qquad A\in\mathcal B,
\]
and $\PP(\mathcal Y)$ for the set of probability measures on $\mathcal Y$. The data are a panel or repeated cross-sections, and $P$ denotes the distribution of the observed variables; in the repeated cross-section case every group--date cell used below has positive probability. Group membership is fixed, and with repeated cross-sections each group is taken to be the same population at every date; changes in its composition then enter the selection bias defined below, and Section~\ref{sec:content} shows how they can violate the identifying assumption.

\begin{assumption}\label{ass:basic}
(i) $0<\Pp(G=1)<1$. (ii) There is no interference across units; $Y_t=Y_t(0)=Y_t(1)$ for $t\le0$ in both groups; and at $t=1$, $Y_1=Y_1(0)$ if $G=0$ and $Y_1=Y_1(1)$ if $G=1$.
\end{assumption}

Part (ii) combines consistency with no anticipation. Under it every untreated distribution except $\mu^0_{1,1}$ is observed: $\mu^0_{g,t}=\mu_{g,t}$ for $t\le0$, and $\mu^0_{0,1}=\mu_{0,1}$. Let $\nu_t=\mu^0_{1,t}-\mu^0_{0,t}$ be the untreated selection bias at date $t$, a signed measure of total mass zero. Fix a finite nonempty set $S\subseteq\{-T_0,\ldots,0\}$ of historical dates and let $\Delta_S=\{\lambda\in\mathbb R^S:\lambda_s\ge0,\ \sum_s\lambda_s=1\}$.

\begin{assumption}[Historical transport]\label{ass:hull}
There is $\lambda\in\Delta_S$ such that $\nu_1=\sum_{s\in S}\lambda_s\nu_s$ as signed measures.
\end{assumption}

With one historical date, Assumption~\ref{ass:hull} is distributional difference-in-differences with the bias fixed at its value on that date. With several dates it lets the bias drift, but only within the range history spans. The equality is between measures, so the same $\lambda$ applies to every outcome event.

\subsection{Economic content: an environment model}\label{sec:content}

A simple environment model makes Assumption~\ref{ass:hull} concrete. Suppose that at each date the economy is in a mixture of latent environments $e\in\{1,\ldots,m\}$, such as recession and expansion, or high and low demand for low-skill labor, with shares $\pi_t$ in the unit simplex $\Delta_m$ of $\mathbb R^m$, common to both groups, and that within environment $e$ the untreated outcome laws of the two groups, $F_{1e}$ and $F_{0e}$, do not change over time. Then $\nu_t=\sum_e\pi_{te}(F_{1e}-F_{0e})$, and Assumption~\ref{ass:hull} holds whenever the post-policy shares $\pi_1$ lie in the convex hull $\co\{\pi_s:s\in S\}$ of the historical shares. Appendix~\ref{S-sec:environments} shows that group-specific environment shares add a composition term that the assumption rules out. The environment model is sufficient for Assumption~\ref{ass:hull}, not necessary: different shares can produce the same selection biases. It has a testable implication of its own. Each group's untreated law at date $t$ is $\sum_e\pi_{te}F_{ge}$, an affine function of $\pi_t$, so $\pi_1\in\co\{\pi_s:s\in S\}$ implies $\mu_{0,1}\in\co\{\mu_{0,s}:s\in S\}$: the comparison group's current distribution must be a mixture of its own past distributions, and so must every statistic of it that is linear in the distribution, such as the nonemployment rate. A comparison-group statistic outside its historical range therefore refutes the environment model, though not Assumption~\ref{ass:hull} itself. The model also suggests where transport deserves a hearing: when the groups differ through stable characteristics, such as education or local labor demand, rather than through a transitory state that triggered treatment. This is the economic translation of Assumption~\ref{ass:hull}, and it organizes the empirical work below. When the environment shares differ between groups, the same model shows that Assumption~\ref{ass:hull} can fail through composition alone; Appendix~\ref{S-sec:environments} gives the resulting composition term and shows that a bound on the difference in environment shares yields the relaxed model of Section~\ref{sec:breakdown}, with the radius equal to that bound (Corollary~\ref{S-cor:composition}).

Covariates enter in the same way as events. If transport is imposed within cells of a discrete pre-determined covariate $X$, with weights $\lambda(x)$, the counterfactual of the treated group is $\sum_x p_1(x)\mu^0_{1,1}(\cdot\mid x)$, where $p_1$ is the treated group's covariate distribution at the policy date, and each conditional counterfactual must be a probability law. The sharp set is then the $p_1$-weighted Minkowski sum of the conditional sharp sets (Proposition~\ref{S-prop:conditional}). Imposing transport only on the unconditional distributions checks coherence of the average alone, exactly as stacking does for events. It can also fail when conditional transport holds: with covariate distributions that are stable over time, the gap between the two is the difference between the groups' covariate distributions times the amount by which the comparison group's conditional distributions drift away from mixtures of their own past (Proposition~\ref{S-prop:conditional}), and covariate distributions that change over time add a further term. Defending transport then requires conditional transport within covariate cells, a stronger assumption that needs its own historical support in every cell.

Each audit in the applications asks one of three questions: can $\pi_1$ lie in $\co\{\pi_s:s\in S\}$, judged by the implication just derived; are the historical dates free of the policy; and is group composition stable?

\begin{table}[t]\centering\small
\caption{Identifying assumptions for the untreated distribution of the treated group}\label{tab:assumptions}
\setlength{\tabcolsep}{4pt}
\begin{tabular}{@{}p{0.21\linewidth}p{0.31\linewidth}p{0.16\linewidth}p{0.25\linewidth}@{}}\toprule
Assumption & Restriction & Extrapolates beyond history & Implication checkable in data\\\midrule
Distributional parallel trends & $\nu_1=\nu_s$ for one base date $s$ & No & Candidate is a distribution; $\nu_s$ equal across $s$ if imposed for all base dates\\
Changes-in-changes & $Y_t(0)=h(U,t)$, $h$ increasing in $U$, law of $U$ stable within group & Yes, through $h$ & Transformation stable across pre-periods\\
Bounded trends & $|\int u\,d\nu_1-\int u\,d\nu_0|$ bounded by a multiple of past changes, one score & Yes, up to the multiple & None\\
Scalar historical range & $\int u\,d\nu_1$ in the range of $\int u\,d\nu_s$, one score & No & None for one score\\
Historical transport & $\nu_1\in\co\{\nu_s\}$, one weight for all events & No & Coherence; benchmark test\\\bottomrule
\end{tabular}
\begin{minipage}{0.97\linewidth}\footnotesize\emph{Notes:} A restriction ``checkable in data'' has an implication that can fail in the observed distributions. Coherence means that some transported candidate is a probability distribution (Section~\ref{sec:matters}). $\nu_t$ is the untreated selection bias, the difference between the treated and comparison groups' untreated outcome laws at date $t$. Changes-in-changes is \citet{AtheyImbens2006}; bounded trends are \citet{ManskiPepper2018} and \citet{RambachanRoth2023}; the scalar historical range is \citet{BanKedagniChenDenteh2026}.\end{minipage}
\end{table}

Table~\ref{tab:assumptions} compares historical transport with other distributional assumptions. Changes-in-changes \citep{AtheyImbens2006} writes $Y_t(0)=h(U,t)$ with $h$ strictly increasing in a scalar $U$ whose distribution within each group is constant over time. It allows the post-policy distribution to move outside anything seen before, provided the movement is a common monotone transformation; historical transport forbids such extrapolation. Conversely, the environment model generally violates changes-in-changes: when the two environments order individuals differently, the outcome is not a single monotone function of a time-invariant rank, yet historical transport holds whenever the post-policy environment shares are bracketed. Bounded trends \citep{ManskiPepper2018,RambachanRoth2023} restrict how far a scalar bias can move relative to its past movements and permit extrapolation up to a chosen multiple; historical transport permits no extrapolation but restricts the whole distribution. The choice between them is therefore an economic one. Historical transport suits settings in which the policy period resembles the historical record but conditions can reorder individuals; changes-in-changes suits settings in which ranks are stable but the policy period may lie outside the historical range. In the NSW application the post-program period lies outside the historical range, and historical transport fails there; changes-in-changes fails as well (Section~\ref{sec:nsw}).

\subsection{The sharp set of untreated laws}

For $\lambda\in\Delta_S$, define the transported candidate and the historical vertices
\begin{equation}\label{eq:candidate}
\mu_\lambda=\mu^0_{0,1}+\sum_{s\in S}\lambda_s\nu_s=\sum_{s\in S}\lambda_s v_s,
\qquad v_s=\mu^0_{0,1}+\nu_s .
\end{equation}
Each $\mu_\lambda$ is identified and has total mass one, but it need not be nonnegative. The coherent weight set is
\[
\Lambda_S(P)=\{\lambda\in\Delta_S:\ \mu_\lambda\in\PP(\mathcal Y)\},
\]
a compact convex subset of $\Delta_S$. For real outcomes write $H_\lambda(y)=\mu_\lambda((-\infty,y])$; then $\lambda\in\Lambda_S(P)$ if and only if $H_\lambda$ is nondecreasing, because the limits at $\pm\infty$ and right continuity are inherited from the observed distribution functions. Let $\mathfrak Q_S(P)$ be the set of joint distributions of potential outcomes, group labels and dates that reproduce $P$ and satisfy Assumptions~\ref{ass:basic}--\ref{ass:hull}, and write $\mu^0_{1,1,Q}$ for the untreated law of the treated at $t=1$ under $Q\in\mathfrak Q_S(P)$.

\begin{theorem}[Sharp identification]\label{thm:sharp-id}
Under Assumption~\ref{ass:basic}(i),
\[
\bigl\{\mu^0_{1,1,Q}:Q\in\mathfrak Q_S(P)\bigr\}
=\Bigl[\mu^0_{0,1}+\co\{\nu_s:s\in S\}\Bigr]\cap\PP(\mathcal Y)
=\{\mu_\lambda:\lambda\in\Lambda_S(P)\}.
\]
In particular, the model is consistent with $P$ if and only if $\Lambda_S(P)\ne\varnothing$.
\end{theorem}

Necessity is immediate. For sufficiency, take a coherent $\mu_\lambda$ and draw the missing untreated outcome of treated units from $\mu_\lambda$ using an independent uniform variable, leaving all observed variables unchanged (Appendix~\ref{S-sec:ident}). The theorem concerns the marginal untreated distribution. It leaves the dependence between $Y_1(0)$ and $Y_1(1)$, and between $Y_1(0)$ and a unit's history, unrestricted, so it identifies quantile treatment effects, which are differences between distributions, and says nothing about the distribution of individual effects. A model that restricted individual earnings dynamics could have a smaller identified set.

How far is the historical hull from the set of distributions? For a signed measure $\mu$ with Jordan decomposition $\mu=\mu^+-\mu^-$, let $b(\mu)=\mu^-(\mathcal Y)$ be its negative mass, and for signed measures $\mu,Q$ of unit mass let $d_{\TV}(\mu,Q)=\sup_{A\in\mathcal B}|\mu(A)-Q(A)|$ be their total-variation (TV) distance.

\begin{theorem}[Distance to probability space]\label{thm:coherence-distance}
For every finite signed measure $\mu$ with $\mu(\mathcal Y)=1$,
\[
\min_{Q\in\PP(\mathcal Y)}d_{\TV}(\mu,Q)=b(\mu)=\sup_{A\in\mathcal B}[-\mu(A)]_+,
\]
where $[x]_+=\max(x,0)$, and the minimum is attained at $Q^\dagger=\mu^+/\{1+b(\mu)\}$. Consequently the minimum repair $d_S=\min_{\lambda\in\Delta_S}b(\mu_\lambda)$ is the total-variation distance between the historical hull and $\PP(\mathcal Y)$, and $d_S=0$ exactly when $\Lambda_S(P)\ne\varnothing$.
\end{theorem}

\begin{proof}
Let $b=b(\mu)$ and let $N$ be a Hahn-negative set, so that $\mu(N)=-b$. For any law $Q$, $d_{\TV}(\mu,Q)\ge|\mu(N)-Q(N)|=b+Q(N)\ge b$. In addition, $\mu-Q^\dagger=\{b/(1+b)\}\mu^+-\mu^-$ is the difference of two mutually singular measures of mass $b$, so $d_{\TV}(\mu,Q^\dagger)=b$. The minimum over $\lambda$ is attained because $\lambda\mapsto b(\mu_\lambda)$ is convex and Lipschitz (Lemma~\ref{S-lem:coherence-lipschitz}).
\end{proof}

The repaired law $Q^\dagger$ is a diagnostic. It usually lies outside the historical hull, so it is not a counterfactual the model allows.

\subsection{Finite grids}\label{sec:grid}

Applications report outcomes on a finite partition $\Pi=\{A_1,\ldots,A_K\}$ of $\mathcal Y$. Let $p_{g,t}\in\mathbb R^K$ collect the cell probabilities of $\mu^0_{g,t}$, let $z_s=p_{0,1}+p_{1,s}-p_{0,s}$ be the cell vector of vertex $v_s$, and let $z_\lambda=\sum_s\lambda_sz_s$. The grid-coherent weights are
\[
\Lambda_{S,\Pi}=\{\lambda\in\Delta_S:\ z_\lambda\ge0\},
\]
and the grid repair is $d_{S,\Pi}=\min_\lambda\sum_k[-z_{\lambda,k}]_+$. A vertex is invalid when some coordinate of $z_s$ is negative. Invalid vertices do not imply an empty set: with two cells, $z_1=(-0.1,1.1)$ and $z_2=(1.1,-0.1)$ are both invalid, yet $\lambda z_1+(1-\lambda)z_2$ is a probability vector for $\lambda\in[1/12,11/12]$. Checking each historical date on its own is therefore not a test of Assumption~\ref{ass:hull}. Every bound below on a linear contrast is a linear program in the $|S|$ weights with $K$ nonnegativity constraints; the repair adds $K$ slack variables.

\section{When Coherence Matters}\label{sec:matters}

\subsection{Policy bounds}

Let $\mathcal H_S=\co\{v_s:s\in S\}$ be the signed hull and $\mathcal H_S^+=\mathcal H_S\cap\PP(\mathcal Y)$ its coherent part. If every vertex is a probability law the two coincide (Proposition~\ref{S-prop:no-gain}); otherwise $\mathcal H_S^+$ is strictly smaller. A strictly smaller set need not move a given bound, and the next result says exactly when it does.

\begin{proposition}[When coherence tightens a linear bound]\label{prop:functional-gain}
Suppose $\Lambda_S(P)\ne\varnothing$. Let $\ell$ be a linear functional that is finite on the vertices, let $a_s=\ell(v_s)$, $a_{\min}=\min_sa_s$, and $S_{\min}=\{s:a_s=a_{\min}\}$. The minimum of $\ell(\mu_\lambda)$ over $\Lambda_S(P)$ equals $a_{\min}$ if and only if some $\lambda\in\Lambda_S(P)$ puts zero weight outside $S_{\min}$. The same holds for the maximum with $S_{\max}$. On a finite grid both conditions are linear feasibility problems.
\end{proposition}

\begin{proof}
For every $\lambda\in\Delta_S$, $\ell(\mu_\lambda)-a_{\min}=\sum_s\lambda_s(a_s-a_{\min})\ge0$, with equality exactly when $\lambda$ is supported on $S_{\min}$. The minimum over the compact set $\Lambda_S(P)$ is attained, so it equals $a_{\min}$ exactly when the face $\{\lambda:\lambda_s=0,\ s\notin S_{\min}\}$ meets $\Lambda_S(P)$. Apply the argument to $-\ell$ for the maximum.
\end{proof}

The gain can reverse a sign. Take three cells, vertices $z_1=(.45,.30,.25)$ and $z_2=(.10,-.35,1.25)$, and an observed treated share of $.20$ in the first cell. With weight $t$ on $z_2$, the second cell has mass $.30-.65t$, so coherence requires $t\le6/13$. The coherent lower bound on the counterfactual-minus-observed first-cell share is $.25-.35\cdot6/13=23/260\approx.088>0$, while the signed hull reports $-.10$, a value attained only by a ``distribution'' with negative mass in the second cell. With threshold-specific weights the negative endpoint is attainable (Appendix~\ref{S-sec:calibration}), so the gain comes from the common weights.

\subsection{One law for several conclusions}\label{sec:joint}

Researchers often report several thresholds or contrasts. Under Assumption~\ref{ass:hull} these must come from one $\lambda$. The next theorem gives the dual certificate that no $\lambda$ works.

\begin{theorem}[A common witness to infeasibility]\label{thm:common-witness}
$\Lambda_S(P)=\varnothing$ if and only if there are events $A_1,\ldots,A_m$, a weight vector $q\in\Delta_m$, and $\gamma>0$ such that $f=\sum_jq_j\1_{A_j}$ satisfies $0\le f\le1$ and
\[
\max_{\lambda\in\Delta_S}\int f\,d\mu_\lambda\le-\gamma .
\]
On a $K$-cell grid the largest such margin equals the grid repair:
\begin{equation}\label{eq:witness-dual}
\min_{\lambda\in\Delta_S}\sum_{k=1}^K[-z_{\lambda,k}]_+
=\max_{f\in[0,1]^K}\min_{s\in S}\bigl(-f^{\mathsf T}z_s\bigr).
\end{equation}
\end{theorem}

\begin{proof}
If $\Lambda_S(P)=\varnothing$, every $\lambda$ gives some event negative mass. The open sets $\{\lambda:\mu_\lambda(A)<0\}$ cover the compact simplex, so finitely many events $A_1,\ldots,A_m$ suffice and $\max_\lambda\min_j\mu_\lambda(A_j)=-\gamma<0$. The payoff $\sum_jq_j\mu_\lambda(A_j)$ is bilinear on $\Delta_m\times\Delta_S$, and the minimax theorem yields $q$ with $\max_\lambda\sum_jq_j\mu_\lambda(A_j)=-\gamma$. Conversely, a nonnegative score has nonnegative expectation under any law. On a grid, $\sum_k[-z_{\lambda,k}]_+=\max_{f\in[0,1]^K}(-f^{\mathsf T}z_\lambda)$, and the minimax theorem with linearity in $\lambda$ gives \eqref{eq:witness-dual}.
\end{proof}

The witness $f$ is an outcome score, for instance a weighted count of low-wage bins, that every historical mixture predicts to have negative mean. It refutes the model as a whole rather than any single pre-trend. It is a population object; Section~\ref{sec:inference} turns it into a test.

\paragraph{Separate thresholds.} Bounding each outcome event separately is weaker than Assumption~\ref{ass:hull}. With categories coded $0,1,2$, valid vertices $v_1=(.2,.6,.2)$ and $v_2=(.4,.1,.5)$, and the mean category $m(p)=p_1+2p_2$, common weights give $m\in[1.0,1.1]$. Imposing the historical range separately on every event admits $(.2,.3,.5)$ and $(.4,.4,.2)$, neither a mixture of $v_1$ and $v_2$, and $m\in[.8,1.3]$. For a single prespecified score, the scalar restriction applied to that score already gives $[1.0,1.1]$; common weights matter when several reported quantities must describe one population.

\subsection{Stacked events and composite counterfactuals}\label{sec:events}

Many applications pool several policy events, such as minimum-wage increases in different states, and report an average effect. Let events $e=1,\ldots,E$ have aggregation weights $\omega\in\Delta_E$ and untreated laws $\mu^0_{e,g,t}$, with vertices $v_{e,s}=\mu^0_{e,0,1}+\mu^0_{e,1,s}-\mu^0_{e,0,s}$, and write $\mu_{e,\lambda}=\sum_s\lambda_sv_{e,s}$, $\bar\mu_\lambda=\sum_e\omega_e\mu_{e,\lambda}$, and $z_{e,\lambda}$ for the cell vector of $\mu_{e,\lambda}$ on a grid. Two models are in use, often without distinction. Under eventwise transport, Assumption~\ref{ass:hull} holds for every event with one common $\lambda$, and the coherent set is $\Lambda^{\rm ev}=\{\lambda:\mu_{e,\lambda}\in\PP(\mathcal Y)\text{ for all }e\}$. Under stacked transport, it is imposed only on the averaged laws, and the coherent set is $\Lambda^{\rm st}=\{\lambda:\bar\mu_\lambda\in\PP(\mathcal Y)\}$.

\begin{theorem}[Stacking can hide infeasibility]\label{thm:stacking}
(i) $\Lambda^{\rm ev}\subseteq\Lambda^{\rm st}$, so for the averaged counterfactual the eventwise sharp set $\{\bar\mu_\lambda:\lambda\in\Lambda^{\rm ev}\}$ is contained in the stacked one. (ii) The inclusion can be strict, and $\Lambda^{\rm st}$ can be nonempty when $\Lambda^{\rm ev}$ is empty. (iii) On a finite grid, $\Lambda^{\rm ev}=\Lambda^{\rm st}$ if and only if $\min_{\lambda\in\Lambda^{\rm st}}z_{e,\lambda,k}\ge0$ for every event $e$ and cell $k$, a set of $EK$ linear programs.
\end{theorem}

\begin{proof}
(i) A convex combination of probability laws is a probability law. (ii) With one historical date, two events and two cells, $z_1=(-0.1,1.1)$ and $z_2=(0.3,0.7)$ average to $(0.1,0.9)$ with equal weights, so $\Lambda^{\rm st}=\Delta_S$ while $\Lambda^{\rm ev}=\varnothing$. (iii) If some $\min_{\Lambda^{\rm st}}z_{e,\lambda,k}<0$, its minimizer is in $\Lambda^{\rm st}\setminus\Lambda^{\rm ev}$; if all minima are nonnegative, every stacked-coherent $\lambda$ is eventwise coherent.
\end{proof}

The theorem is elementary, but its consequence for practice is not. Stacking is the standard way to summarize several policy events \citep{CengizEtAl2019,CallawaySantAnna2021}, and a stacked distributional bound is usually read as a bound on the average of the events' effects. When $\Lambda^{\rm st}\ne\varnothing=\Lambda^{\rm ev}$, that reading fails under the eventwise model: every historical weight the stacked model admits gives at least one event a counterfactual with negative probabilities, so the averaged counterfactual is a composite that no event-by-event application of the model produces. This is a statement about models, not a logical impossibility. Pooling is the right target when the object of interest is the treated population as a whole and the analyst is willing to assume transport only for that population, for instance because state-level histories are too noisy to discipline state-level counterfactuals. The bound then concerns the pooled population only and says nothing about any event. Our recommendation is to compute the $EK$ programs in part (iii) before interpreting a stacked distributional bound, and to report the eventwise bound whenever the two differ. In the minimum-wage application the stacked set is nonempty, but no common weight makes all five event-level counterfactuals valid, and, under the third model described next, neither do weights chosen separately for each state.

Weights that differ across events give a third model, the one to use when events are believed to draw on their histories differently: each event has its own $\lambda_e\in\Delta_S$ with $\mu_{e,\lambda_e}\in\PP(\mathcal Y)$, and the sharp set for the averaged law is the Minkowski average $\sum_e\omega_e\{\mu_{e,\lambda_e}:\mu_{e,\lambda_e}\in\PP(\mathcal Y)\}$, so bounds on an averaged contrast are averages of event-level bounds. It is nonempty exactly when every event's own coherent set is, it contains the eventwise set, and it need not contain or be contained in the stacked one (Proposition~\ref{S-prop:event-specific}). Between the first two lies a family indexed by $\delta\ge0$: event weights $\lambda_e$ with $\|\lambda_e-\bar\lambda\|_1\le\delta$ for a common $\bar\lambda\in\Delta_S$. Its sharp bounds are linear programs, widen with $\delta$, equal the eventwise bounds at $\delta=0$ and the event-specific ones at $\delta=2$, and the smallest $\delta$ at which every event is coherent is itself a linear program (Proposition~\ref{S-prop:approx-common}). It measures how much weight heterogeneity the data require.

The choice among these models changes conclusions, not only interpretations. In a population-level simulation in which each of three events has its own true weight and a valid counterfactual (Appendix~\ref{S-sec:pooling-sim}), the event-specific bound always contains the true average effect. The stacked bound excludes it in 29 percent of designs, and the common-weight eventwise model is empty or excludes it in 72 percent. Stacking is thus a legitimate assumption only when the events are believed to share one weight on their pooled history; it is not a harmless summary of event-level transport.

\subsection{Projection}\label{sec:projection}

A tempting repair projects each invalid vertex onto the simplex and mixes the projections. This replaces the identifying restriction with a different one.

\begin{proposition}[Projection does not commute with coherent transport]\label{prop:projection-noncommute}
On three categories there are probability vectors $a,b_s,c_s$ ($s=1,2$) such that the vertices $v_s=a+b_s-c_s$ have a nonempty coherent hull that is disjoint from the convex hull of their Euclidean projections onto the simplex. Projecting the whole signed hull can strictly enlarge the coherent hull.
\end{proposition}

\begin{proof}
Take $a=(\tfrac15,\tfrac25,\tfrac25)$, $b_1=(0,\tfrac3{10},\tfrac7{10})$, $c_1=(\tfrac12,0,\tfrac12)$, $b_2=(\tfrac12,\tfrac3{10},\tfrac15)$, $c_2=(0,0,1)$. Then $v_1=(-\tfrac3{10},\tfrac7{10},\tfrac35)$ and $v_2=(\tfrac7{10},\tfrac7{10},-\tfrac25)$. With weight $t$ on $v_1$ the mixture is $(\tfrac7{10}-t,\tfrac7{10},t-\tfrac25)$, coherent for $t\in[\tfrac25,\tfrac7{10}]$, with middle mass always $\tfrac7{10}$. The projections are $(0,\tfrac{11}{20},\tfrac9{20})$ and $(\tfrac12,\tfrac12,0)$, whose mixtures have middle mass in $[\tfrac12,\tfrac{11}{20}]$. Projection fixes coherent mixtures but sends $v_1$ outside the coherent hull.
\end{proof}

Three repairs are in use and should be distinguished: Euclidean projection of each candidate onto the simplex, monotone rearrangement of an estimated distribution function \citep{ChernozhukovFernandezValGalichon2010}, and minimum-distance projection of the whole hull. The first two repair each candidate before mixing, and the repair of an invalid vertex depends on the chosen distance; the third, in total variation, is the relaxation of Section~\ref{sec:breakdown} at radius $d_S$. Projection can still serve as a declared sensitivity analysis. The relaxation of Section~\ref{sec:breakdown} repairs the mixture instead, by the least total variation needed, and its bounds are sharp for an explicit model (Proposition~\ref{prop:breakdown}). We recommend it over projection when a repaired model is wanted.

\subsection{Grids}\label{sec:refinement}

Assumption~\ref{ass:hull} restricts a distribution, and any grid sees only part of it. Let $\Pi_n$ be nested finite partitions, with grid-coherent sets $\Lambda_{S,n}$ and grid repairs $d_{S,n}$.

\begin{proposition}[Refinement]\label{prop:partition-refinement}
$\Lambda_S(P)\subseteq\Lambda_{S,n+1}\subseteq\Lambda_{S,n}$ and $d_{S,n}\le d_{S,n+1}\le d_S$. If the partitions generate $\mathcal B$, then $\Lambda_S(P)=\bigcap_n\Lambda_{S,n}$ and $d_{S,n}\uparrow d_S$.
\end{proposition}

Incoherence on any grid refutes the model, bounds on a contrast that is measurable on a grid weakly tighten when the grid is refined, and coherence on one grid says nothing about finer ones, nor about the law on $\mathcal Y$. A coarse grid can hide a lot.

\begin{proposition}[Coarsening]\label{prop:partition-mesh}
(i) If some cell of $\Pi$ contains two points, then $d_S\le1$ for every observed law, and the bound is attained, with one historical date, by an observed law for which $d_{S,\Pi}=0$. (ii) Suppose $\mathcal Y=[a,b]$, each vertex has a density that is Lipschitz with constant $L$, and every cell of $\Pi$ is an interval of length at most $h$. Then
\[
0\le d_S-d_{S,\Pi}\le\frac{L(b-a)h}{8},
\]
and the constant $1/8$ cannot be improved.
\end{proposition}

In part (i), with point masses $\delta_x,\delta_y$ at two points of one cell, $\mu^0_{0,1}=\mu^0_{1,s}=\delta_x$ and $\mu^0_{0,s}=\delta_y$ give the candidate $2\delta_x-\delta_y$, which looks coherent on the grid but has negative mass one. Part (ii) shows what smoothness buys. Smoothness does not make coherence irrelevant: on $[0,1]$, valid historical densities $1\pm\sin(2\pi x)$ produce the vertex density $1+2\sin(2\pi x)$, mixtures with the uniform are coherent only for weights up to $1/2$, and this halves the upper bound on $\int\sin(2\pi x)\,d\mu$ on every sufficiently fine grid (Appendix~\ref{S-sec:smooth}). Wage distributions with heaping at round numbers and a spike at the minimum wage need not have smooth densities, so part (ii) is a guide, not a guarantee, for the application.

The grid is chosen by the analyst, and one may worry that it can be chosen to obtain a desired answer. Proposition~\ref{prop:partition-refinement} limits the scope for this. Coarsening can never narrow a bound; it can only conceal infeasibility. The risk is therefore false acceptance of the model, and that risk can be removed by testing coherence on every interval grid at once.

\begin{proposition}[Grid choice]\label{prop:grid-choice}
Let $\mathcal Y\subseteq\mathbb R$, let $H_s(y)=v_s((-\infty,y])$ be the vertex distribution functions and $H_\lambda=\sum_s\lambda_sH_s$, let $\widehat H_s$ estimate $H_s$ with $\widehat H_\lambda=\sum_s\lambda_s\widehat H_s$, and suppose $\max_s\norm{\widehat H_s-H_s}_\infty\le\varepsilon$, where $\norm{\cdot}_\infty$ is the supremum norm. Then for every partition $\Pi$ of $\mathcal Y$ into intervals, including partitions chosen after seeing the data,
\[
\Lambda_{S,\Pi}\subseteq\{\lambda\in\Delta_S:\ \widehat z^{\,\Pi}_{\lambda}\ge-2\varepsilon\},
\]
where $\widehat z^{\,\Pi}_\lambda$ are the estimated cell masses of $\widehat H_\lambda$. In particular, a coherence test calibrated on the supremum-norm event has the same size whatever grid is used.
\end{proposition}

\begin{proof}
For an interval cell $(u,w]$, $|\widehat H_\lambda(w)-\widehat H_\lambda(u)-H_\lambda(w)+H_\lambda(u)|\le2\norm{\widehat H_\lambda-H_\lambda}_\infty\le2\varepsilon$, using Lemma~\ref{lem:vertex} below for the last step; open and half-open cells are handled by left limits.
\end{proof}

Since $\Lambda_S(P)\subseteq\Lambda_{S,\Pi}$ for every partition and the containment in Proposition~\ref{prop:grid-choice} holds for all interval partitions on one event, on that event $\Lambda_S(P)$ lies in the intersection of the right-hand sides,
\[
\widehat\Lambda_\infty(\varepsilon)=\bigl\{\lambda\in\Delta_S:\ \widehat H_\lambda(w)-\widehat H_\lambda(u)\ge-2\varepsilon\ \text{for all }u<w,\ \ -\varepsilon\le\widehat H_\lambda\le1+\varepsilon\bigr\},
\]
the set of weights whose estimated distribution function never falls by more than $2\varepsilon$ and stays within $\varepsilon$ of $[0,1]$; the range condition holds because every coherent $H_\lambda$ takes values in $[0,1]$. By the exact formula for the supremum-norm distance to the set of distribution functions (Theorem~\ref{S-thm:cdf-distance}), these are precisely the weights for which $\widehat H_\lambda$ lies within $\varepsilon$ of some distribution function, so $\widehat\Lambda_\infty(\varepsilon)$ is the set of weights admitted by the coherence test of Theorem~\ref{thm:coherent-inference}. A confidence region for a contrast, obtained by optimizing it over $\widehat\Lambda_\infty(\varepsilon)$ and adding the band for the contrast itself as in Section~\ref{sec:inference}, is therefore valid, and it is at least as narrow as the region built in the same way on any single grid. This gives a concrete procedure. (1) Fix in advance a reporting grid that contains every cutoff to be reported and the outcome's natural cells, such as zero earnings, and report the plug-in bounds on it. (2) Run the studentized coherence test of Section~\ref{sec:inference} on that grid. It is a falsification test: incoherence on any grid refutes the model, so a rejection on the prespecified grid is a rejection of the model, whatever coarser grids show. Its critical value is calibrated for that grid only. (3) Report the confidence region built on $\widehat\Lambda_\infty(\varepsilon)$, with $\varepsilon$ calibrated on the supremum norm; it involves no grid choice. Because $\widehat H_\lambda$ is a step function with jumps at the observed outcome values, $\widehat\Lambda_\infty(\varepsilon)$ is computed exactly by one linear program: adding one variable per support point for the running maximum of $\widehat H_\lambda$ turns the condition on all pairs $u<w$ into a number of linear inequalities proportional to the number of support points (Appendix~\ref{S-sec:gridfree}). The program has $|S|+n$ variables and $5n-1$ sparse inequalities for $n$ support points; with the 22,339 distinct earnings values of the NSW and CPS samples each program takes about seven seconds with a standard solver. For larger $n$, imposing the conditions at a subset of support points enlarges the set, so the region stays valid. We do not recommend choosing the grid that gives the narrowest bound among those not rejected by the test, because that rule selects the grid on the outcome of the test.

\section{Quantiles and Policy Rankings}\label{sec:quantile}

For real outcomes, let $\cF_S(P)=\{H_\lambda:\lambda\in\Lambda_S(P)\}$ be nonempty, and define the envelopes $\underline F_S(y)=\min_{\lambda\in\Lambda_S(P)}H_\lambda(y)$ and $\overline F_S(y)=\max_{\lambda\in\Lambda_S(P)}H_\lambda(y)$ and the quantile function $Q_F(\tau)=\inf\{y:F(y)\ge\tau\}$.

\begin{theorem}[Distributional and quantile sharpness]\label{thm:quantile}
(i) $\underline F_S$ and $\overline F_S$ are distribution functions. (ii) $\underline F_S\in\cF_S(P)$ if and only if $\bigcap_y\argmin_{\lambda\in\Lambda_S(P)}H_\lambda(y)\ne\varnothing$, and similarly for $\overline F_S$. (iii) For $\tau\in(0,1)$,
\[
\min_{\lambda\in\Lambda_S(P)}Q_{H_\lambda}(\tau)=Q_{\overline F_S}(\tau),\qquad
\sup_{\lambda\in\Lambda_S(P)}Q_{H_\lambda}(\tau)=Q_{\underline F_S}(\tau),
\]
and the supremum is attained when $\Lambda_S(P)$ is a polytope, in particular on any finite grid.
\end{theorem}

Part (ii) separates pointwise from functional sharpness: one mixture may attain the lower envelope at some thresholds and another elsewhere, so the envelope need not be an admissible distribution. Part (iii) gives the quantile extrema, but the set $\cQ_S(\tau)=\{Q_{H_\lambda}(\tau):\lambda\in\Lambda_S(P)\}$ need not be an interval. If $H_p=p\delta_0+(1-p)\delta_1$ with $p\in[0.4,0.6]$, the median is $1$ for $p<1/2$ and $0$ for $p\ge1/2$, so $\cQ_S(1/2)=\{0,1\}$; the interval $[0,1]$ is a valid but conservative report. Proposition~\ref{S-prop:quantile-interval} gives conditions under which $\cQ_S(\tau)$ is an interval. These are bounds on quantiles of the counterfactual distribution; differences of quantiles are not quantiles of individual treatment effects, which the model does not identify.

Many policy questions are linear in the counterfactual. For a bounded score $u$ and the observed treated distribution $F_T=\mu_{1,1}$, let $W_u(\lambda)=\int u\,dF_T-\int u\,dH_\lambda$.

\begin{proposition}[Policy contrasts and dominance]\label{prop:policy}
The sharp set of $W_u$ is the interval with endpoints $\int u\,dF_T-\max_{\Lambda_S(P)}\int u\,dH_\lambda$ and $\int u\,dF_T-\min_{\Lambda_S(P)}\int u\,dH_\lambda$, both attained. The treated distribution first-order stochastically dominates every admissible counterfactual if and only if $F_T\le\underline F_S$ pointwise.
\end{proposition}

The proof is immediate from Theorem~\ref{thm:sharp-id} and linearity. The optimizing mixture may differ across scores, so intervals for several scores are projections of a joint set; a statement about several scores at once requires optimizing over $\Lambda_S(P)$ jointly.

\section{Relaxation and Breakdown}\label{sec:breakdown}

An empirical bound is only as convincing as its robustness to departures from Assumption~\ref{ass:hull}. We use the neighborhood
\[
\mathcal M_S(\tau)=\{Q\in\PP(\mathcal Y):\ d_{\TV}(Q,\mu_\lambda)\le\tau\ \text{for some }\lambda\in\Delta_S\},\qquad\tau\ge0 .
\]
By Theorem~\ref{thm:coherence-distance}, $\mathcal M_S(\tau)$ is nonempty exactly when $\tau\ge d_S$, and $\mathcal M_S(0)$ is the sharp set. When the reference $\mu_\lambda$ is a probability law, $\tau$ is the largest share of the population whose untreated outcome may differ from a historically transported one. When $\mu_\lambda$ is signed, part of $\tau$ is spent removing negative mass, and $\tau$ should be read as a distance rather than as a share of people.

Total variation is one choice among several: a Kolmogorov or Wasserstein ball around the historical hull is also polyhedral on a grid, so the programs below carry over. Two radii should be kept apart. The \textbf{minimum repair} $d_S$ is where the relaxed model first becomes nonempty; the \textbf{sign breakdown radius} defined next is where a conclusion is lost. Consider the effect $\theta=g-Q(A)$ on an outcome share, where $A$ is a nonempty proper event, $g=\mu_{1,1}(A)$ is the treated share, and $Q$ is the counterfactual. The sign breakdown radius $\tau^{\rm sign}$ is the smallest $\tau$ at which $\mathcal M_S(\tau)$ contains a law with $\theta\ge0$ (or $\theta\le0$, for the opposite conclusion). For a signed vector $z$ on a grid and a set of cells $A$, write $z^+(A)=\sum_{k\in A}[z_k]_+$ and $b(z)=\sum_k[-z_k]_+$.

\begin{proposition}[Exact breakdown radius]\label{prop:breakdown}
For a unit-mass signed $z$ on a finite grid, a nonempty proper set of cells $A$, and $g\in[0,1]$,
\[
\min\{d_{\TV}(q,z):\ q\text{ a probability vector},\ q(A)\le g\}=\max\{b(z),\ z^+(A)-g\}.
\]
Hence the radius at which a nonnegative effect becomes admissible is
\[
\tau^{\rm sign}=\min_{\lambda\in\Delta_S}\max\bigl\{b(z_\lambda),\ z_\lambda^+(A)-g\bigr\},
\]
a linear program in which both terms are evaluated at the same $\lambda$. For estimates $\widehat z_s$ and $\widehat g$, if $\max_s d_{\TV}(\widehat z_s,z_s)\le\varepsilon$ and $|\widehat g-g|\le\delta$, the plug-in repair is within $\varepsilon$ of $d_{S,\Pi}$ and $|\widehat\tau^{\rm sign}-\tau^{\rm sign}|\le\varepsilon+\delta$. Consequently, if this event has probability at least $1-\alpha$, then $[\widehat\tau^{\rm sign}-\varepsilon-\delta,\ \widehat\tau^{\rm sign}+\varepsilon+\delta]$ is a $1-\alpha$ confidence interval for $\tau^{\rm sign}$.
\end{proposition}

\begin{proof}
Lower bound: $d_{\TV}(q,z)\ge b(z)$ by Theorem~\ref{thm:coherence-distance}, and with $B=A\cap\{z>0\}$, $d_{\TV}(q,z)\ge z(B)-q(B)\ge z^+(A)-g$. Upper bound: remove mass $b(z)$ from the positive cells, starting with those in $A$, and set negative cells to zero; the result is a law at distance $b(z)$ with $q(A)\le g$ whenever $z^+(A)-b(z)\le g$. Otherwise remove $z^+(A)-g$ from $A$ and add $z^+(A)-g-b(z)$ to $A^c$, which is possible because $A$ is proper; the distance is $z^+(A)-g$. Stability: $b$ and $z\mapsto z^+(A)$ are $1$-Lipschitz in total variation, the total-variation error of a mixture is at most the largest vertex error, and minimization preserves Lipschitz bounds.
\end{proof}

\begin{figure}[t]\centering
\includegraphics[width=\linewidth]{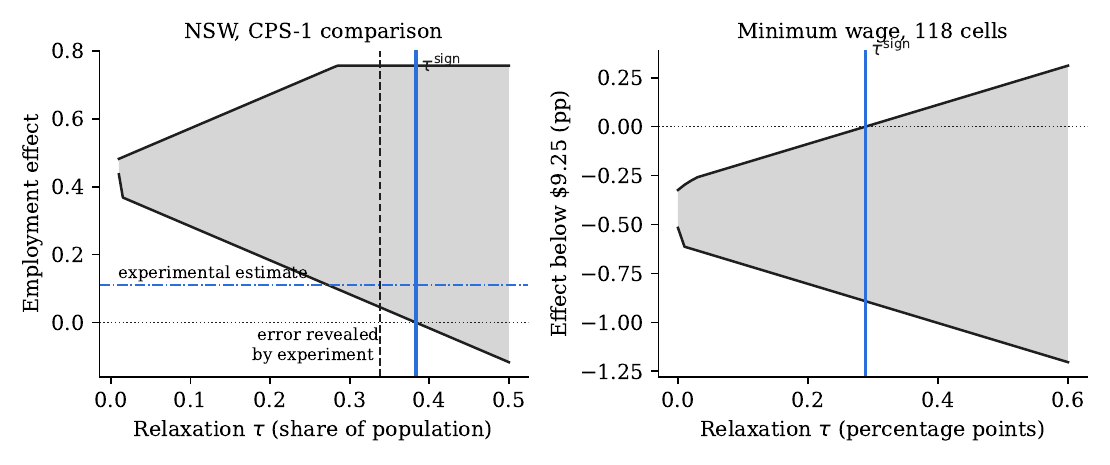}
\caption{Relaxed bounds as the departure from historical transport grows}\label{fig:breakdown}
\begin{minipage}{0.97\linewidth}\footnotesize\emph{Notes:} Shaded: sharp interval for the effect under $\mathcal M_S(\tau)$. Left: NSW, CPS-1 comparison group, seven cells; the interval is empty below the minimum repair $\tau=0.009$. Dashed: total-variation distance from the experimental control distribution to the historical hull. Right: minimum-wage application, 118 cells; the coherent set is nonempty at $\tau=0$.\end{minipage}
\end{figure}

Minimizing the two terms separately understates the radius (Appendix~\ref{S-sec:breakdown-supp}). The confidence interval inherits the coverage of the band: exact with a finite-sample band for the cell masses, asymptotic with a bootstrap one. Figure~\ref{fig:breakdown} shows how the relaxed bound widens with $\tau$ in the NSW and minimum-wage applications. The radius has to be judged against departures one finds plausible, and the NSW application supplies a calibration: there the experiment reveals a departure of about a third of the probability mass. A radius larger than the realized departure means only that the sign survives; it does not validate the unrelaxed bound, whose size can be badly wrong. Nor should the difference $\tau^{\rm sign}-d_S$ be read as the departure needed after the model has been repaired: the two programs choose their weights separately.

Total variation treats moving a worker from zero earnings to \$30,000 like moving her to \$5,000. When such moves are implausible, a Wasserstein ball, which charges for the distance moved, is the natural alternative, and on a grid it keeps the linear-programming form. In the NSW data with CPS it reverses the comparison above: the employment effect loses its sign once earnings may move by an average of \$1,053, while the experiment reveals a departure of \$2,630 (Appendix~\ref{S-sec:w1}). Restrictions on which moves are admissible, such as monotone treatment response, which requires the counterfactual distribution function to lie above the treated one, are also linear and can be added to every program in the paper.

\paragraph{Exclusion restrictions.} Economic theory sometimes says where a policy cannot act. A minimum wage near \$9 should not change the share of the population employed at \$18 an hour. If $\mu^1_{1,1}(B)=\mu^0_{1,1}(B)$ for events $B$ in a known class, the sharp set becomes $\{\mu_\lambda:\lambda\in\Lambda_S(P),\ \mu_\lambda(B)=\mu_{1,1}(B)\text{ for all such }B\}$, which adds linear equalities on a grid (Proposition~\ref{S-prop:anchoring}). The restriction is an additional identifying assumption, not a fact: wage spillovers and changes in composition can move the upper tail. It both sharpens the bounds and can be tested, since an empty set means that transport and the exclusion cannot both hold, and it can be checked in placebo periods.

\section{Inference}\label{sec:inference}

\subsection{Sampling and what the theory covers}

We consider three settings and keep them apart. In the first, the treated, comparison and (if available) benchmark samples are independent, and within each the units are independent draws with complete outcome histories; outcomes of the same unit may be dependent across dates. Here the procedures below are valid in finite samples when their critical values come from the Dvoretzky--Kiefer--Wolfowitz inequality or from the sample-splitting construction of Appendix~\ref{S-sec:split-outer}. In the second, the same structure holds with bootstrap or multiplier critical values, which are valid as the sample sizes grow for a fixed data-generating process; Appendix~\ref{S-sec:uniform} gives uniform validity under an explicit coupling condition. In the third, outcomes are cells of estimated distributions without a known joint sampling distribution, as with state--quarter wage bins built from the CPS. The theory does not cover this case, and we report no confidence levels for it. When units are sampled in independent clusters, such as states or villages, with dependence of any form within them, the coverage theorem below applies on a fixed grid once the band is calibrated over clusters rather than units: Appendix~\ref{S-sec:cluster} gives a finite-sample radius and a uniformly valid Gaussian calibration. Weighted survey estimators and designs with a handful of treated clusters remain outside the theory; with five treated states the finite-sample cluster radius is about one, the scale of the cells. Table~\ref{tab:guarantees} summarizes what each procedure guarantees.

\begin{table}[t]\centering\small
\caption{What the inference procedures guarantee}\label{tab:guarantees}
\setlength{\tabcolsep}{4pt}
\begin{tabular}{@{}p{0.30\linewidth}p{0.25\linewidth}p{0.40\linewidth}@{}}\toprule
Procedure & Critical value & Guarantee\\\midrule
Confidence sets, coherence test, benchmark test & DKW--Massart radii & Finite-sample level, independent units and samples\\
Confidence sets for one score & Sample split & Finite-sample level, bounded scores, independent pilot and estimation samples\\
Confidence sets, grid-free region, breakdown interval & Bootstrap of complete histories & Asymptotic for a fixed distribution; uniform under the coupling condition of Appendix~\ref{S-sec:uniform}\\
Studentized coherence test & Bootstrap, estimated standard errors & Asymptotic; conservative in simulations\\
Confidence sets and coherence test, clustered data, fixed grid & Hoeffding over clusters & Finite-sample level, independent clusters, equal cluster weights\\
\quad same & Gaussian, cluster covariance & Uniform asymptotic level, bounded reports, cluster variances bounded below\\
Cells of estimated distributions (minimum wages) & None & None; results are descriptive\\\bottomrule
\end{tabular}
\end{table}

For dates $t$ and groups $g$, let $F_{g,t}$ be the distribution function of $Y_t$ given $G=g$, $\widehat F_{g,t}$ its empirical counterpart from $N_g$ units, $H_s=F_{0,1}+F_{1,s}-F_{0,s}$, $\widehat H_s$ the plug-in, and $\widehat H_\lambda=\sum_s\lambda_s\widehat H_s$.

Imposing exact coherence on the plug-in is a mistake. With continuous outcomes, each $\widehat F_{0,s}$ almost surely jumps where no positively signed empirical distribution function jumps, so $\widehat H_\lambda$ decreases somewhere for every $\lambda$ and the empirical coherent set is empty (Proposition~\ref{S-prop:empirical-incoherence}). Even on a fixed grid, when the population coherent set is a single point, the plug-in set is empty about half the time (Section~\ref{sec:sims}). Point-estimate incoherence is not evidence against the model.

\subsection{Vertex reduction and confidence sets}

\begin{lemma}[Vertex reduction]\label{lem:vertex}
For any seminorm $\rho$ and signed measures $e_s$, $s\in S$,
\[
\sup_{\lambda\in\Delta_S}\rho\Bigl(\sum_{s\in S}\lambda_se_s\Bigr)=\max_{s\in S}\rho(e_s).
\]
\end{lemma}

Indeed, $\rho(\sum_s\lambda_se_s)\le\sum_s\lambda_s\rho(e_s)\le\max_s\rho(e_s)$, with equality at a vertex. The lemma means that no optimal historical weight ever needs to be estimated, which matters because estimated optima are irregular at ties. Let $\Dcal$ be the set of distribution functions on $\mathbb R$ and define the coherence tube
\[
\mathfrak C_N(\varepsilon)=\bigcup_{\lambda\in\Delta_S}\bigl\{F\in\Dcal:\ \norm{F-\widehat H_\lambda}_\infty\le\varepsilon\bigr\}.
\]

\begin{theorem}[Coherent inference]\label{thm:coherent-inference}
If $\Pp\bigl(\max_{s\in S}\norm{\widehat H_s-H_s}_\infty\le\varepsilon_N\bigr)\ge1-\alpha$ for a measurable $\varepsilon_N$, then $\Pp\{\cF_S(P)\subseteq\mathfrak C_N(\varepsilon_N)\}\ge1-\alpha$. The coherence test that rejects when $\min_{\lambda}\inf_{F\in\Dcal}\norm{\widehat H_\lambda-F}_\infty>\varepsilon_N$ has size at most $\alpha$ whenever $\Lambda_S(P)\ne\varnothing$.
\end{theorem}

\begin{proof}
On the event, every $H_\lambda\in\cF_S(P)$ satisfies $\norm{H_\lambda-\widehat H_\lambda}_\infty\le\max_s\norm{H_s-\widehat H_s}_\infty\le\varepsilon_N$ by Lemma~\ref{lem:vertex}. If $\lambda_0\in\Lambda_S(P)$, then $H_{\lambda_0}\in\Dcal$ lies within $\varepsilon_N$ of $\widehat H_{\lambda_0}$ on the same event. Measurability is shown in Appendix~\ref{S-sec:emp}.
\end{proof}

The radius $\varepsilon_N$ can be calibrated in two ways. With $M=1+2|S|$ primitive distribution functions, the Dvoretzky--Kiefer--Wolfowitz inequality with Massart's constant and a union bound give the finite-sample radius
\[
\varepsilon_N^{\rm DKW}=\max_{s\in S}\bigl\{r_{0,1}+r_{1,s}+r_{0,s}\bigr\},\qquad r_{g,t}=\Bigl\{\frac{\log(2M/\alpha)}{2N_{g,t}}\Bigr\}^{1/2},
\]
where $N_{g,t}$ is the number of observations behind $\widehat F_{g,t}$ (the group size, with complete histories), valid conditional on sample sizes and without restrictions on dependence across dates \citep{DvoretzkyKieferWolfowitz1956,Massart1990}. A sharper radius is the bootstrap quantile of $\max_s\norm{\widehat H^*_s-\widehat H_s}_\infty$, resampling complete histories within each sample (Appendix~\ref{S-sec:emp}). On a grid, the same logic applies cell by cell. Studentizing helps: if $\sigma_{s,k}$ is the standard error of $\widehat z_{s,k}$ and $c$ bounds $\max_{s,k}|\widehat z_{s,k}-z_{s,k}|/\sigma_{s,k}$ with probability $1-\alpha$, then every $\lambda\in\Lambda_{S,\Pi}$ satisfies $\widehat z_{\lambda,k}\ge-c\sum_s\lambda_s\sigma_{s,k}$ for all $k$, a linear constraint in $\lambda$. With a bootstrap critical value and estimated standard errors, this guarantee is asymptotic. The studentized coherence test rejects when no $\lambda$ satisfies it, and its $p$-value is the bootstrap probability that the maximal studentized error exceeds the smallest $c$ for which some $\lambda$ does. Rare cells, where infeasibility usually appears, then receive narrow bands. This is the test we recommend as the default diagnostic whenever no benchmark is available. Coherence is the statement that some $\lambda\in\Delta_S$ satisfies the $K$ moment inequalities $z_{\lambda,k}\ge0$, so generic procedures for moment inequalities with a nuisance parameter \citep{ChernozhukovHongTamer2007,AndrewsSoares2010,RomanoShaikhWolf2014} apply to the same restrictions and can be less conservative near the boundary. What the structure here adds is the vertex reduction, which makes the band for $|S|$ candidates a band for all mixtures, and finite-sample critical values. Because the cells are finitely many, or the indicator class of half-lines in the grid-free case, the empirical processes involved are Donsker, and bootstrap validity follows from standard arguments under independent sampling (Appendix~\ref{S-sec:emp}). Two facts about power hold without further assumptions (Proposition~\ref{S-prop:cluster-power}). First, the test may use the one-sided critical value of $\max_{s,k}[-(\widehat z_{s,k}-z_{s,k})]_+$, which is never larger than the two-sided one. Second, the statistic has an exact local limit (Theorem~\ref{S-thm:local-limit}). The deficit is directionally differentiable, and along local alternatives $Z_0+HN^{-1/2}$ the scaled statistic converges to the value $D(H+W)$ of a small linear program in the cells that bind at the coherent weights, with the weight free to move within the simplex. Three facts follow. The null limit is stochastically smaller than the maximum over all vertex cells, which is why vertex critical values are conservative. When the coherent set contains a weight with no binding cell, every local perturbation is still coherent and the limit is zero: such points lie in the interior of the null, where the test is conservative, and local power requires that every coherent weight have a binding cell. And critical values estimated from the local program, as in generalized moment selection, are valid pointwise and more powerful, but their uniform validity needs tuning conditions we have not established; in our cluster simulations a crude version over-rejects (Appendix~\ref{S-sec:cluster}). Confidence regions for a contrast follow by optimizing it over the same set of $\lambda$, adding the band for the contrast itself; for quantile effects the treated distribution must be covered jointly (Corollary~\ref{S-cor:qte-projection}).

\subsection{Testing transport against a benchmark}

Coherence is a necessary implication of Assumption~\ref{ass:hull}. A much stronger check is available when an independent source, such as a randomized control group, identifies the missing distribution. Let $F_R$ be that benchmark distribution function and $\widehat F_R$ its estimate from an independent sample.

\begin{proposition}[Benchmark test and confidence bound on the transport error]\label{prop:benchmark-test}
Let $D=\min_{\lambda\in\Delta_S}\norm{F_R-H_\lambda}_\infty$ and $\widehat D=\min_{\lambda\in\Delta_S}\norm{\widehat F_R-\widehat H_\lambda}_\infty$. If
\[
\Pp\Bigl(\max_{s\in S}\bigl\|(\widehat F_R-\widehat H_s)-(F_R-H_s)\bigr\|_\infty\le c\Bigr)\ge1-\alpha,
\]
then $\Pp(|\widehat D-D|\le c)\ge1-\alpha$. Under Assumption~\ref{ass:hull}, $D=0$; the test that rejects when $\widehat D>c$ has size at most $\alpha$, and $\widehat D-c$ is a $1-\alpha$ lower confidence bound on $D$. The choice $c=r_R+\varepsilon_N^{\rm DKW}$, where $r_R$ is the radius for the benchmark sample size $N_R$ and every radius is computed using $M+1$ in place of $M$, is valid in finite samples.
\end{proposition}

\begin{proof}
For each $\lambda$, $\bigl|\norm{\widehat F_R-\widehat H_\lambda}_\infty-\norm{F_R-H_\lambda}_\infty\bigr|\le\norm{\sum_s\lambda_s\{(\widehat F_R-\widehat H_s)-(F_R-H_s)\}}_\infty\le c$ on the event, by Lemma~\ref{lem:vertex}. Minimizing both sides over $\lambda$ preserves the inequality.
\end{proof}

$\widehat D$ is a linear program over the pooled sample points. The test is consistent against any fixed $D>0$, and the lower confidence bound turns a rejection into a statement about magnitude. The benchmark must identify the treated group's untreated distribution: randomization within the treated population's sampling frame does that, while a selected experimental subsample identifies it only for the population from which the subsample is drawn. Without an experiment, a period in which the policy cannot yet have acted supplies a natural benchmark: the treated group's observed distribution in that period is its untreated distribution, so the proposition applies with that period as the outcome date and its own historical dates. Such a benchmark tests transport at a different date and horizon, not at the policy date. A rejection is evidence that the historical biases are unstable, and therefore strong evidence against relying on the assumption; a non-rejection is weaker support.

\subsection{Simulations}\label{sec:sims}

We study size, power and coverage in three designs on six outcome cells, calibrated to earnings data with a mass at zero (Appendix~\ref{S-sec:sims}), using sample sizes from $(N_1,N_0)=(200,2{,}000)$, similar to NSW, to $(5{,}000,20{,}000)$. In the first design historical transport holds, one of three historical vertices is invalid, and coherence narrows the population bound on the zero-earnings share by 39 percent. All tests have rejection rates below one percent (Monte Carlo standard errors are at most 1.6 points with 1,000 replications), so they are conservative at this design, the confidence regions cover the identified interval in at least 99.9 percent of replications, and the coherence gain shows up in the confidence regions once samples are large (a 19 percent shorter region at the largest size). In the second design the benchmark lies at CDF distance 0.03, 0.07 or 0.12 from the hull; the bootstrap benchmark test has power 0.85 at distance 0.07 with 1,000 treated units and 0.62 at distance 0.12 with 200, while the DKW test is valid but much more conservative. In the third design the coherent set is a single point or empty with population repair between 0.005 and 0.04. The naive rule ``reject if the plug-in coherent set is empty'' rejects half the time when the model holds; the studentized test rejects in less than one percent of samples at the boundary and has power 0.69 against a repair of 0.02 with 1,000 treated units, while the unstudentized test has no power there. With clustered data, the one-sided Gaussian calibration controls size in every configuration we tried and is 4 to 8 points more powerful than the two-sided one, while a contact-set rule gains power but rejects 6.7 percent of the time at a 5 percent level (Appendix~\ref{S-sec:cluster}). Two practical conclusions follow: coherence tests should be studentized, and with samples of a few hundred treated units, coherence is mainly a diagnostic; it sharpens confidence regions only when the samples are large.

\subsection{What pre-treatment data cannot do}

Pre-treatment outcomes reveal how historical biases vary and can refute a model whose implied set is empty. They cannot reveal which post-policy bias will occur. If two nested models generate the same observed distribution but different missing distributions, no test distinguishes them, and no confidence set can be valid for the larger model while shrinking to the identified set of the smaller one (Appendix~\ref{S-sec:adapt}). Adding historical dates weakens Assumption~\ref{ass:hull} and weakly widens the identified set. Choosing the window that gives the narrowest interval therefore trades credibility for apparent precision; the choice of $S$ is an identifying decision to be argued on economic grounds.

\section{Application I: An Experimental Benchmark}\label{sec:nsw}

\subsection{Purpose and design}

This application asks whether historical transport recovers a distribution that is known, and whether the diagnostics of Section~\ref{sec:inference} would have warned an analyst who did not know it. With hindsight the failure is expected, which is what makes the setting useful: the question is whether the tests find it, and for the right reason. The NSW demonstration randomized disadvantaged workers into subsidized jobs. Assignment took place between 1975 and 1977, and \citet{LaLonde1986} restricted the male sample to people assigned from January 1976 on, so that 1975 earnings precede assignment for everyone \citep{DehejiaWahba1999}. We use the \citet{DehejiaWahba1999} subsample of these men with earnings recorded for 1974, 1975 and 1978: 185 participants and 260 experimental controls. The comparison groups are the CPS-1 (15,992 men) and PSID-1 (2,490 men) samples of \citet{LaLonde1986}. Earnings are in 1982 dollars. All files come from the replication archive of \citet{ImbensXu2025}.

We play the role of an analyst without the experiment. The treated group is the 185 participants, the comparison group is CPS-1 or PSID-1, the historical dates are 1974 and 1975, and the outcome date is 1978. The vertices are $H_s=F^{C}_{78}+F^{T}_{s}-F^{C}_{s}$ for $s\in\{74,75\}$, where $F^T_s$ and $F^C_s$ are the participants' and the comparison group's earnings distributions in year $s$. Random assignment makes the experimental controls' 1978 distribution equal to the participants' untreated distribution, so it is the benchmark $F_R$; it is never used to build the bounds. The target is the untreated 1978 earnings distribution of the men in this experimental subsample, not of all NSW applicants or of the survey populations. Participants and experimental controls are balanced before assignment: on the grid below their total-variation distance is 0.06 in 1974 and 0.09 in 1975. Inference treats each sample as a random sample from its population. CPS and PSID have complex designs, and our results, like the rest of this literature, abstract from them.

The grid, fixed in advance, has seven cells: zero earnings, $(0,5]$, $(5,10]$, $(10,15]$, $(15,20]$, $(20,30]$ and above 30 thousand dollars. The main outcome is the probability of positive 1978 earnings. Its experimental effect is 0.111, with 95 percent confidence interval $[0.026,0.195]$. All bootstraps resample complete individual histories within each sample, with 2,000 draws.

\subsection{Results}

\begin{table}[t]\centering\small
\caption{Historical transport against the NSW experimental benchmark}\label{tab:nsw}
\setlength{\tabcolsep}{4pt}
\begin{tabular}{@{}lcccc@{}}\toprule
 & CPS-1 & CPS-1 & PSID-1 & PSID-1 \\
 & & reweighted & & reweighted \\\midrule
\multicolumn{5}{@{}l}{\emph{Coherence (seven cells)}}\\
\quad Invalid vertices (of 2) & 2 & 1 & 2 & 2 \\
\quad Plug-in minimum repair (percentage points) & 0.88 & 0 & 3.73 & 0 \\
\quad Studentized coherence test, $p$-value & 0.84 & -- & 0.013 & -- \\[2pt]
\multicolumn{5}{@{}l}{\emph{Effect on probability of positive earnings}}\\
\quad Signed-hull interval & $[0.38,0.48]$ & $[0.32,0.41]$ & $[0.37,0.49]$ & $[0.38,0.51]$\\
\quad Coherent interval, plug-in & empty & $[0.35,0.41]$ & empty & $[0.40,0.40]$\\
\quad 95\% outer region & $[0.23,0.64]$ & $[0.15,0.58]$ & $[0.21,0.65]$ & $[0.16,0.74]$\\
\quad Changes-in-changes (base 1975) & 0.41 & & 0.51 & \\
\quad Experimental benchmark & \multicolumn{4}{c}{$0.111\ [0.026,0.195]$}\\[2pt]
\multicolumn{5}{@{}l}{\emph{Distance between the benchmark and the historical hull}}\\
\quad Total variation, seven cells & 0.34 & 0.29 & 0.35 & 0.32\\
\quad CDF distance $\widehat D$ & 0.32 & 0.25 & 0.35 & 0.38\\
\quad Bootstrap / DKW 5\% critical value & 0.12 / 0.25 & 0.14 / -- & 0.12 / 0.29 & 0.18 / --\\
\quad 95\% lower bound on $D$ (bootstrap / DKW) & 0.20 / 0.07 & 0.11 / -- & 0.23 / 0.06 & 0.20 / --\\
\quad Bootstrap $p$-value & $<0.001$ & 0.001 & $<0.001$ & 0.001\\\bottomrule
\end{tabular}
\begin{minipage}{0.97\linewidth}\footnotesize\emph{Notes:} 185 participants, 260 experimental controls, 15,992 CPS-1 and 2,490 PSID-1 men; historical dates 1974 and 1975; outcome 1978. ``Reweighted'' weights comparison men by the odds of a logit propensity score in age, education, their squares, and indicators for black, Hispanic, married and no degree; the bootstrap re-estimates the weights in each of 999 draws, and no DKW value is reported because the weights are estimated. The outer region optimizes the effect over historical weights allowed by simultaneous bootstrap bands for the cell masses and the treated share. $\widehat D$ is the supremum distance between the experimental-control 1978 distribution function and the nearest transported one. Changes-in-changes uses the \citet{AtheyImbens2006} bounds, which coincide here.\end{minipage}
\end{table}

\begin{figure}[t]\centering
\includegraphics[width=\linewidth]{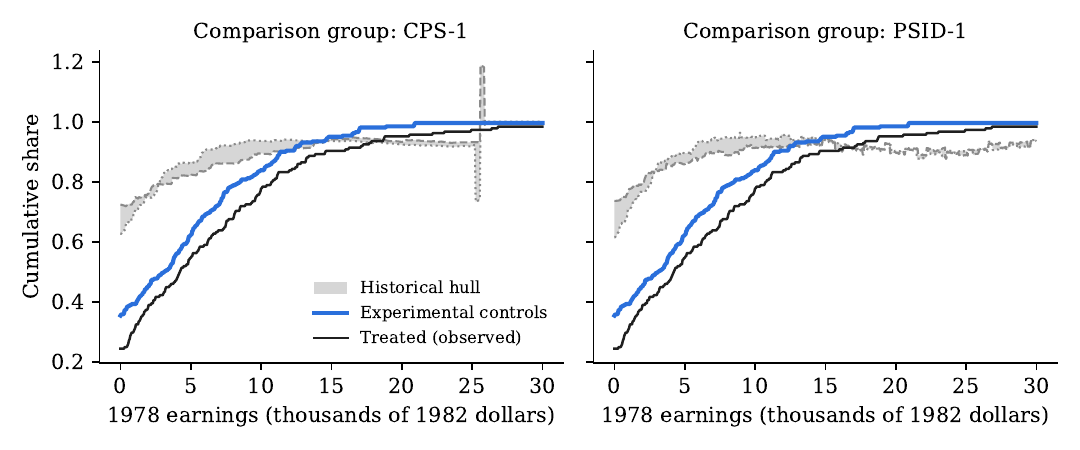}
\caption{1978 earnings: experimental controls and the historical hull}\label{fig:nsw}
\begin{minipage}{0.97\linewidth}\footnotesize\emph{Notes:} Shaded: distribution functions of the mixtures $\lambda H_{74}+(1-\lambda)H_{75}$ (dashed: $H_{74}$; dotted: $H_{75}$). Values above one and the downward jump near the CPS top-code show that the transported candidates are not distributions.\end{minipage}
\end{figure}

Table~\ref{tab:nsw} and Figure~\ref{fig:nsw} summarize the results.

\emph{Transport fails, and the benchmark test says so.} The experimental control distribution lies 0.32 (CPS) and 0.35 (PSID) from the nearest transported distribution, and the bootstrap and finite-sample tests both reject. The lower confidence bounds say that, at some earnings level, every transported distribution function misses the truth by at least 20 percentage points. In total variation on the seven cells, about a third of the probability mass would have to be moved. For the policy question, historical transport puts the employment effect between 0.38 and 0.49; even the 95 percent outer regions exclude the experimental estimate of 0.11. Distributional difference-in-differences with a single base year gives the endpoints of the signed interval (0.48 from 1974, 0.38 from 1975), and changes-in-changes gives 0.41 (CPS) and 0.51 (PSID). Every method that uses pre-program gaps overstates the effect by a factor of three or more.

\emph{Coherence detects part of the problem without the experiment.} Both historical vertices put negative mass on some earnings cell for both comparison groups, and no mixture repairs this; the plug-in repair is 0.9 points for CPS and 3.7 for PSID, located between \$10,000 and \$20,000 where the survey distributions shift between 1974--75 and 1978. As Section~\ref{sec:inference} warns, plug-in incoherence alone proves nothing. The studentized test, which uses only pre-program and comparison-group data, rejects coherence for PSID ($p=0.013$) but not for CPS ($p=0.84$). An analyst without the experiment could therefore have rejected transport from PSID on the data alone. Grid choice does not drive these conclusions: the grid-free region of Section~\ref{sec:refinement}, calibrated on a joint sup-norm band and computed exactly, is $[0.21,0.62]$ for CPS and $[0.19,0.67]$ for PSID, no wider than the regions obtained with the same band on 7, 12 or 23 cells (Appendix~\ref{S-sec:gridfree}), and it excludes the experimental estimate.

\emph{Do the tests reject when transport holds?} Replacing the survey sample by a random half of the experimental controls, and using the other half as the benchmark, makes every historical bias zero by design. Across 200 random splits the benchmark test rejects once and the studentized coherence test never, and the median CDF distance falls from 0.32--0.35 to 0.10 (Appendix~\ref{S-sec:positive-control}). The rejections above are not an artifact of the small samples or sparse cells.

\emph{Why transport fails.} NSW admitted people who were jobless at the time and had worked little in the preceding months. Eligibility selected on a transitory state, the pre-program dip of \citet{Ashenfelter1978}. In 1974 and 1975, 71 and 60 percent of participants had no earnings, against 12 and 11 percent in CPS-1, so the historical nonemployment gaps are 0.59 and 0.49. In 1978 the gap between experimental controls (35 percent with no earnings) and CPS-1 (14 percent) is 0.22. The post-program bias is far outside the range of the historical biases; in the language of Section~\ref{sec:content}, the post-program environment is not a mixture of the pre-program ones. PSID-1 gives the same picture (0.62 and 0.50 against 0.24). The failure is not a quirk of the method: when selection into treatment works through the pre-period outcome, history cannot bracket the post-policy bias.

\emph{Composition and a second target group.} Reweighting the survey samples to the participants' demographic characteristics makes the seven-cell hulls coherent, but transport is still rejected and the implied effects stay near 0.35 to 0.40. For the women receiving Aid to Families with Dependent Children (AFDC) in the demonstration \citep{CalonicoSmith2017}, selected on long-term welfare receipt, the transported employment effect is 0.27 against an experimental 0.01, and the benchmark test rejects; with one historical date this checks single-period distributional difference-in-differences (Appendix~\ref{S-sec:nsw-supp}).

\emph{Breakdown.} For CPS-1, the relaxed model is nonempty from $\tau=0.009$ and first admits a zero employment effect at $\tau^{\rm sign}=0.38$, with 95 percent confidence interval $[0.26,0.51]$. The departure from transport revealed by the experiment, the total-variation distance from the benchmark to the hull, is 0.34. The sign of the effect survives this departure, but its size does not: a relaxation large enough to contain the truth admits every effect between 0.05 and 0.76 (Figure~\ref{fig:breakdown}, left).

What this application shows is that historical transport can fail by a large margin in a canonical setting, that the reason is economic and detectable in pre-program data, and that the tests of Section~\ref{sec:inference} detect it, with and without an experiment, while not rejecting when transport holds by design.

\section{Application II: The 1993 Expansion of the EITC}\label{sec:eitc}

\subsection{Purpose and design}

This application asks what the method delivers where the assumption deserves a hearing. The Omnibus Budget Reconciliation Act of 1993, proposed in February and enacted in August 1993, raised the Earned Income Tax Credit for families with children over tax years 1994--1996, most for families with two or more children. The standard design compares single women with and without children \citep{EissaLiebman1996,MeyerRosenbaum2001}. The groups differ by the presence of children, a characteristic that does not respond to a recent labor-market shock, so the selection mechanism that defeats transport in the NSW is absent. Transport here says that the 1994--96 earnings gap between mothers and childless women, absent the expansion, would have resembled a mixture of the 1991 and 1992 gaps.

We use a public extract of March Current Population Survey records for unmarried women aged 20--54 without a high-school diploma, tax years 1991--1996 (13,746 women; 1,138 mothers and 867 childless women in 1996). The outcome is annual earnings in the tax year, in 1997 dollars, with its atom at zero; we call positive earnings employment. (The extract labels its year variable as the tax year; March survey income refers to the preceding calendar year.) The extract also records each woman's state of residence. We fixed the design in writing before computing any outcome by group and year: the treated group (women with children; secondary, two or more children), the comparison group (women without children), the historical years (1991 and 1992, which end before the proposal; 1993 as a sensitivity check), the grid (zero earnings, and approximately the credit's phase-in, plateau and phase-out ranges and beyond for two-child families in 1996, in 1997 dollars: $(0,9{,}000]$, $(9{,}000,12{,}000]$, $(12{,}000,28{,}000]$, above), the contrasts, the diagnostics, and three success criteria: the coherence test does not reject, a placebo benchmark does not reject, and the 1996 employment bound excludes zero. The extract has no survey weights or design variables. Table~\ref{tab:eitc} treats records as independent; Section~\ref{sec:eitc-states} resamples whole states instead.

\begin{table}[t]\centering\small
\caption{The 1993 EITC expansion: bounds and diagnostics, women with children}\label{tab:eitc}
\setlength{\tabcolsep}{5pt}
\begin{tabular}{@{}lccc@{}}\toprule
 & 1994 & 1995 & 1996\\\midrule
Employment effect, bound & $[-0.010,0.014]$ & $[0.047,0.070]$ & $[0.115,0.139]$\\
\quad 95\% outer region & $[-0.07,0.08]$ & $[-0.02,0.13]$ & $[0.05,0.20]$\\
\quad Grid-free 95\% region & $[-0.16,0.16]$ & $[-0.10,0.22]$ & $[-0.04,0.29]$\\
Phase-out share effect, bound & $[-0.031,-0.008]$ & $[-0.009,0.014]$ & $[0.053,0.076]$\\
Sign breakdown radius, employment & 0 & 0.047 & 0.115\\
Comparison law in its own hull: distance & 0.032 & 0.044 & 0.077\\
\quad bootstrap $p$-value & 0.76 & 0.38 & 0.01\\\midrule
\multicolumn{4}{@{}l}{Coherence: plug-in repair 0; studentized $p=1.00$ (5 cells, 1996; 21 cells, all years)}\\
\multicolumn{4}{@{}l}{Placebo benchmark at 1993: CDF distance 0.032, bootstrap critical value 0.081, $p=0.96$}\\
\multicolumn{4}{@{}l}{Common weight across 1994--96: feasible on 5 cells}\\\bottomrule
\end{tabular}
\begin{minipage}{0.97\linewidth}\footnotesize\emph{Notes:} Effects are mothers' observed shares minus their counterfactual shares under transport from 1991 and 1992, on the five-cell grid; the bound is the same with and without coherence on this grid. Outer regions: vertex reduction (Lemma~\ref{lem:vertex}) with a joint bootstrap band for the two vertex effects. Grid-free region: Section~\ref{sec:refinement}. Comparison-law check: CDF sup-norm distance from the childless women's earnings distribution to the hull of their 1991 and 1992 distributions. Results for women with two or more children are in Appendix~\ref{S-sec:eitc-supp}.\end{minipage}
\end{table}

\subsection{Results}

The diagnostics pass (Table~\ref{tab:eitc}). Every transported candidate is a distribution on the prespecified grid, and the studentized test does not reject on it or on a 21-cell grid. The placebo benchmark, which transports the 1991--92 gaps to 1993, a year whose credit the act did not change, misses the mothers' observed 1993 earnings distribution by 0.032 in CDF distance, well inside the bootstrap critical value. The three success criteria are met.

The bounds are narrow and move as the credit phased in: no detectable change in 1994, an employment gain of 4.7 to 7.0 points in 1995, and 11.5 to 13.9 points in 1996, against pre-program employment of 53 percent. For mothers of two or more children, who gained most, the 1996 bound is 13.4 to 17.1 points. Sampling uncertainty is large relative to these widths: the 95 percent outer region excludes zero in 1996 but not in 1995, and the grid-free region, which pays for avoiding a grid, excludes zero in no year. Changes-in-changes gives 0.15 to 0.16 for 1996, above the transport bound.

The protocol of Section~\ref{sec:conclusion} flags 1996. In 1994 and 1995 the childless women's earnings distribution is a mixture of its own 1991 and 1992 distributions, up to sampling error; in 1996 it is not ($p=0.01$), as their employment rate falls to 64.6 percent from 67.7 and 69.5. In the environment model, 1996 is an extrapolation of history, and it is also the year in which welfare reform and a strong labor market changed single mothers' incentives \citep{Kleven2019}. We therefore read the 1996 bound as an effect of everything that changed for single mothers relative to childless women, not of the credit alone. The 1995 bound passes the diagnostics, but it too is an effect of the credit only if nothing else that changed between 1992 and 1995 affected mothers relative to childless women. State welfare waivers, which spread in these years, did; \citet{MeyerRosenbaum2001} attribute a large share, though not all, of the rise in single mothers' employment over 1984--96 to the EITC and other tax changes. If the waivers raised mothers' employment, the 1995 bound overstates what the credit alone did under transport. Section~\ref{sec:eitc-states} drops the waiver states.

\subsection{Welfare waivers and state clustering}\label{sec:eitc-states}

The state identifiers allow two further checks, which were not in the written plan. First, resampling whole states rather than women, which allows any dependence within a state across years and groups, changes no endpoint of the outer regions of Table~\ref{tab:eitc} by more than 0.01: $[-0.01,0.13]$ in 1995 and $[0.05,0.20]$ in 1996. Second, we drop states with welfare waivers. The year in which each state's first statewide AFDC waiver was approved is taken from \citet[Table A3]{SchanzenbachStrain2021}, before examining any outcome by state. One design drops the 27 states with an approval by 1995, which could affect outcomes in 1994 and 1995; the other drops all 36 states with an approval by 1996, the robustness sample of \citet{SchanzenbachStrain2021}. Approval is not implementation, and dropping states changes the population, so these designs test the sensitivity of the bounds to the waivers rather than separating the credit from welfare reform.

\begin{table}[t]\centering\small
\caption{The EITC expansion after dropping welfare-waiver states, women with children}\label{tab:eitc-waiver}
\setlength{\tabcolsep}{5pt}
\begin{tabular}{@{}lccc@{}}\toprule
 & All states & No approval by 1995 & No approval by 1996\\\midrule
States & 51 & 24 & 15\\
Employment effect 1995, bound & $[0.047,0.070]$ & $[0.055,0.102]$ & $[-0.010,0.079]$\\
\quad 95\% outer region & $[-0.01,0.13]$ & $[-0.04,0.20]$ & $[-0.16,0.22]$\\
Employment effect 1996, bound & $[0.115,0.139]$ & $[0.084,0.118]$ & $[0.031,0.120]$\\
\quad 95\% outer region & $[0.05,0.20]$ & $[0.01,0.20]$ & $[-0.06,0.21]$\\
Comparison-law check 1996, $p$-value & 0.02 & 0.20 & 0.56\\\bottomrule
\end{tabular}
\begin{minipage}{0.97\linewidth}\footnotesize\emph{Notes:} Bounds as in Table~\ref{tab:eitc}, computed on the women living in the retained states. Outer regions and $p$-values resample whole states. Waiver approval years from \citet[Table A3]{SchanzenbachStrain2021}. Results for 1994 and for mothers of two or more children are in Appendix~\ref{S-sec:eitc-supp}.\end{minipage}
\end{table}

Table~\ref{tab:eitc-waiver} reports the results. In the 24 states without an approval by 1995 every diagnostic passes, including the comparison-law check for 1996, and the employment bounds are 5.5 to 10.2 points in 1995 and 8.4 to 11.8 points in 1996. Here coherence binds: the signed bound for 1996 reaches 13.1 points, and no probability distribution in the hull supports a gain above 11.8. In the 15 states with no approval through 1996 the two historical gaps differ more, and the bounds widen to $-1.0$ to 7.9 points in 1995 and 3.1 to 12.0 points in 1996. With 15 or 24 states the sampling uncertainty is large, and only the 1996 region in the 24-state design excludes zero. Three conclusions follow. The 1996 bound keeps its sign in every design, and the 1995 bound keeps it unless every state with any approval through 1996 is dropped. The 1996 failure of the comparison-law check is not detected outside the waiver states, although the check has less power in the smaller samples. And the waivers do not account for the employment gains under transport, but with so few states the data cannot apportion them between the credit and the waivers. \citet{SchanzenbachStrain2021} likewise find that their regression estimate barely changes when the waiver states are dropped.

\subsection{What common weights add}

On the prespecified grid, coherence does not bind: every candidate is already a distribution, so the gain from the probability restriction is nil there, while on the 21-cell grid it narrows the 1996 employment bound to $[0.128,0.131]$ (on that grid the plug-in coherent set is empty in other years and for mothers of two or more children, but the studentized test does not reject in any of these cases). The common weight across outcome events matters even so. Scalar bounds computed contrast by contrast give an employment effect in $[0.047,0.070]$ and an effect on the share of mothers earning in the phase-out range in $[-0.009,0.014]$ for 1995, and so allow any combination. Under transport the two move together along one segment: every counterfactual with an employment gain above 6.2 points implies a fall in the phase-out share, and every one with a rise in the phase-out share implies an employment gain below 6.2 points. Whether the credit drew mothers into work at low earnings or also shifted earnings toward the phase-out range cannot be read from separate bounds, and it is answered once the employment gain is pinned down.

\section{Application III: State Minimum Wages in 2014}\label{sec:mw}

\subsection{Purpose and design}

This application shows the pooling problem of Section~\ref{sec:events} in a standard stacked design, and how much probability feasibility can change a bound on a realistic grid. The bounds are descriptive. We do not claim causal effects of minimum wages: the audit below shows that the transport assumption is not credible here, and no sampling distribution is available for the cells.

We use the state--quarter panel of \citet{CengizEtAl2019}, as distributed with \citet{RothSantAnna2023}. For each of 51 jurisdictions and each quarter of 2007--2015 it gives the number of workers per capita in 117 quarter-dollar wage bins (2016 dollars), with a residual cell equal to one minus the employment-to-population ratio. The District of Columbia, where jobs are counted at the workplace and population at residence, has a negative residual and is dropped. These vectors are job counts per resident rather than the distribution of an individual's outcome: multiple jobholding and commuting break the individual interpretation. The coherence restriction does not depend on it. A counterfactual count of jobs in a wage bin cannot be negative under any interpretation, and in every event-level counterfactual below the negative mass lies in wage bins, never in the residual cell (Appendix~\ref{S-sec:mw-supp}). Five states raised their minimum wage in 2014 after no increase since 2010: New Jersey (effective January 1), Delaware (June 1), California (July 1), Minnesota (August 1) and Michigan (September 1). Each is evaluated at its first fully treated quarter, with comparison states that had no increase between 2010 and that quarter. The historical dates are 4, 8, 12 and 16 quarters earlier, which keeps seasons aligned. The outcome is the change in the population share employed below \$9.25, which falls either when jobs disappear or when wages rise above the cutoff.

The events are stacked with equal weights, and transport is imposed on the averaged distribution. By Theorem~\ref{thm:stacking} this is weaker than eventwise transport, and the resulting bound concerns the pooled population of the five events only.

\subsection{Pooling and coherence}

\begin{table}[t]\centering\small
\caption{Effect on the population share employed below a cutoff (percentage points)}\label{tab:mw-main}
\begin{tabular}{@{}lcc@{}}\toprule
 & Signed hull & Coherent hull \\\midrule
Below \$9.25, seven cells & $[-0.614,-0.228]$ & $[-0.614,-0.228]$\\
Below \$9.25, 118 cells & $[-0.614,-0.228]$ & $[-0.517,-0.324]$\\[3pt]
\multicolumn{3}{@{}l}{\emph{Other prespecified cutoffs, 118 cells}}\\
\quad Below \$7.25 & & $[-0.416,-0.284]$\\
\quad Below \$11.25 & & $[-0.442,\ \ 0.450]$\\
\quad Below \$14.25 & & $[-0.680,\ \ 0.170]$\\
\quad Below \$18.25 & & $[-0.366,\ \ 0.716]$\\[3pt]
Minimum repair: stacked / eventwise & \multicolumn{2}{c}{$0$ / $0.303$}\\
Sign breakdown radius $\tau^{\rm sign}$, below \$9.25 & \multicolumn{2}{c}{$0.288$}\\
One-year-earlier placebo: TV distance to observed law & \multicolumn{2}{c}{$4.92$}\\\bottomrule
\end{tabular}
\begin{minipage}{0.97\linewidth}\footnotesize\emph{Notes:} The treated share below \$9.25 is 7.136 percent. The seven cells are the residual and the intervals delimited by the five cutoffs. Repairs, radii and distances are in percentage points of the population. Intervals for different cutoffs need not be attained by the same historical weights.\end{minipage}
\end{table}

On the seven-cell grid every vertex is a probability vector and coherence has no effect (Table~\ref{tab:mw-main}). On the full grid all four vertices have negative bins, although the stacked hull still contains coherent mixtures, and the bound below \$9.25 shrinks from $[-0.61,-0.23]$ to $[-0.52,-0.32]$, half its width; the midpoint barely moves. This is Proposition~\ref{prop:functional-gain} at work: the weights that attain the unrestricted endpoints put negative mass on bins that the reported outcome never mentions. The stacked counterfactual that delivers this bound is a composite. No common weight makes all five event-level counterfactuals valid; at least 0.3 percent of the population must be moved to achieve that. Weights chosen separately for each state do not help: each state's own coherent set is empty, and the smallest negative mass of a state-level counterfactual ranges from 0.03 points (Michigan) to 0.52 (California). The stacked set is nonempty only because negative bins of different states offset each other in the average. About two thirds of the negative bins lie below \$8.00, where cells are sparse and sampling noise is largest, and none exceeds 0.13 percent of the population; without a sampling distribution for the cells we cannot say how much of the infeasibility is noise and how much is a failure of transport. The bound may still be defended as a statement about the pooled population, but not as an average of state effects. The rest of the table adds two warnings: the sign breaks down at $\tau=0.29$ points, and one year earlier, before any policy, the nearest transported distribution is 4.9 points in total variation from the one observed.

\subsection{An audit of the transport assumption}

In the environment model of Section~\ref{sec:content}, the bound is causal only if the 2014 environment $\pi_1$ lies in $\co\{\pi_s:s\in S\}$, the historical quarters are free of the policy, and composition is stable. Each check fails; details and the full table are in Appendix~\ref{S-sec:mw-supp} (Table~\ref{S-tab:mw-audit}). These checks concern the stacked model; the event-level models already fail on coherence alone.

\emph{Bracketing.} By Section~\ref{sec:content}, the environment model requires the comparison states' 2014 distribution to be a mixture of their own historical ones. It is not: comparison-state nonemployment was 43.8 to 44.4 percent at the four historical dates and 43.4 percent at the policy date, 0.41 points below the historical minimum (95 percent interval $[0.35,0.47]$, resampling comparison states, which ignores common shocks across states), and on the full 118-cell grid the 2014 comparison distribution lies 12 points in total variation from the hull of its own history (13 points one year earlier, partly because nominal wages drift across bins defined in real dollars). In the environment model, the 2014 labor market is an extrapolation of history, not a mixture of it.

\emph{Anticipation and aggregation weights.} California's increase was signed in September 2013 and New Jersey's amendment passed the legislature in 2012--13, both before or inside the most recent historical quarter. Dropping that quarter makes the coherent set empty. Weighting events by population, which gives California 60 percent of the weight, moves the coherent bound to $[-0.01,0.08]$. Equal weights describe the average state event and population weights the average resident of a treated state; the data do not choose between them.

\emph{Composition and comparison groups.} Imposing that the increases left unchanged the share employed at \$18.25 or more is compatible with the data at the policy date and narrows the bound to $[-0.51,-0.37]$; one year earlier, with no policy change, the observed upper-tail share lies outside the hull's range, as a composition shift would produce. Never-treated or non-contiguous comparison states move the bound by about 0.1 point in opposite directions.

\emph{Placebo assignments.} Giving the actual event dates to five randomly drawn never-treated states yields a coherent set in 55 percent of draws and, among those, an upper endpoint at or below the actual one 25 percent of the time (15 percent when placebo states are drawn from the larger half of the pool); unconditionally, 13 percent of all draws produce such a bound. The placebo distribution is wide, so this exercise has low power against a real effect: it cannot show that the bound is spurious. It does show that bounds like the actual one arise in a sizable share of draws without any policy.

The cells are survey aggregates without a published joint sampling distribution, and with five treated states and overlapping event windows the cluster calibration of Appendix~\ref{S-sec:cluster} cannot supply one: the finite-sample radius exceeds one, and with 118 cells the error bound of the Gaussian approximation, of order $\{\log^7(JKC)/C\}^{1/6}$, is not small for any realistic number of states. We therefore report no confidence sets here; the breakdown radius of 0.29 points and the placebo distance of 4.9 points are descriptive comparisons.

\subsection{Lessons}

Probability feasibility can cut the width of a distributional bound in half on a realistic grid, through bins far from the reported outcome. Here the bound does not survive the audit: the policy date lies outside the historical range, the coherent set depends on a quarter exposed to anticipation and on how events are weighted, and event-level counterfactuals are infeasible. A credible analysis would end histories before the earliest announcement or enactment, use comparison states that neither anticipate nor neighbor the treatment, hold composition fixed with worker-level data, impose transport event by event, and obtain a sampling distribution for the cells (Appendix~\ref{S-sec:mw-supp}).

\section{Conclusion}\label{sec:conclusion}

Using pre-policy gaps to bound a post-policy counterfactual distribution replaces one untestable equality with a range informed by history. That range must consist of distributions, and reported conclusions must come from one of them. The sharp set is the historical hull intersected with the probability simplex. We have characterized when this restriction changes a conclusion, how to certify that it is violated, why stacked events and coarse grids can conceal violations, how far a relaxation must go to overturn a sign, and how to carry sampling error into tests and confidence sets, including a test against an experimental benchmark and a confidence region that requires no grid choice.

The applications show both sides. Coherence is necessary and can sharpen bounds considerably, but it is not a validation. In the environment model, historical transport holds when the post-policy environment lies within the historical one. In the NSW it does not, because eligibility selected on a transitory pre-program state; in the minimum-wage data the 2014 labor market lay outside its recent history and the pooled bound rests on impossible state counterfactuals. For the EITC, where the groups differ by a stable characteristic, the diagnostics pass through 1995 and the bounds are informative. They stay positive after the states with early welfare waivers are dropped, but with few states they cannot apportion the gain between the credit and the waivers. The same diagnostics identify 1996 as the year in which the comparison group leaves its historical range, a departure not detected outside the waiver states. For applied work this suggests a short protocol. Choose settings in which selection reflects stable differences between groups rather than a recent outcome. Fix the historical dates, comparison group, grid and outcomes before looking at post-policy data, and end the historical window before the policy is announced, not merely before it starts. Run the studentized coherence test and, with several events, the event-level check of Theorem~\ref{thm:stacking}. Check that the comparison group's current distribution is a mixture of its own past ones and that composition is stable. With clustered data, calibrate the bands over clusters, not units. Report the grid-free confidence region and the breakdown radius, and, wherever a benchmark exists, the transport test.

The EITC application has no experimental benchmark. The natural validation is an experiment in which eligibility depends on stable characteristics rather than on a recent outcome, with an untreated observational comparison group observed at several pre-program dates under the same measurement as the treated group. Randomized phase-ins of conditional cash transfers, such as PROGRESA \citep{Schultz2004,ToddWolpin2006}, where eligibility is set by a household poverty index, fit this description if the outcome is measured consistently in at least two baseline rounds. The experimental controls would be held out as the benchmark, and the grid, dates, outcomes and comparison group fixed before the benchmark is examined.

\section*{Data availability}
The NSW and survey comparison samples are from the public replication archive of \citet{ImbensXu2025}, which redistributes the files of \citet{DehejiaWahba1999} and \citet{CalonicoSmith2017}. The state--quarter wage panel is the version of the \citet{CengizEtAl2019} data distributed with \citet{RothSantAnna2023}; the original data and programs are in the Harvard Dataverse deposit of \citet{CengizEtAl2019} (doi:10.7910/DVN/TJCTC7). The replication package reproduces every number in the paper from these files.

\clearpage
\appendix
\renewcommand{\thesection}{S\arabic{section}}
\titlelabel{\thetitle\quad}
\setcounter{section}{0}\setcounter{table}{0}\setcounter{figure}{0}\setcounter{equation}{0}
\renewcommand{\theequation}{S.\arabic{equation}}
\renewcommand{\thetable}{S\arabic{table}}
\renewcommand{\thefigure}{S\arabic{figure}}
\counterwithin{assumption}{section}\counterwithin{theorem}{section}\counterwithin{proposition}{section}
\counterwithin{corollary}{section}\counterwithin{lemma}{section}\counterwithin{remark}{section}\counterwithin{example}{section}
\begin{center}\Large\bfseries Appendix\end{center}
\noindent Results in the appendix are numbered S$x$.$y$; references without the prefix S point to the main text.

\section{The Model: Interpretation and Sharpness}\label{S-sec:supp-model}
\subsection{An environment interpretation}\label{S-sec:environments}
Assumption~\ref{ass:hull} can arise from an aggregate
mixture of environments, but the environment restriction is substantive.
Let $R$ take values in the finite set of historical dates $S$. At the
post-treatment date, suppose the hypothetical untreated outcome has a
conditional law $M_{g,s}$ in group $g\in\{0,1\}$ given $R=s$,
and suppose $\Pr(R=s\mid g)=\lambda_s$ for both groups. Assume
\[
 M_{1,s}-M_{0,s}=\mu^0_{1,s}-\mu^0_{0,s},\qquad s\in S,
\]
where the measures on the right are the historical observed untreated
laws. Then, by the law of total probability, the post-treatment
untreated group difference is
\[
 \mu^0_{1,1}-\mu^0_{0,1}
 =\sum_{s\in S}\lambda_s(M_{1,s}-M_{0,s})
 =\sum_{s\in S}\lambda_s(\mu^0_{1,s}-\mu^0_{0,s}).
\]
This is only a sufficient economic interpretation. It is neither implied by probability coherence nor testable from pretreatment fit alone; in particular, it requires the same environment weights for both groups. To see what equal regime shares exclude,
write $\lambda_{g,s}=\Pr(R=s\mid g)$ and retain the conditional
laws $M_{g,s}$. Without equal shares, the untreated difference is
\[
 \sum_s\lambda_{1,s}(M_{1,s}-M_{0,s})
 +\sum_s(\lambda_{1,s}-\lambda_{0,s})M_{0,s}.
\]
The second term is a signed composition discrepancy. Equal group shares eliminate it, whereas common aggregate shocks generally do not. For every outcome event its absolute value
is bounded by $\tfrac12\sum_s|\lambda_{1,s}-\lambda_{0,s}|$,
because the mixing-weight discrepancies sum to zero and each
$M_{0,s}(A)$ is between zero and one. This bound has no empirical
content unless the group-specific environment-share differences can be
bounded on economic grounds. It can fail if the distribution of $R$
differs across the groups, even when every conditional law is valid.
Nor does it assert that the observed historical date is itself an
unobserved random post-treatment regime. The interpretation requires a
substantive argument connecting the environments to historical dates.

\begin{corollary}[Bounded composition discrepancy]\label{S-cor:composition}
In this model let $\lambda_1=(\lambda_{1,s})_s$ and suppose $\frac12\sum_s|\lambda_{1,s}-\lambda_{0,s}|\le\eta$. Then $\mu^0_{1,1}\in\mathcal M_S(\eta)$, the relaxed model of Section~\ref{sec:breakdown}. Hence every bound computed from $\mathcal M_S(\eta)$ is a valid outer bound under group-specific environment shares that differ by at most $\eta$, and the radius $\tau$ of that section can be read as a bound on the difference in environment shares.
\end{corollary}

\begin{proof}
The untreated difference equals $\sum_s\lambda_{1,s}\nu_s+\kappa$ with $\kappa=\sum_s(\lambda_{1,s}-\lambda_{0,s})M_{0,s}$. Since $\kappa(\mathcal Y)=0$ and $|\kappa(A)|\le\eta$ for every $A$, as shown above, $d_{\TV}(\mu^0_{1,1},\mu_{\lambda_1})=\sup_A|\kappa(A)|\le\eta$ by Lemma~\ref{S-lem:equal-mass-tv}, and $\mu^0_{1,1}$ is a probability law.
\end{proof}

The corollary does not claim sharpness: the composition model restricts $\kappa$ to combinations of the unobserved laws $M_{0,s}$, which $\mathcal M_S(\eta)$ does not impose.

Let $(\mathcal Y,\mathcal B)$ be a measurable space. For a finite signed measure $\eta$, write $|\eta|$ for its variation measure and $\eta=\eta^+-\eta^-$ for its Jordan decomposition, and $b(\mu)=\mu^-(\mathcal Y)$.

\subsection{Equal-mass total variation}

\begin{lemma}[Equal-mass representation]\label{S-lem:equal-mass-tv}
Suppose $\eta(\mathcal Y)=0$.  Then
\[
\sup_{A\in\mathcal B}|\eta(A)|
=\eta^+(\mathcal Y)
=\eta^-(\mathcal Y)
=\frac12|\eta|(\mathcal Y).
\]
\end{lemma}

\begin{proof}
Because $\eta(\mathcal Y)=0$, the positive and negative masses coincide.  If $P$ is a Hahn-positive set and $N=P^c$ its negative complement, then $\eta(P)=\eta^+(\mathcal Y)$ and $\eta(N)=-\eta^-(\mathcal Y)$.  For arbitrary $A$, $-\eta^-(\mathcal Y)\leq\eta(A)\leq\eta^+(\mathcal Y)$.  The claim follows.
\end{proof}

This lemma fixes the convention used in the main paper:
\[
d_{\TV}(\mu,\nu)=\sup_A|\mu(A)-\nu(A)|
\]
whenever $\mu$ and $\nu$ have equal total mass.

\subsection{Continuity of the coherence deficit}

For a finite collection of zero-mass signed measures $\nu_1,\ldots,\nu_J$ and a unit-mass baseline $\mu_0$, define $\mu_\lambda=\mu_0+\sum_j\lambda_j\nu_j$.

\begin{lemma}[Lipschitz continuity]\label{S-lem:coherence-lipschitz}
The map $\lambda\mapsto b(\mu_\lambda)$ is convex and continuous on $\mathbb R^J$.  More precisely,
\[
|b(\mu_\lambda)-b(\mu_{\lambda'})|
\leq d_{\TV}(\mu_\lambda,\mu_{\lambda'})
\leq\sum_{j=1}^J|\lambda_j-\lambda'_j|\sup_A|\nu_j(A)|.
\]
\end{lemma}

\begin{proof}
Distance to a fixed set in a metric space is one-Lipschitz.  By the probability-coherence theorem, $b(\mu)$ is precisely the distance from $\mu$ to probability space under $d_{\TV}$ on unit-mass measures.  The second inequality is the triangle inequality.  Convexity follows either from convexity of distance to the convex set $\PP(\mathcal Y)$ under the signed-measure norm or directly from the eventwise representation.
\end{proof}

Consequently a minimum coherence deficit over a finite-dimensional simplex is attained.

\subsection{Negative mass and minimal replacement}

Write $b(\mu)=\mu^-(\mathcal Y)$ for the negative mass of a unit-mass signed measure, and set $\Delta_S=\{\lambda\in\mathbb R^S:\lambda_s\ge0,\ \sum_s\lambda_s=1\}$. The minimum fraction of unrestricted replacement mass measures the severity of incoherence; it is not a new identifying assumption. For a signed measure $\mu$ of total mass one, define
\[
 r(\mu)=\inf\bigl\{\varepsilon\in[0,1]:
 (1-\varepsilon)\mu+\varepsilon R\in\mathsf P(\mathcal Y)
 \text{ for some }R\in\mathsf P(\mathcal Y)\bigr\}.
\]
Here $\varepsilon$ is the replacement fraction and $R$ is a probability law; neither is an identified policy quantity.

\begin{proposition}[Exact minimum replacement fraction]\label{S-prop:replacement-fraction}
For any finite signed measure $\mu$ of total mass one,
\[
 r(\mu)=\frac{b(\mu)}{1+b(\mu)},
\]
where $b(\mu)$ is its negative mass. If $b(\mu)>0$, the minimum is attained with $R=\mu^-/b(\mu)$ and the resulting probability law is $\mu^+/(1+b(\mu))$. For a finite historical family of unit-mass signed measures $\{\mu_s:s\in S\}$, let $\mu_\lambda=\sum_s\lambda_s\mu_s$. Then the smallest replacement fraction over all historical weights is
\[
 r_S=\min_{\lambda\in\Delta_S}r(\mu_\lambda)
 =\frac{d_S^{\TV}}{1+d_S^{\TV}},
 \qquad d_S^{\TV}=\min_{\lambda\in\Delta_S}b(\mu_\lambda).
\]
For a fixed finite outcome partition $\Pi$, define $d_{S,\Pi}^{\TV}=\min_{\lambda\in\Delta_S}\sum_{A\in\Pi}[-\mu_\lambda(A)]_+$; the analogous fraction is $r_{S,\Pi}=d_{S,\Pi}^{\TV}/(1+d_{S,\Pi}^{\TV})$. For a nested sequence of finite partitions whose generated sigma-field agrees with the outcome sigma-field up to null sets under $\nu=\sum_{s\in S}|\mu_s|$, $r_{S,\Pi_n}$ increases to $r_S$.
\end{proposition}

\begin{proof}
Set $b=b(\mu)$ and choose a Hahn-negative set $N$ with $\mu(N)=-b$. If $(1-\varepsilon)\mu+\varepsilon R$ is a probability measure, nonnegativity on $N$ implies $\varepsilon R(N)\ge(1-\varepsilon)b$. Since $R(N)\le1$, necessarily $\varepsilon\ge b/(1+b)$. At equality, for $b>0$, choose $R=\mu^-/b$. The signed combination becomes $\mu^+/(1+b)$, which is a probability measure. For $b=0$, $\mu$ is already a probability law and the minimum is zero. The map $b\mapsto b/(1+b)$ is continuous and strictly increasing on $[0,\infty)$; therefore minimization over the historical simplex commutes with this transform. For the partition limit, put $f_n(\lambda)=\sum_{A\in\Pi_n}[-\mu_\lambda(A)]_+$ and $f(\lambda)=b(\mu_\lambda)$. Refinement gives $f_n\le f_{n+1}\le f$. For each fixed $\lambda$, $f_n(\lambda)\uparrow f(\lambda)$: use the finite dominating measure $\nu$ from the statement and write $h_\lambda=d\mu_\lambda/d\nu$. Since the partitions are nested and generate $\mathcal B$ modulo $\nu$-null sets, the conditional expectations $\mathbb E_\nu[h_\lambda\mid\sigma(\Pi_n)]$ converge to $h_\lambda$ in $L^1(\nu)$ (normalizing $\nu$ when necessary). Consequently the integrals of their negative parts converge to $\int[-h_\lambda]_+\,d\nu=f(\lambda)$. Each $f_n$ is continuous on the compact simplex $\Delta_S$, and $f$ is continuous there because $|b(\mu_\lambda)-b(\mu_{\lambda'})|\le\|\mu_\lambda-\mu_{\lambda'}\|_{\TV,\mathrm{var}}\le\sum_s|\lambda_s-\lambda_s'|\,|\mu_s|(\mathcal Y)$, where $\|\eta\|_{\TV,\mathrm{var}}:=|\eta|(\mathcal Y)$. Dini's theorem therefore gives uniform convergence $f_n\to f$ on $\Delta_S$, and hence $\min_\lambda f_n(\lambda)\uparrow\min_\lambda f(\lambda)$. Continuity of $x\mapsto x/(1+x)$ yields the claimed convergence of $r_{S,\Pi_n}$.
\end{proof}

Although each historical vertex may be invalid, a mixture of them can be a probability law. The replacement fraction therefore has to be minimized over the same common weights used by the identifying restriction. It measures the amount of negative signed mass, not the fraction of workers whose outcomes changed. Replacing mass outside the historical hull is a diagnostic operation and does not identify a new counterfactual law.

\subsection{Sharp identification}\label{S-sec:ident}

\begin{proposition}[Fixed-history restrictions and coherent mixtures]
\label{S-prop:fixed-history-comparison}
Suppose untreated group--date laws are observed for pretreatment dates,
and the post-treatment control law is observed. Write
$\mu^0_{g,t}$ for the untreated potential-outcome distribution in
group $g\in\{0,1\}$ at date $t$, with $g=1$ denoting treated units
and $t=1$ the post-treatment date. Fix a finite nonempty
set $S$ of pretreatment dates.
For $s\in S$, let $v_s=\mu^0_{0,1}+\mu^0_{1,s}-\mu^0_{0,s}$ be the
counterfactual implied by fixing the post-treatment untreated bias at its
value in period $s$. Define
$\mathcal H_S^+=\operatorname{co}\{v_s:s\in S\}\cap\mathsf P(\mathcal Y)$,
where $\mathsf P(\mathcal Y)$ is the set of probability laws.
Then:
\begin{enumerate}
\item[(i)] The fixed-history restriction at $s$ is compatible with an untreated
counterfactual law if and only if $v_s\in\mathsf P(\mathcal Y)$.
When compatible, its identified law $v_s$ belongs to $\mathcal H_S^+$.
\item[(ii)] It is possible that $\mathcal H_S^+\ne\varnothing$ although
$v_s\notin\mathsf P(\mathcal Y)$ for \emph{every} $s\in S$.
\item[(iii)] If every $v_s$ is a probability law, then
$\mathcal H_S^+=\operatorname{co}\{v_s:s\in S\}$.
\end{enumerate}
These statements compare candidate sets under the historical restrictions;
they do not rank the credibility of those restrictions.
\end{proposition}
\begin{proof}
Part (i) follows because fixing the bias at $s$ gives the single candidate
$v_s$, which must be a probability law. For (ii), take two outcome categories,
$S=\{s_1,s_2\}$, a post-period control law $(1/2,1/2)$, and historical
treated and control laws $(1/5,4/5)$ and $(4/5,1/5)$ at $s_1$, with their
roles reversed at $s_2$. The two transported candidates are
$v_{s_1}=(-1/10,11/10)$ and $v_{s_2}=(11/10,-1/10)$.
Neither is a probability law, but their equally weighted mixture is
$(1/2,1/2)$. Part (iii) follows from convexity of the probability laws.
\end{proof}

\paragraph{Standard-Borel realization.}

\begin{lemma}[Realization of an arbitrary marginal law]\label{S-lem:realization}
Let $\mathcal Y$ be standard Borel and let $\mu\in\PP(\mathcal Y)$.  There exists a Borel map $q:[0,1]\to\mathcal Y$ such that $q(U)\sim\mu$ for $U\sim U[0,1]$.
\end{lemma}

\begin{proof}
If $\mathcal Y$ is countable, enumerate it as $\{y_j:j\geq1\}$ (with the evident finite modification), partition $[0,1]$ into consecutive half-open intervals of lengths $\mu(\{y_j\})$, and set $q(u)=y_j$ on the $j$th interval.  The remaining null set, if any, may be mapped to an arbitrary point of $\mathcal Y$.

If $\mathcal Y$ is uncountable, the Borel isomorphism theorem supplies a bijection $\phi:\mathcal Y\to[0,1]$ for which both $\phi$ and $\phi^{-1}$ are Borel measurable.  Let $\widetilde\mu=\mu\circ\phi^{-1}$ and let
\[
r(u)=\inf\{x\in[0,1]:\widetilde\mu([0,x])\geq u\},
\qquad u\in(0,1).
\]
The map $r$ is Borel measurable and the generalized probability-integral transform gives $r(U)\sim\widetilde\mu$.  Defining $q=\phi^{-1}\circ r$ on $(0,1)$, with arbitrary values at the endpoints, yields $q(U)\sim\mu$.  This covers every standard Borel outcome space and does not require an order or vector-space structure on $\mathcal Y$.
\end{proof}

\paragraph{Sharpness construction.}

Let $P$ be the observed-data law for either a panel $O^P=(G,(Y_t)_t)$ or repeated cross-sections $O^R=(G,T,Y)$. For a finite nonempty historical set $S$, let $\mathfrak Q_S(P)$ be the class of full-data laws that induce $P$ and satisfy Assumptions~\ref{ass:basic}--\ref{ass:hull}.

\begin{theorem}[Sharp coherent identification]\label{S-thm:sharp-id}
Suppose the outcome space is standard Borel and $P$ satisfies Assumption~\ref{ass:basic}(i). For repeated cross-sections, suppose every entering group--date cell has positive probability. Then
\[
\left\{\mu^0_{1,1,Q}:Q\in\mathfrak Q_S(P)\right\}
=
\left[\mu^0_{0,1}+\co\{\mu^0_{1,s}-\mu^0_{0,s}:s\in S\}\right]\cap\PP(\mathcal Y).
\]
\end{theorem}

\begin{proof}
Necessity follows from Assumption~\ref{ass:hull} and from the requirement that a potential-outcome marginal be a probability law.

For sufficiency, fix a law $\mu^\star$ on the right-hand side and choose $\lambda^\star\in\Delta_S$ that represents it. First consider a panel. Let $(\Omega_P,\mathcal A_P,P)$ carry the complete observed-data vector and define the product extension
\[
(\widetilde\Omega,\widetilde{\mathcal A},\widetilde P)
=(\Omega_P\times[0,1],\mathcal A_P\otimes\mathcal B[0,1],P\otimes\mathrm{Leb}).
\]
Write $U$ for the second coordinate. Lemma~\ref{S-lem:realization} supplies a measurable $q$ with $q(U)\sim\mu^\star$. Retain $G$ and every observed outcome as coordinate projections. For every $t\leq0$, define $Y_t(0)=Y_t(1)=Y_t$. At date 1 define
\[
Y_1(0)=
\begin{cases}Y_1,&G=0,\\q(U),&G=1,\end{cases}
\qquad
Y_1(1)=
\begin{cases}q(U),&G=0,\\Y_1,&G=1.\end{cases}
\]
These variables are measurable because the group events and all component maps are measurable; no vector-space structure on $\mathcal Y$ is used.

The observed-data projection of $\widetilde P$ is exactly $P$. The definitions also give no anticipation and treatment consistency pathwise. Since $U$ is independent of $(G,\text{observed data})$ and both groups have positive probability,
\[
\mu^0_{1,1,\widetilde P}=\mu^\star,
\qquad \mu^0_{g,t,\widetilde P}=\mu^0_{g,t,P}\quad(t\leq0),
\qquad \mu^0_{0,1,\widetilde P}=\mu^0_{0,1,P}.
\]
Finally, the representation of $\mu^\star$ gives
\[
\mu^0_{1,1,\widetilde P}-\mu^0_{0,1,\widetilde P}
=\sum_{s\in S}\lambda_s^\star
 (\mu^0_{1,s,\widetilde P}-\mu^0_{0,s,\widetilde P}),
\]
as an equality of signed measures. Thus the extension belongs to $\mathfrak Q_S(P)$ and realizes the chosen target law.

For repeated cross-sections, let $\mathcal T$ be the finite set of observed dates,
$p_g=P(G=g)$, $p_{t\mid g}=P(T=t\mid G=g)$, and
$K_{g,t}=\mathcal L_P(Y\mid G=g,T=t)$ for each entering cell.  The
positive-cell condition makes each $K_{g,t}$ uniquely determined by $P$.
For cells outside the maintained restriction whose probability is zero,
choose any probability law on $\mathcal Y$.  Consider the product space
\[
\{0,1\}\times\mathcal T\times[0,1]^{2|\mathcal T|+1}
\]
with its product sigma-field.  Draw $G$ with probabilities $(p_0,p_1)$.
Conditional on $G=g$, draw $T$ with probabilities $(p_{t\mid g})_{t\in\mathcal T}$;
take all uniform coordinates independent of $(G,T)$.  Lemma~\ref{S-lem:realization}
gives measurable functions $q_{g,t}$ such that
$Z_{g,t}=q_{g,t}(U_{g,t})\sim K_{g,t}$ for every $g,t$.  Set
$Y=Z_{G,T}$.  For every $g,t$ and $A\in\mathcal B$,
\[
\widetilde P(G=g,T=t,Y\in A)=p_gp_{t\mid g}K_{g,t}(A)=P(G=g,T=t,Y\in A).
\]
The sets $\{g\}\times\{t\}\times A$ form a generating pi-system for
$\sigma(G,T,Y)$, so these equalities and uniqueness of probability measures
show that the constructed observed-data law is exactly $P$.  Set
$Y_t(0)=Y_t(1)=Z_{G,t}$ for $t\leq0$.  At date 1, set
$Y_1(0)=Z_{0,1}$ on $\{G=0\}$ and $Y_1(1)=Z_{1,1}$ on $\{G=1\}$.
On $\{G=1\}$, set $Y_1(0)=q_\star(U_\star)$, where the final uniform
coordinate is independent of $(G,T,(U_{g,t})_{g,t})$ and
$q_\star(U_\star)\sim\mu^\star$.  Assign $Y_1(1)$ on $\{G=0\}$
arbitrarily, for instance $Z_{0,1}$.  Every potential-outcome coordinate
is measurable.  Consistency and no anticipation hold pathwise; positive
group and entering-cell probabilities identify the untreated historical
conditional laws as $K_{g,s}$, the control post-treatment law as $K_{0,1}$,
and the missing treated untreated law as $\mu^\star$.  Therefore the same
signed-measure equality used in the panel case verifies Assumption~\ref{ass:hull}.  Because each
unit contributes only one observed date, this construction places no
restriction on any cross-date dependence observable under repeated
cross-sections.
\end{proof}

Independence is used only to exhibit one admissible completion. The theorem identifies a marginal law; it neither assumes nor concludes that the missing untreated outcome is independent of the treated outcome. Any copula between $Y_1(0)$ and $Y_1(1)$ for treated units is admissible if it preserves the required marginals and observed law.

\paragraph{Real-outcome feasible weights.}

For real outcomes write
\[
H_\lambda(y)=F^0_{0,1}(y)+\sum_{s\in S}\lambda_s\{F^0_{1,s}(y)-F^0_{0,s}(y)\}.
\]

\begin{proposition}[Compact coherent-weight set]\label{S-prop:weight-compact}
For finite $S$,
\[
\Lambda_S(P)=\Delta_S\cap\bigcap_{a<b}\{\lambda:H_\lambda(b)-H_\lambda(a)\geq0\}
\]
is closed, convex, and compact.
\end{proposition}

\begin{proof}
The tails and right continuity of $H_\lambda$ follow from the primitive CDFs, and a right-continuous function with CDF tails is a CDF if and only if it is nondecreasing.  For fixed $a<b$, the displayed increment is affine in $\lambda$, so each restriction is a closed half-space.  Intersecting these half-spaces with the compact simplex gives the claim.
\end{proof}

\begin{proposition}[Compact sharp CDF set]\label{S-prop:cdfset-compact}
The map $\lambda\mapsto H_\lambda$ is continuous from $\Delta_S$ into $(\ell^\infty(\mathbb R),\|\cdot\|_\infty)$ and therefore
\[
\Fcal_S(P)=\{H_\lambda:\lambda\in\Lambda_S(P)\}
\]
is compact and convex.
\end{proposition}

\begin{proof}
For $\lambda,\lambda'\in\Delta_S$,
\[
\|H_\lambda-H_{\lambda'}\|_\infty
\leq\sum_{s\in S}|\lambda_s-\lambda_s'|\,\|F^0_{1,s}-F^0_{0,s}\|_\infty.
\]
Thus the map is Lipschitz.  Convexity follows from linearity in $\lambda$ and convexity of $\Lambda_S(P)$.
\end{proof}

\section{When Coherence Matters: Proofs and Complements}\label{S-sec:supp-matters}

\subsection{The full signed hull}

Let $v_s$, $\mathcal H_S$ and $\mathcal H_S^+$ be as in Section~\ref{sec:matters}.

\begin{proposition}[No-gain criterion for the full law]\label{S-prop:no-gain}
$\mathcal H_S^+=\mathcal H_S$ if and only if every $v_s$ is a probability law. In that case coherence leaves the range of every functional on the hull unchanged. If some $v_s$ is invalid, the inclusion $\mathcal H_S^+\subsetneq\mathcal H_S$ is strict, but the range of a given functional may still be unchanged.
\end{proposition}
\begin{proof}
The probability laws form a convex set, so valid vertices give a valid hull. Conversely, equality puts every vertex in $\PP(\mathcal Y)$. Set equality implies equality of all functional images; strict inclusion does not. For instance, on three cells with $v_1=(-0.1,0.6,0.5)$ and $v_2=(0.2,0.3,0.5)$, mixtures with weight at most $2/3$ on $v_1$ are coherent, yet the third-cell probability equals $0.5$ for every candidate.
\end{proof}

The width lost by ignoring coherence decomposes into the two endpoint gains characterized in Proposition~\ref{prop:functional-gain}:
\begin{proposition}[Exact width cost of ignoring coherence]
\label{S-prop:premium-decomposition}
Suppose the coherent set $\Lambda_S$ is nonempty and let $\ell$ be a
linear functional finite on the historical vertices. With
$a_s=\ell(v_s)$, define $L^+=\min_{\lambda\in\Lambda_S}
\sum_s\lambda_sa_s$ and $U^+=\max_{\lambda\in\Lambda_S}
\sum_s\lambda_sa_s$. Then the reduction in the width of the
unrestricted signed-hull range is
\[
 \mathsf G_\ell=\bigl(\max_s a_s-U^+\bigr)
                 +\bigl(L^+-\min_s a_s\bigr)\ge0.
\]
Both summands have a separate optimizing-face characterization given
in Proposition~\ref{prop:functional-gain}. In particular,
$\mathsf G_\ell>0$ if and only if at least one unrestricted
optimizing face has empty intersection with $\Lambda_S$.
There is no strictly positive universal gain merely because one
historical vertex is incoherent or because the historical bias
densities are linearly independent.
\end{proposition}
\begin{proof}
The unrestricted range has endpoints $\min_s a_s$ and $\max_s a_s$,
whereas the coherent range has endpoints $L^+,U^+$. Subtracting
widths gives the display, and each summand is nonnegative because
$\Lambda_S\subseteq\Delta_S$. Proposition~\ref{prop:functional-gain}
characterizes when the corresponding summand vanishes. As a
counterexample to any universal strict-gain claim, let the policy
score have identical expectations under all historical vertices;
then both ranges are singletons even if some vertices are invalid.
\end{proof}

\subsection{The three-cell calibration}\label{S-sec:calibration}
Consider three ordered wage categories, with a post-policy control law
$(.35,.35,.30)$, an observed treated law $(.20,.40,.40)$ and two transported
historical candidates
\[
v_1=(.45,.30,.25),\qquad v_2=(.10,-.35,1.25).
\]
Each candidate is obtained by adding a genuine historical treated--control
signed-measure difference to the current control law, but $v_2$ is not itself
a probability law. With weight $t$ on $v_2$ the candidate is
\[
p(t)=(.45-.35t,.30-.65t,.25+t),\qquad 0\le t\le1.
\]
Nonnegative probabilities require $t\le6/13$. The smallest coherent
counterfactual-minus-observed share below the illustrative wage floor is
\[
\min_{0\le t\le 6/13}\{.45-.35t-.20\}=23/260\simeq .08846.
\]
Optimizing over the signed hull without checking probability instead gives
$-.10$, a mathematically inadmissible lower bound. Under a less restrictive
model that permits threshold-specific historical weights \emph{and} imposes
CDF monotonicity, the same negative lower bound is attained, as verified next. Thus a policy ranking depends
both on probability coherence and on requiring one historical mixture across
thresholds. The difference between the lower bounds is $49/260\simeq0.18846$, or 18.85 percentage points. The signed-hull endpoint is inadmissible under the maintained model. This constructed example changes the lower-bound sign when invalid candidates are excluded.

\paragraph{Threshold-specific weights.}
For the example's first threshold, use the second historical candidate:
$F(y_1)=.10$. For the second threshold, use weight $t=.65$ on the
second candidate: $F(y_2)=.75-.65=.10$. The resulting two
coordinates are nondecreasing and determine the law
$(.10,0,.90)$. This law gives counterfactual-minus-observed
first-category share $-.10$. It does not satisfy the maintained
common-weight transport restriction, because its first and second
threshold weights differ. Thus the illustration distinguishes the
cross-threshold restriction from marginal CDF monotonicity.

\subsection{Witness programs and the CDF-distance screen}
\begin{proposition}[Finite-support margin and linear programs]\label{S-prop:witness-lp}
Suppose $\mathcal Y=\{y_1,\ldots,y_K\}$ and $p_s=(M_s\{y_k\})_{k=1}^K$. Let $p_\lambda=\sum_s\lambda_sp_s$. Then
\begin{equation}\label{S-eq:witness-lp-dual}
d_S^{\TV}:=\min_{\lambda\in\Delta_S}\sum_{k=1}^K[-p_{\lambda,k}]_+
=\max_{f\in[0,1]^K}\min_{s\in S}(-f^{\mathsf T}p_s).
\end{equation}
Both extrema are attained. The left side is the value of the linear program
\[
\min_{\lambda,r}\ \sum_{k=1}^K r_k
\quad\text{subject to}\quad
\lambda\in\Delta_S,\quad r_k\ge0,\quad r_k\ge-\sum_s\lambda_sp_{s,k}\quad(k=1,\ldots,K),
\]
and the right side is the value of
\[
\max_{f,t}\ t
\quad\text{subject to}\quad
0\le f_k\le1\quad(k=1,\ldots,K),\quad
t\le-\sum_{k=1}^K f_kp_{s,k}\quad(s\in S).
\]
In particular, the historical hull is incoherent precisely when $d_S^{\TV}>0$.
\end{proposition}

\begin{proof}
For each $\lambda$, coordinatewise maximization over the cube gives
\[
\sum_k[-p_{\lambda,k}]_+=\max_{f\in[0,1]^K}(-f^{\mathsf T}p_\lambda).
\]
The simplex and cube are compact convex sets and the payoff is continuous and bilinear. Thus its minimum over $\lambda$ and maximum over $f$ commute. Since a linear functional on a simplex attains its minimum at a vertex,
\[
\min_{\lambda\in\Delta_S}(-f^{\mathsf T}p_\lambda)=\min_{s\in S}(-f^{\mathsf T}p_s),
\]
proving \eqref{S-eq:witness-lp-dual}. For fixed $\lambda$, the smallest feasible $r_k$ is $[-p_{\lambda,k}]_+$, which establishes the first program. For fixed $f$, the largest feasible $t$ is $\min_s(-f^{\mathsf T}p_s)$, which establishes the second. Compactness attains the extrema in \eqref{S-eq:witness-lp-dual}; although $t$ is formally unbounded below in the displayed program, an optimizer exists by its equivalent compact formulation. Finally, a signed unit-mass vector is a probability vector exactly when all of its negative parts vanish; continuity and compactness ensure that an empty coherent set has $d_S^{\TV}>0$.
\end{proof}

Here $d_S^{\TV}$ is a population total-variation distance to coherence. It should not be substituted for the empirical CDF sup-norm screen of the main paper: empirical and nonatomic laws are separated in total variation, and the sampling analysis uses a different norm.

\subsection{A sampling-adjusted finite-support certificate}\label{S-sec:adjusted-witness}

Suppose the outcome takes values in an ordered, prespecified $K$-point support. The empirical vertex vectors $\widehat p_s$ and population vectors $p_s$ each have total mass one. Put $F_s(k)=\sum_{i=1}^kp_{s,i}$ and $\widehat F_s(k)=\sum_{i=1}^k\widehat p_{s,i}$ for $k<K$. Let $\eta=\max_{s,k<K}|\widehat F_s(k)-F_s(k)|$. For $f\in[0,1]^K$, write $V(f)=\sum_{k<K}|f_k-f_{k+1}|$.

\begin{proposition}[Uniform score correction and observable program]\label{S-prop:adjusted-witness-supp}
For every realization and every $\varepsilon\geq\eta$,
\begin{equation}\label{S-eq:adjusted-witness-supp}
0\leq \widehat L_S(\varepsilon)
:=\max_{f\in[0,1]^K}\left\{\min_s(-f^{\mathsf T}\widehat p_s)-\varepsilon V(f)\right\}
\leq d_S^{\TV}.
\end{equation}
The statistic is the optimal value of the linear program
\[
\max_{f,a,t}\quad t-\varepsilon\sum_{k=1}^{K-1}a_k
\]
subject to $0\leq f_k\leq1$, $a_k\geq f_k-f_{k+1}$, $a_k\geq f_{k+1}-f_k$, $a_k\geq0$, and $t\leq-f^{\mathsf T}\widehat p_s$ for every $s$. If a measurable, possibly data-dependent $\varepsilon$ satisfies $\Pr\{\eta\leq\varepsilon\}\geq1-\alpha$, rejection on $\{\widehat L_S(\varepsilon)>0\}$ has size at most $\alpha$ under coherence. If $\widehat L_S(\varepsilon)>0$, then the empirical candidate hull has CDF sup-norm distance strictly greater than $\varepsilon$ from the CDF polytope.
\end{proposition}

\begin{proof}
Set $e_{s,k}=\widehat F_s(k)-F_s(k)$ and note $e_{s,0}=e_{s,K}=0$, the latter by equal total mass. Discrete summation by parts yields the exact identity
\[
f^{\mathsf T}(\widehat p_s-p_s)
=\sum_{k=1}^K f_k(e_{s,k}-e_{s,k-1})
=\sum_{k=1}^{K-1}(f_k-f_{k+1})e_{s,k}.
\]
It follows that $|f^{\mathsf T}(\widehat p_s-p_s)|\leq\eta V(f)$ simultaneously for all vertices and all scores. Hence, whenever $\varepsilon\geq\eta$,
\[
\min_s(-f^{\mathsf T}\widehat p_s)-\varepsilon V(f)
\leq\min_s(-f^{\mathsf T}p_s)\leq d_S^{\TV},
\]
where the last inequality uses Proposition~\ref{S-prop:witness-lp}. Taking a maximum establishes the upper bound in \eqref{S-eq:adjusted-witness-supp}; $f=0$ establishes its lower bound. For fixed $f$, the largest feasible $t$ is $\min_s(-f^{\mathsf T}\widehat p_s)$, and the smallest feasible $a_k$ is $|f_k-f_{k+1}|$. If $\varepsilon=0$, a minimal $a$ remains available. This proves the program and attainment follows from continuity on the compact score cube.

Under the coherent null, $d_S^{\TV}=0$. On $\{\eta\leq\varepsilon\}$, inequality \eqref{S-eq:adjusted-witness-supp} therefore rules out rejection regardless of how $f$ was selected. If instead a probability vector $q$ has CDF sup-norm distance at most $\varepsilon$ from some empirical mixture $\widehat p_\lambda$, the same summation-by-parts identity applied to $\widehat p_\lambda-q$ gives $-f^{\mathsf T}\widehat p_\lambda\leq\varepsilon V(f)$, since $f^{\mathsf T}q\geq0$. The inequality $\min_s(-f^{\mathsf T}\widehat p_s)\leq-f^{\mathsf T}\widehat p_\lambda$ implies $\widehat L_S(\varepsilon)=0$. By attainment in the finite-support distance program, a strictly positive certificate therefore implies distance strictly above $\varepsilon$.
\end{proof}

The certificate in fact has an exact converse relative to the finite-support distance screen, proved next. If the support size is fixed, $d_S^{\TV}>0$, $\eta=o_P(1)$, and $\varepsilon=o_P(1)$, choose a population maximizer $f^\star$ in Proposition~\ref{S-prop:witness-lp}. Then $V(f^\star)\leq K-1$ and
\[
\widehat L_S(\varepsilon)\geq d_S^{\TV}-(\eta+\varepsilon)V(f^\star)>0
\]
with probability tending to one. This consistency claim concerns the fixed coarsening only; violations that disappear under coarsening are outside its scope.

\begin{theorem}[Exact duality with the CDF distance screen]\label{S-thm:exact-screen-witness}
Suppose $K\geq2$, $S$ is finite and nonempty, and each empirical vector $\widehat p_s\in\mathbb R^K$ has total mass one; its coordinates may be signed. Write $\widehat F_\lambda(k)=\sum_{i=1}^k\widehat p_{\lambda,i}$, with $\widehat p_\lambda=\sum_s\lambda_s\widehat p_s$, and let $\Dcal_K$ be the CDF polytope on these $K$ bins. Then
\begin{equation}\label{S-eq:exact-screen-witness}
\widehat d_{S,K}:=\min_{\lambda\in\Delta_S,\,Q\in\PP_K}
\max_{k<K}|\widehat F_\lambda(k)-F_Q(k)|
=\max_{\substack{f\in[0,1]^K\\V(f)\leq1}}
\min_{s\in S}(-f^{\mathsf T}\widehat p_s).
\end{equation}
For every $\varepsilon\geq0$, the adjusted score program in Proposition~\ref{S-prop:adjusted-witness-supp} obeys
\[
\widehat L_S(\varepsilon)>0
\quad\Longleftrightarrow\quad
\widehat d_{S,K}>\varepsilon.
\]
The strict inequalities are essential: when the tube touches the CDF polytope at distance $\varepsilon$, neither rule rejects.
\end{theorem}

\begin{proof}
For any zero-mass vector $a\in\mathbb R^K$, put $A_k=\sum_{i=1}^ka_i$, so $A_0=A_K=0$. Let
\[
\mathcal V_0=\{f\in\mathbb R^K:f_K=0,\ V(f)\leq1\}.
\]
This is a nonempty compact convex set: from $f_K=0$ and $V(f)\leq1$ it follows that $|f_k|\leq1$. Summation by parts gives $f^{\mathsf T}a=\sum_{k<K}(f_k-f_{k+1})A_k$. The adjacent differences of scores in $\mathcal V_0$ range over the closed unit ball of $\ell_1^{K-1}$, because they uniquely determine a score after $f_K=0$ is fixed. Therefore
\begin{equation}\label{S-eq:variation-duality}
\max_{f\in\mathcal V_0}f^{\mathsf T}a
=\max_{k<K}|A_k|.
\end{equation}
The simplex $\Delta_S$, the probability simplex $\PP_K$, and $\mathcal V_0$ are compact convex sets. By \eqref{S-eq:variation-duality} and finite-dimensional bilinear minimax applied to $(\lambda,Q)$ and $f$,
\[
\widehat d_{S,K}
=\max_{f\in\mathcal V_0}\min_{\lambda,Q}
f^{\mathsf T}(Q-\widehat p_\lambda)
=\max_{f\in\mathcal V_0}
\left\{\min_k f_k-\max_{s\in S}f^{\mathsf T}\widehat p_s\right\}.
\]
Every $\widehat p_s$ has total mass one. Replacing $f$ by $g=f-(\min_k f_k)\mathbf1$ leaves the expression in braces and $V(f)$ unchanged. The new score has minimum zero, and its maximum is at most its variation, hence at most one. It follows that the preceding maximum is no larger than the right side of \eqref{S-eq:exact-screen-witness}. Conversely, every $g\in[0,1]^K$ with $V(g)\leq1$ can be replaced by $f=g-g_K\mathbf1\in\mathcal V_0$, and the expression in braces becomes $\min_k g_k-\max_s g^{\mathsf T}\widehat p_s\leq\min_s(-g^{\mathsf T}\widehat p_s)$. In fact an optimizer of the latter can always be shifted down by its minimum, increasing its value by that minimum and preserving feasibility. Thus an optimizer with $\min_k g_k=0$ exists, and both maxima coincide. This proves \eqref{S-eq:exact-screen-witness}.

If $\widehat d_{S,K}>\varepsilon$, a maximizing score $g$ in \eqref{S-eq:exact-screen-witness} has $V(g)\leq1$ and $\min_s(-g^{\mathsf T}\widehat p_s)=\widehat d_{S,K}$. Its value in the adjusted score program is at least $\widehat d_{S,K}-\varepsilon V(g)\geq\widehat d_{S,K}-\varepsilon>0$. Conversely, Proposition~\ref{S-prop:adjusted-witness-supp} shows that a positive adjusted score precludes a probability vector within CDF distance $\varepsilon$ of the empirical hull. The minimum defining $\widehat d_{S,K}$ is attained on compact sets, so it must be strictly greater than $\varepsilon$. At equality the latter implication rules out a positive adjusted score, while the zero score gives $\widehat L_S(\varepsilon)=0$.
\end{proof}

Equation~\eqref{S-eq:exact-screen-witness} also gives a direct dual LP for the original distance calculation: maximize $t$ subject to $t\leq-f^{\mathsf T}\widehat p_s$ for every $s$, $0\leq f_k\leq1$, and auxiliary variables $a_k\geq|f_k-f_{k+1}|$ with $\sum_{k<K}a_k\leq1$. An optimal score identifies which bins, possibly in combination, make the candidate hull incompatible with any CDF. This identity applies to the fixed grid only. It preserves the paper's original CDF-distance gate and does not change its finite-sample or design-asymptotic calibration.

\subsection{Comparison with robust difference-in-differences under scalar bias stability}
\label{S-sec:supp-bkcd-comparison}
\citet{BanKedagniChenDenteh2026}
combines scalar selection-bias stability with several distinct relaxations:
parallel-trends robustness under no anticipation, bounded anticipation
under stronger parallel trends, a jointly relaxed model, and
minimum data-consistent relaxations. Their parameters are scalar
average treatment effects on the treated. Their sharpness constructions
concern that scalar target, not the entire missing marginal distribution.
Appendix~F.1.4 of their revision also uses the term
``distributionally robust'' for a sensitivity envelope obtained by varying
probability laws of \emph{scalar baseline selection biases} under decision losses.
That construction recovers their scalar ATT interval; it is not a sharp
identified set of untreated \emph{outcome distributions}. The difference is
one of the objects being distributed, not of terminology alone.
This paper retains no anticipation and imposes the historical restriction
on signed measures. The two projects have different identified targets
and their identifying conditions should not be conflated.

The following elementary equivalence gives the exact boundary of the
claim that a law-level restriction is stronger than scalar bias-hull
restrictions. Let $\mathcal B_b(\mathcal Y)$ denote bounded measurable
real functions on $(\mathcal Y,\mathcal B)$ and let $S$ be nonempty and
finite. All measures in the proposition are finite signed measures.

\begin{proposition}[Complete scalar family and the measure hull]
\label{S-prop:supp-all-score-hull}\label{S-prop:all-scores}
Given signed measures $\nu_*,\nu_s$, $s\in S$, the following are
equivalent:
\begin{enumerate}
\item There is one $\lambda\in\Delta_S$ with
$\nu_*=\sum_{s\in S}\lambda_s\nu_s$ as signed measures.
\item For every $f\in\mathcal B_b(\mathcal Y)$,
\[
\min_{s\in S}\int f\,d\nu_s
\le \int f\,d\nu_*
\le \max_{s\in S}\int f\,d\nu_s.
\]
\end{enumerate}
The analogous statement for a fixed finite collection of indicators
is false in general.
\end{proposition}
\begin{proof}
The first assertion implies the second by integration. Conversely,
for any finite collection of measurable sets $A_1,\ldots,A_m$, set
$v_*=(\nu_*(A_j))_{j=1}^m$ and
$v_s=(\nu_s(A_j))_{j=1}^m$. If $v_*$ were outside the compact convex
hull of the $v_s$, finite-dimensional strict separation would give
$a\in\mathbb R^m$ with
$a^\top v_*>\max_{s\in S} a^\top v_s$ (reverse $a$ if needed).
The bounded measurable score $f=\sum_{j=1}^m a_j\mathbf 1_{A_j}$
would contradict the second assertion. Thus the set
\[
L_{A_1,\ldots,A_m}=
\left\{\lambda\in\Delta_S:
\nu_*(A_j)=\sum_{s\in S}\lambda_s\nu_s(A_j)
\text{ for } j=1,\ldots,m\right\}
\]
is nonempty for every finite list. These sets are closed in the
compact simplex $\Delta_S$, and have the finite-intersection property.
Their intersection over every measurable $A$ is therefore nonempty.
Any $\lambda$ in the intersection gives equality on every measurable
set and hence equality of signed measures. For a fixed finite
collection of indicators, the intersection need not contain weights
that match events outside that collection; eventwise weights may also
vary across the reported indicators. The finite-list bounds do not
imply the full signed-measure equality.
\end{proof}

For example, take three outcome categories and historical signed-bias
mass vectors $\nu_1=(0.2,-0.2,0)$ and
$\nu_2=(0,0.2,-0.2)$. The candidate
$\nu_*=(0.15,0,-0.15)$ has each coordinate between the corresponding
historical minima and maxima. Yet its first coordinate requires weight
$0.75$ on $\nu_1$, whereas its third coordinate requires weight $0.25$
on $\nu_1$. Thus separately valid bounds on all three category
indicators admit a bias vector that no common historical mixture can
generate. All three bias vectors can be constructed as differences of
probability vectors by using a uniform control distribution on these
categories. The missing restriction is common weight compatibility, not
a failure of the individual indicator bounds.

The proposition does not imply that the scalar framework of
\citet{BanKedagniChenDenteh2026} computes distributional sharp sets
or probability-valid historical mixtures: its primary targets and
assumptions differ. It does show that the phrase ``stronger than
scalar bounds'' must be qualified by the family of scalar scores
being compared. For a fixed bounded policy score, unrestricted
historical-hull optimization already produces the corresponding
scalar extrema. Probability coherence can tighten those extrema
only if an optimizer is excluded from the probability-valid hull.
The authors' anticipation and misspecification-robust constructions
are not implied by the no-anticipation distributional model here;
they require separate identifying restrictions and inference.

\subsection{Exact common-witness duality and minimum-TV relaxation}
Let $S$ be finite, and let $\mu_s$ be finite signed unit-mass measures on a common measurable space. Write $\mu_\lambda=\sum_s\lambda_s\mu_s$ for $\lambda\in\Delta_S$ and $b(\mu)=\mu^-(\mathcal Y)$. The following equality does not require finite outcome support.
\begin{proposition}[Exact signed-measure witness and minimum relaxation]\label{S-prop:general-duality-repair}
Let $\rho=\sum_s|\mu_s|$. Then
\[
\rho_*=\min_{\lambda\in\Delta_S}b(\mu_\lambda)
=\max_{f\in L^\infty(\rho),\;0\le f\le1}\min_{s\in S}\left\{-\int f\,d\mu_s\right\}.
\]
The maximum is attained. Moreover, $\rho_*$ is the smallest $\tau\ge0$ for which there exist $\lambda\in\Delta_S$ and $Q\in\mathsf P(\mathcal Y)$ with $d_{\TV}(Q,\mu_\lambda)\le\tau$. For a partition into $J$ cells, this latter feasibility question is a linear program.
\end{proposition}
\begin{proof}
The Hahn decomposition gives $b(\mu)=\max_{0\le f\le1}-\int f\,d\mu$ for each signed unit-mass $\mu$. The simplex $\Delta_S$ is compact and convex; the order interval $\{f\in L^\infty(\rho):0\le f\le1\}$ is weak-star compact and convex. Each signed $\mu_s$ has an $L^1(\rho)$ density, so the bilinear payoff is continuous in $\lambda$ and weak-star continuous in $f$. The minimax theorem exchanges $\min_\lambda$ and $\max_f$. Minimizing a linear functional over the simplex selects one vertex, yielding the displayed identity. Compactness gives attainment. Finally, the probability-coherence theorem identifies $\min_{Q\in\mathsf P}d_{\TV}(Q,\mu_\lambda)$ with $b(\mu_\lambda)$. Minimizing over $\lambda$ proves the relaxation statement. On $J$ cells, introduce $z=\sum_s\lambda_s z_s$, $q\ge0$ with $\sum_jq_j=1$, and nonnegative slacks $u_j\ge q_j-z_j$, $u_j\ge z_j-q_j$. The constraint $\sum_ju_j\le2\tau$ enforces the total-variation neighborhood. All constraints are linear.
\end{proof}

A zero value of $\rho_*$ means that at least one coherent mixture exists; individual historical vertices may still be incoherent and policy endpoints may still tighten. A positive value means that the maintained historical-hull model is falsified at the population distribution under examination. The minimum-TV relaxation is then a transparent *distance to model consistency*, not evidence that the substantive counterfactual has in fact moved by that distance. 

\subsection{Stacked and eventwise coherence}\label{S-sec:stacking}

We now make explicit what is required to apply the historical restriction to an average of events occurring in different calendar quarters. All objects are probability laws or finite signed measures on one measurable outcome space $(\mathcal Y,\mathcal B)$. Let $e=1,\ldots,E$ index policy events. At event $e$, write $C_e$ for the law of the comparison population at the outcome date and $D_{eh}=T_{eh}-C_{eh}$ for the historical treated-minus-comparison signed measure at offset $h\in H$, where $T_{eh}$ and $C_{eh}$ are probability laws and $H$ is a fixed finite set of offsets. In particular, every $D_{eh}$ has mass zero. Let $\Delta_H$ be the simplex of nonnegative historical weights. Take nonnegative event weights $w_e$ summing to one, fixed independently of the outcome. In the wage application, $E=5$ and $w_e=1/5$. Define
\[
 Q_{\lambda}=\sum_{e=1}^Ew_e\biggl(C_e+\sum_{h\in H}\lambda_hD_{eh}\biggr),
 \qquad \lambda\in\Delta_H.
\]
The aggregate expression is a candidate probability law only when its cells are nonnegative. Event-specific candidates need not be probabilities for every $\lambda$, even when the aggregate candidate is.

\begin{proposition}[Heterogeneous event weights need not aggregate]\label{S-prop:aggregation-heterogeneity}
Suppose an event's untreated law satisfies
\[
 Q_e^0=C_e+\sum_{h\in H}\lambda_{eh}D_{eh}+R_e,
 \qquad \lambda_e\in\Delta_H,
\]
where $R_e$ is a finite signed measure of mass zero. Define the stacked untreated law $Q^0=\sum_ew_eQ_e^0$ and, for a signed mass-zero measure $\nu$, $\|\nu\|_{\TV}=\sup_{A\in\mathcal B}|\nu(A)|$. Then for every $\lambda\in\Delta_H$,
\begin{equation}\label{S-eq:stacking-drift-bound}
 \|Q^0-Q_{\lambda}\|_{\TV}
 \le \sum_{e=1}^Ew_e\left\{\|R_e\|_{\TV}
 +\sum_{h\in H}|\lambda_{eh}-\lambda_h|\,\|D_{eh}\|_{\TV}\right\}.
\end{equation}
Even when $R_e=0$ for all events and each event satisfies an exact historical-mixture restriction, there may be no common $\lambda$ with $Q^0=Q_{\lambda}$.
\end{proposition}

\begin{proof}
Subtract the two displays and use linearity:
$Q^0-Q_{\lambda}=\sum_ew_e\{R_e+\sum_h(\lambda_{eh}-\lambda_h)D_{eh}\}$.
The triangle inequality for the signed-measure total-variation seminorm gives \eqref{S-eq:stacking-drift-bound}.
To see strict failure, take two equally weighted events, two historical offsets, and the binary outcome space $\{0,1\}$. In each event let the comparison law be $\delta_0$ at the outcome date and at both offsets, $C_e=C_{e1}=C_{e2}=\delta_0$. Set $(T_{11},T_{12})=(\delta_0,\delta_1)$ and $(T_{21},T_{22})=(\delta_1,\delta_0)$. The untreated event law is $Q_e^0=\delta_0$ for both events, with event-specific weights $\lambda_1=(1,0)$ and $\lambda_2=(0,1)$ and $R_1=R_2=0$. Their stacked law is $\delta_0$. But both aggregated historical vertices are $\tfrac12\delta_0+\tfrac12\delta_1$, so every common-weight $Q_\lambda$ is that same mixture and cannot equal $Q^0$. All component laws in this construction are valid probabilities. Thus probability feasibility does not repair the event-weight aggregation error.
\end{proof}

The proposition is a bound, not an estimated restriction: the event-level weight vectors $\lambda_e$ and residuals $R_e$ are not observed. An economic argument that all events share the same historical response mechanism would impose $\lambda_e=\lambda$, while a weaker argument could supply substantive upper bounds on their differences. A good pre-policy fit of the \emph{stacked} histories alone establishes neither condition.

\begin{proposition}[Exact finite-grid test for eventwise coherence]
\label{S-prop:stacked-eventwise-lp-test}\label{S-prop:eventwise-test}
Fix $E,H,J<\infty$, event weights $w_e\geq0$ with
$\sum_ew_e=1$, and unit-mass signed vectors
$v_{eh}\in\mathbb R^J$. For a common historical vector
$\lambda\in\Delta_H$, set
\[
 p_e(\lambda)=\sum_{h=1}^H\lambda_hv_{eh},\qquad
 \bar p(\lambda)=\sum_{e=1}^Ew_ep_e(\lambda).
\]
Let
\begin{align*}
 \Lambda_{\rm stack}
 &=\{\lambda\in\Delta_H:\bar p_j(\lambda)\geq0
       \text{ for all }j\},\\
 \Lambda_{\rm event}
 &=\{\lambda\in\Lambda_{\rm stack}:p_{e,j}(\lambda)\geq0
       \text{ for all }e,j\}.
\end{align*}
Suppose $\Lambda_{\rm stack}\ne\varnothing$. Define
\[
 m_{ej}=\min_{\lambda\in\Lambda_{\rm stack}}
             \sum_{h=1}^H\lambda_hv_{eh,j}.
\]
Then $\Lambda_{\rm event}=\Lambda_{\rm stack}$ if and only if
$m_{ej}\geq0$ for every event--cell pair $(e,j)$. Every $m_{ej}$ is
the attained optimum of a linear program with $H$ simplex variables
and $J$ stacked nonnegativity inequalities. If some $m_{ej}<0$,
its minimizing weights are an explicit witness to strict inclusion
$\Lambda_{\rm event}\subsetneq\Lambda_{\rm stack}$.
If each $v_{eh}$ is a probability vector, equality holds without
solving these programs. The converse to this sufficient condition is
false.
\end{proposition}
\begin{proof}
Each $p_e(\lambda)$ and $\bar p(\lambda)$ has total mass one.
The stacked feasible set is a nonempty compact polytope, so each
linear objective defining $m_{ej}$ attains its minimum. Every
$\lambda\in\Lambda_{\rm stack}$ satisfies all eventwise
inequalities exactly when each minimum over that set is nonnegative.
If a minimum is negative, its optimizer belongs to
$\Lambda_{\rm stack}$ but violates one of the defining eventwise
inequalities. If every individual historical vertex is a probability
vector, convexity implies $p_e(\lambda)\geq0$ for all
$\lambda\in\Delta_H$. For failure of the converse, take two events
with equal aggregation weights, one history, and vectors
$v_{11}=(-1/10,11/10)$ and $v_{21}=(2/10,8/10)$: their stacked
average is $(1/20,19/20)$ although the first event vector is not
probability-valid. Conversely, equality of the two weight sets can
hold even with an invalid vertex: with two histories and identical
events, $v_{e1}=(-1/10,11/10)$ and $v_{e2}=(2/10,8/10)$, stacked
nonnegativity is the same as eventwise nonnegativity, despite the
invalid first historical vertex.
\end{proof}

This test is a population characterization for a fixed exhaustive
partition. Applying it to estimated cell masses without joint sampling
error control is a plug-in diagnostic. It makes no assertion that
coherence on a coarse partition persists under refinement.

\begin{proposition}[Event-specific weights]\label{S-prop:event-specific}
Suppose each event $e$ has its own weight $\lambda_e\in\Delta_S$ with $\mu_{e,\lambda_e}\in\PP(\mathcal Y)$, and write $\Lambda_e=\{\lambda\in\Delta_S:\mu_{e,\lambda}\in\PP(\mathcal Y)\}$. (i) The sharp set for the averaged untreated law is the Minkowski average $\mathcal Q^{\rm sp}=\sum_e\omega_e\{\mu_{e,\lambda}:\lambda\in\Lambda_e\}$, nonempty if and only if every $\Lambda_e$ is nonempty, and for a bounded score $u$ its range is $\sum_e\omega_e[\min_{\Lambda_e}\int u\,d\mu_{e,\lambda},\max_{\Lambda_e}\int u\,d\mu_{e,\lambda}]$. (ii) The eventwise set $\{\bar\mu_\lambda:\lambda\in\Lambda^{\rm ev}\}$ is contained in $\mathcal Q^{\rm sp}$. (iii) Neither $\mathcal Q^{\rm sp}$ nor the stacked set $\{\bar\mu_\lambda:\lambda\in\Lambda^{\rm st}\}$ need contain the other.
\end{proposition}

\begin{proof}
(i) The events' weights are unrestricted across events, so the set of attainable averaged laws is the weighted sum of the event-level sets, and a linear score is optimized event by event. (ii) Take $\lambda_e=\lambda$ for all $e$. (iii) In the two-cell example of Theorem~\ref{thm:stacking}(ii), $\Lambda_1=\varnothing$, so $\mathcal Q^{\rm sp}$ is empty while the stacked set is not. Conversely, on $\{0,1\}$ with two events, two dates, equal weights, and vertices $v_{1,1}=v_{2,2}=\delta_0$ and $v_{1,2}=v_{2,1}=\delta_1$, every vertex is a probability law, so $\mathcal Q^{\rm sp}$ is the set of all laws on $\{0,1\}$, whereas both averaged vertices equal $\frac12(\delta_0+\delta_1)$ and the stacked set is that single law.
\end{proof}

In the minimum-wage application each $\Lambda_e$ is empty on the 118-cell grid: the smallest negative mass of an event-level mixture is 0.05, 0.52, 0.23, 0.03 and 0.15 percentage points for New Jersey, California, Delaware, Michigan and Minnesota. The stacked set is nonempty only because negative bins of different events offset each other in the average.

\begin{proposition}[Approximately common weights]\label{S-prop:approx-common}
For $\delta\ge0$ let $\Lambda^\delta$ be the set of arrays $(\bar\lambda,\lambda_1,\ldots,\lambda_E)\in\Delta_S^{E+1}$ with $\|\lambda_e-\bar\lambda\|_1\le\delta$ and $\mu_{e,\lambda_e}\in\PP(\mathcal Y)$ for every $e$. (i) The sharp set for the averaged untreated law is $\mathcal Q^\delta=\{\sum_e\omega_e\mu_{e,\lambda_e}:(\bar\lambda,\lambda_1,\ldots,\lambda_E)\in\Lambda^\delta\}$; on a finite grid its bounds for a linear contrast are linear programs. (ii) $\mathcal Q^\delta$ is nondecreasing in $\delta$, equals the eventwise set at $\delta=0$, and equals the event-specific set $\mathcal Q^{\rm sp}$ for $\delta\ge2$. (iii) $\delta^*=\min\{\delta:\Lambda^\delta\ne\varnothing\}$ is attained, is the value of a linear program on a finite grid, and is finite if and only if every $\Lambda_e$ is nonempty.
\end{proposition}

\begin{proof}
(i) For a fixed feasible array, Theorem~\ref{thm:sharp-id} applied event by event gives a data-generating process consistent with the observed distributions whose event-$e$ untreated law is $\mu_{e,\lambda_e}$; the events concern different populations, so these processes can be combined. Conversely every law in the model has this form. On a grid, introduce $t_{e,s}\ge|\lambda_{e,s}-\bar\lambda_s|$ through the two linear inequalities $t_{e,s}\ge\pm(\lambda_{e,s}-\bar\lambda_s)$ and impose $\sum_st_{e,s}\le\delta$; coherence is $Z_e^\top\lambda_e\ge0$; the objective is linear. (ii) The constraint sets are nested in $\delta$. At $\delta=0$ every $\lambda_e$ equals $\bar\lambda$. For any two points of $\Delta_S$, $\|\lambda_e-\bar\lambda\|_1\le\|\lambda_e\|_1+\|\bar\lambda\|_1=2$, so the constraint is void for $\delta\ge2$. (iii) $\delta^*=\min\max_e\|\lambda_e-\bar\lambda\|_1$ over $\bar\lambda\in\Delta_S$ and $\lambda_e\in\Lambda_e$, a minimum of a continuous function over a compact set, empty exactly when some $\Lambda_e$ is; with the variables $t_{e,s}$ and an epigraph variable it is a linear program.
\end{proof}

In the minimum-wage application every $\Lambda_e$ is empty, so $\delta^*=\infty$: no degree of weight heterogeneity reconciles the five events with historical transport.

\subsection{Covariates}

Let $X$ be a pre-determined covariate with finite support $\mathcal X$, unaffected by treatment, and write $p_{g,t}(x)=\Pr(X=x\mid G=g,\text{date }t)$ and $\mu_{g,t}(\cdot\mid x)$ for conditional laws. Conditional transport states that for every $x$ there is $\lambda(x)\in\Delta_S$ with
\[
\mu^0_{1,1}(\cdot\mid x)-\mu^0_{0,1}(\cdot\mid x)=\sum_s\lambda_s(x)\{\mu^0_{1,s}(\cdot\mid x)-\mu^0_{0,s}(\cdot\mid x)\}.
\]

\begin{proposition}[Conditional transport]\label{S-prop:conditional}
(i) Under conditional transport with weights free across $x$, the sharp set for the treated group's untreated law at the policy date is the Minkowski sum $\sum_xp_{1,1}(x)\mathcal F_S(P\mid x)$, where $\mathcal F_S(P\mid x)$ is the sharp set of Theorem~\ref{thm:sharp-id} computed from the conditional laws; it is nonempty if and only if every conditional set is. With a common weight $\lambda(x)=\lambda$ the sharp set is $\{\sum_xp_{1,1}(x)\mu_{\lambda}(\cdot\mid x):\lambda\in\bigcap_x\Lambda_S(P\mid x)\}$. (ii) Suppose the covariate distribution in each group does not change over time, $p_{g,t}=p_g$, and conditional transport holds with a common weight $\lambda$. Then
\[
\nu_1-\sum_s\lambda_s\nu_s=\sum_x\{p_1(x)-p_0(x)\}\Bigl\{\mu^0_{0,1}(\cdot\mid x)-\sum_s\lambda_s\mu^0_{0,s}(\cdot\mid x)\Bigr\},
\]
so unconditional transport with weight $\lambda$ holds if the two groups have the same covariate distribution or if every conditional comparison law is the same mixture of its own past, and can fail otherwise.
\end{proposition}

\begin{proof}
(i) The treated group's untreated law is $\sum_xp_{1,1}(x)\mu^0_{1,1}(\cdot\mid x)$, with $p_{1,1}$ identified because $X$ is unaffected by treatment. The model restricts the conditional laws cell by cell, so the argument of Proposition~\ref{S-prop:event-specific}(i), with covariate cells in place of events, applies. (ii) Write $\nu_t=\sum_x\{p_1(x)\mu^0_{1,t}(\cdot\mid x)-p_0(x)\mu^0_{0,t}(\cdot\mid x)\}$ and add and subtract $p_1(x)\mu^0_{0,t}(\cdot\mid x)$: $\nu_t=\sum_xp_1(x)\nu_t(x)+\sum_x\{p_1(x)-p_0(x)\}\mu^0_{0,t}(\cdot\mid x)$, where $\nu_t(x)$ is the conditional bias. Conditional transport gives $\sum_xp_1(x)\nu_1(x)=\sum_s\lambda_s\sum_xp_1(x)\nu_s(x)$; subtracting $\sum_s\lambda_s\nu_s$ leaves the stated term.
\end{proof}
\subsection{Refinement}
\begin{proof}[Proof of Proposition~\ref{prop:partition-refinement}]
Let $S$ be finite, let $\Pi_n$ be nested finite measurable partitions whose
sigma-fields $\mathcal A_n=\sigma(\Pi_n)$ generate the outcome sigma-field
$\mathcal B$, and let $\mu_s$ ($s\in S$) be finite signed measures of unit mass.
For $\lambda\in\Delta_S$, put $\mu_\lambda=\sum_s\lambda_s\mu_s$ and
$b(\mu_\lambda)=\sup_{A\in\mathcal B}[-\mu_\lambda(A)]_+$.
Define the partition deficit
$ b_n(\mu_\lambda)=\sum_{A\in\Pi_n}[-\mu_\lambda(A)]_+$.
Each coarse cell is a union of finer cells, hence
$0\le b_n(\mu_\lambda)\le b_{n+1}(\mu_\lambda)\le b(\mu_\lambda)$.
The corresponding feasible-weight sets are decreasing and contain the
full-outcome feasible set.

Let $\rho=\sum_{s\in S}|\mu_s|$ and $g_s=d\mu_s/d\rho$.
Since every $\mu_s$ is finite, $\rho$ is finite. Conditional expectation
relative to $\rho/\rho(\mathcal Y)$ gives
\[
 b_n(\mu_\lambda)=\int\Big[-\sum_s\lambda_s\mathbb E_\rho(g_s\mid\mathcal A_n)\Big]_+d\rho.
\]
The increasing sigma-fields generate $\mathcal B$, so conditional-expectation
convergence gives $\|\mathbb E_\rho(g_s\mid\mathcal A_n)-g_s\|_{L^1(\rho)}\to0$
for every $s$. The negative-part map is one-Lipschitz, and therefore
\[
 \sup_{\lambda\in\Delta_S}|b_n(\mu_\lambda)-b(\mu_\lambda)|
 \le\max_{s\in S}\|\mathbb E_\rho(g_s\mid\mathcal A_n)-g_s\|_{L^1(\rho)}\longrightarrow0.
\]
Taking minima over $\Delta_S$ yields convergence of the optimized
partition deficits. Finally, a fixed signed measure that is nonnegative on
every cell of every partition is nonnegative on the algebra generated by
these cells. That algebra is dense in measurable sets for the metric
$\rho(A\triangle B)$; approximating an arbitrary measurable event shows
that its signed mass is nonnegative. Hence the intersection of the
partition-feasible weight sets equals the full-outcome feasible set.
\end{proof}

\subsection{Coarsening}\label{S-sec:coarsening}

\begin{proof}[Proof of Proposition~\ref{prop:partition-mesh}]
(i) Write $\mu_\lambda=P_{0,1}+\sum_s\lambda_sP_{1,s}-\sum_s\lambda_sP_{0,s}$ with $P_{g,t}=\mu^0_{g,t}$. For every event $A$, $\mu_\lambda(A)\ge-\sum_s\lambda_sP_{0,s}(A)\ge-1$, so $b(\mu_\lambda)\le1$. For attainment, let $x\ne y$ lie in one cell of $\Pi$, take one historical date, and set $P_{0,1}=P_{1,s}=\delta_x$, $P_{0,s}=\delta_y$. The candidate $2\delta_x-\delta_y$ has cell masses equal to those of $\delta_x$, so its grid repair is zero, while its negative mass is one.

(ii) Fix a mixture with density $g$ and a cell $A$ of length $\ell$, and let $\bar g$ be the average of $g$ on $A$. The coarsening loss on $A$ is
\[
\int_A[-g]_+-\Bigl[-\int_Ag\Bigr]_+=\tfrac12\Bigl(\int_A|g|-\Bigl|\int_Ag\Bigr|\Bigr)\le\tfrac12\int_A|g-\bar g| ,
\]
because $\int_A|g|\le\int_A|g-\bar g|+|\int_A g|$. The function $h=g-\bar g$ has mean zero on $A$ and is $L$-Lipschitz, since a mixture of $L$-Lipschitz densities is $L$-Lipschitz. We claim $\int_A|h|\le L\ell^2/4$. Let $h^\ast$ be the nonincreasing rearrangement of $h$ on an interval of length $\ell$. Rearrangement preserves the distribution of values, so $h^\ast$ has mean zero and $\int|h^\ast|=\int_A|h|$, and in one dimension it does not increase the Lipschitz constant. Let $c$ be a point where $h^\ast$ changes sign. Then $0\le h^\ast(x)\le L(c-x)$ for $x<c$ and $0\ge h^\ast(x)\ge-L(x-c)$ for $x>c$, so $\int h^\ast_+\le Lc^2/2$ and $\int h^\ast_-\le L(\ell-c)^2/2$. Mean zero makes the two integrals equal, hence each is at most $L\min\{c,\ell-c\}^2/2\le L\ell^2/8$, and $\int|h^\ast|\le L\ell^2/4$. The cell loss is therefore at most $L\ell^2/8$. Summing over cells and using $\sum_A\ell_A^2\le h(b-a)$ gives $b(\mu_\lambda)-b_\Pi(\mu_\lambda)\le Lh(b-a)/8$ for every $\lambda$, and minimization over a common simplex preserves the inequality: $d_S-d_{S,\Pi}\le b(\mu_{\lambda_\Pi})-b_\Pi(\mu_{\lambda_\Pi})$, where $\lambda_\Pi$ minimizes the grid repair. For the constant, $g(x)=L(x-\ell/2)$ on $[0,\ell]$ has cell loss $\int_0^{\ell/2}L(\ell/2-x)\,dx=L\ell^2/8$.
\end{proof}

A consequence is that refining the grid measures the population deficit consistently without forcing it to zero: if $d_S>0$, every partition with mesh $h<8d_S/\{L(b-a)\}$ has a strictly positive grid deficit. Part (i) shows why some smoothness is needed for any such statement.

\subsection{Smooth outcomes}\label{S-sec:smooth}

The probability restriction is not an artifact of discrete outcomes.
\begin{proposition}[Smooth distributions can have persistent coherence gains]
\label{S-prop:smooth-persistent}
There exist genuinely observed pre-policy probability laws on $[0,1]$,
a post-policy control law with a bounded density, and two transported
historical candidates with smooth cumulative signed-measure functions such
that the following hold:
(i) the coherent historical-mixture set is nonempty;
(ii) for a bounded, continuous policy score, the coherent and unrestricted
upper bounds differ by $1/2$;
(iii) on uniform outcome partitions with mesh tending to zero, the
corresponding finite-grid upper-bound difference converges to $1/2$.
Hence neither bounded densities nor Lipschitz cumulative distribution
functions imply vanishing coherence gains on fine grids.
\end{proposition}
\begin{proof}
Let the post-policy control density be $1$ on $[0,1]$. One historical
treated--control difference is zero, obtained from two uniform laws.
For the other, take genuine historical treated and control densities
$1+\sin(2\pi x)$ and $1-\sin(2\pi x)$, respectively. Both are
nonnegative and integrate to one. The transported candidates therefore
have densities $g_0(x)=1$ and $g_1(x)=1+2\sin(2\pi x)$.
A weight $\lambda$ on the second candidate yields
$g_\lambda(x)=1+2\lambda\sin(2\pi x)$.
Since $\min_{x\in[0,1]}\sin(2\pi x)=-1$, the candidate is a
probability density exactly when $0\le\lambda\le1/2$.
For $\ell(g)=\int_0^1\sin(2\pi x)g(x)\,dx$, direct integration gives
$\ell(g_\lambda)=\lambda$. Thus the signed-hull upper bound is $1$
and the coherent upper bound is $1/2$.

For equal-length partitions $\Pi_K$, let $\lambda_K$ be the largest
weight satisfying nonnegativity of every cell mass. The cell averages
of $\sin(2\pi x)$ attain values approaching $-1$ as $K\to\infty$;
therefore $\lambda_K\to1/2$. Define the partition score as the
within-cell average of $\sin(2\pi x)$. The corresponding linear
functional converges uniformly on the two-candidate hull to $\ell$,
because the scores converge uniformly to $\sin(2\pi x)$. Its
unrestricted upper endpoint tends to $1$, while its coherent upper
endpoint tends to $1/2$. The finite-grid gain thus tends to $1/2$.
All underlying observed densities are bounded and smooth, and all
associated CDFs are Lipschitz-continuous.
\end{proof}

\begin{proposition}[Sharp density-amplitude boundary]\label{S-prop:density-amplitude}
Let $f$ be a probability density with respect to a common $\sigma$-finite
measure $m$ on $\mathcal Y$, and let $\nu$ be an integrable signed density
with $\int\nu\,dm=0$. For $a\ge0$, put $g_a=f+a\nu$ and define
\[
 a_\star=\operatorname*{ess\,inf}_{\{y:\nu(y)<0\}}
                   \frac{f(y)}{-\nu(y)},
\]
with the essential infimum over the empty set defined as $+\infty$.
Then $g_a$ is a probability density if and only if $0\le a\le a_\star$.
If $\nu\not\equiv0$, the negative set has positive measure and
$a_\star<+\infty$. More generally, for historical
signed-density vertices $g_s=f+\nu_s$, $s\in S$, the full-law coherent
weight set is exactly
\[
 \Lambda_S=\left\{\lambda\in\Delta_S:
             f(y)+\sum_{s\in S}\lambda_s\nu_s(y)\ge0
             \quad m\text{-almost everywhere}\right\}.
\]
The entire historical hull is coherent if and only if
$\nu_s\ge-f$ almost everywhere for every $s\in S$.
\end{proposition}
For a zero-integral integrable signed density $\nu$ and baseline
probability density $f$, every $f+a\nu$ has total mass one.
Nonnegativity on $\{\nu\ge0\}$ is automatic for $a\ge0$; on
$\{\nu<0\}$ it requires $a\le f/(-\nu)$ almost everywhere.
Taking the essential infimum proves the exact amplitude threshold.
If $\nu\not\equiv0$ and has zero integral, its negative set has
positive measure. The ratio $f/(-\nu)$ is finite almost everywhere
there, so its essential infimum is finite. For several vertices,
nonnegativity of the entire convex hull is equivalent to that of
each vertex, by inspection of the extreme simplex weights.

The threshold can be zero when the baseline density vanishes where a
negative historical shift occurs. It is not determined by the maximum
absolute shift: positive shifts do not violate coherence. The amplitude condition
characterizes law validity, not whether any particular policy endpoint moves.

\section{Quantiles and Policy Rankings}\label{S-sec:supp-quantile}

Assume throughout that $S$ is finite and $\Lambda_S(P)\neq\varnothing$.

\begin{proposition}[Quantile endpoint attainment under polyhedral coherence]\label{S-prop:polyhedral-quantile-extrema}
Suppose the coherent weight set $\Lambda_S(P)$ is a nonempty polytope, and let $V$ be its finite set of extreme points. For every $\tau\in(0,1)$,
\[
\min_{\lambda\in\Lambda_S(P)}Q_{H_\lambda}(\tau)=\min_{v\in V}Q_{H_v}(\tau),
\qquad
\max_{\lambda\in\Lambda_S(P)}Q_{H_\lambda}(\tau)=\max_{v\in V}Q_{H_v}(\tau).
\]
In particular, both endpoints are attained when the outcome support is finite.
\end{proposition}
\begin{proof}
Every feasible $\lambda$ is a convex combination of the finitely many vertices $v\in V$, so $H_\lambda$ is the same convex combination of the vertex CDFs. Set $a=\min_{v\in V}Q_{H_v}(\tau)$ and $b=\max_{v\in V}Q_{H_v}(\tau)$. For any $y<a$, each vertex CDF obeys $H_v(y)<\tau$, hence so does $H_\lambda(y)$; therefore $Q_{H_\lambda}(\tau)\ge a$. At $y=b$ each vertex CDF satisfies $H_v(b)\ge\tau$, so every mixture satisfies $H_\lambda(b)\ge\tau$ and $Q_{H_\lambda}(\tau)\le b$. The vertex quantiles attaining $a$ and $b$ belong to the feasible set, proving both equalities. On finite outcome support, coherence is defined by finitely many linear cell constraints, so $\Lambda_S(P)$ is a polytope.
\end{proof}
The proposition concerns the extrema of a single quantile, not simultaneous attainment of the CDF envelope by a single mixture. It also does not imply that every value between the two quantile extrema is attainable.

\subsection{Envelope regularity}

Define
\[
\underline F(y)=\min_{\lambda\in\Lambda_S}H_\lambda(y),\qquad
\overline F(y)=\max_{\lambda\in\Lambda_S}H_\lambda(y).
\]

\begin{lemma}[Uniform right continuity over the historical simplex]\label{S-lem:uniform-rc}
If $y_n\downarrow y$, then
\[
\sup_{\lambda\in\Delta_S}|H_\lambda(y_n)-H_\lambda(y)|\to0.
\]
The analogous uniform tail limits hold as $y\to\pm\infty$.
\end{lemma}

\begin{proof}
By the triangle inequality,
\begin{align*}
|H_\lambda(y_n)-H_\lambda(y)|
&\leq |F^0_{0,1}(y_n)-F^0_{0,1}(y)|\\
&\quad+\sum_{s\in S}\lambda_s\{|F^0_{1,s}(y_n)-F^0_{1,s}(y)|+|F^0_{0,s}(y_n)-F^0_{0,s}(y)|\}.
\end{align*}
The right side is bounded by the sum over the finite collection of primitive CDF increments, which tends to zero.  The tail claim is identical.
\end{proof}

\begin{proposition}[CDF envelopes]\label{S-prop:cdf-envelopes}
Both $\underline F$ and $\overline F$ are CDFs.
\end{proposition}

\begin{proof}
Pointwise minima and maxima of nondecreasing functions are nondecreasing.  Right continuity and the tail limits follow from the preceding uniform lemma.
\end{proof}

\subsection{Generalized inverse}

For a CDF $F$, define $Q_F(\tau)=\inf\{y:F(y)\geq\tau\}$.

\begin{lemma}[Generalized-inverse equivalence]\label{S-lem:geninv}
For $\tau\in(0,1)$,
\[
Q_F(\tau)\leq y\quad\Longleftrightarrow\quad F(y)\geq\tau.
\]
\end{lemma}

\begin{proof}
If $F(y)\geq\tau$, the defining set contains $y$, so $Q_F(\tau)\leq y$.  Conversely, suppose $F(y)<\tau$.  By right continuity, there exists $\delta>0$ such that $F(y+\delta)<\tau$.  Monotonicity then implies $F(x)<\tau$ for every $x\leq y+\delta$, so the defining set is contained in $(y+\delta,\infty)$ and $Q_F(\tau)\geq y+\delta>y$.  Taking the contrapositive proves the forward implication.
\end{proof}

\begin{proposition}[Sharp quantile extrema]\label{S-prop:quantile-extrema}
For every $\tau\in(0,1)$,
\[
\inf_{\lambda\in\Lambda_S}Q_{H_\lambda}(\tau)=Q_{\overline F}(\tau),
\qquad
\sup_{\lambda\in\Lambda_S}Q_{H_\lambda}(\tau)=Q_{\underline F}(\tau).
\]
The infimum is attained.  Without further restrictions, the supremum is not asserted to be attained.
\end{proposition}

\begin{proof}
Let $q_\lambda=Q_{H_\lambda}(\tau)$.  By Lemma~\ref{S-lem:geninv},
\[
\{y:\exists\lambda,\ q_\lambda\leq y\}
=\{y:\max_\lambda H_\lambda(y)\geq\tau\}
=\{y:\overline F(y)\geq\tau\}.
\]
The infimum of the left set is $\inf_\lambda q_\lambda$, and the infimum of the right set is $Q_{\overline F}(\tau)$.  To establish attainment, let $\lambda_n\to\lambda$.  The affine representation and finiteness of $S$ imply $\lVert H_{\lambda_n}-H_\lambda\rVert_\infty\to0$.  If $x<Q_{H_\lambda}(\tau)$, then $H_\lambda(x)<\tau$, so $H_{\lambda_n}(x)<\tau$ eventually.  Thus
\[
\liminf_{n\to\infty}Q_{H_{\lambda_n}}(\tau)\geq Q_{H_\lambda}(\tau),
\]
and the generalized-inverse map is lower semicontinuous on $\Lambda_S$.  Since $\Lambda_S$ is compact, it attains its minimum.  This proves the first identity and the claimed minimum statement.

Similarly,
\[
\{y:q_\lambda\leq y\ \text{for every }\lambda\}
=\{y:\min_\lambda H_\lambda(y)\geq\tau\}
=\{y:\underline F(y)\geq\tau\}.
\]
The infimum of the left set is $\sup_\lambda q_\lambda$, while the infimum of the right set is $Q_{\underline F}(\tau)$.  This proves the second identity.  Lower semicontinuity does not supply a maximizer, and the generalized-inverse map is not upper semicontinuous without additional restrictions; accordingly, this endpoint is stated as a supremum.
\end{proof}

\subsection{Continuity under strict crossing}

\begin{proposition}[Continuity of the quantile map]\label{S-prop:quantile-cont}\label{S-prop:quantile-interval}
Fix $\tau\in(0,1)$ and let $I=[a,b]$ be compact.  Suppose $\Lambda_S$ is nonempty, $(\lambda,y)\mapsto H_\lambda(y)$ is continuous on $\Lambda_S\times I$, $H_\lambda(a)<\tau<H_\lambda(b)$ for every $\lambda\in\Lambda_S$, and the equation $H_\lambda(y)=\tau$ has a unique solution $q(\lambda)\in(a,b)$.  Then $Q_{H_\lambda}(\tau)=q(\lambda)$, the map $\lambda\mapsto q(\lambda)$ is continuous, and its image is a compact interval whose endpoints are attained.
\end{proposition}

\begin{proof}
Fix $\lambda$.  Monotonicity and continuity of $H_\lambda$, together with uniqueness, imply $H_\lambda(y)<\tau$ for $y<q(\lambda)$ and $H_\lambda(q(\lambda))=\tau$; hence the generalized inverse equals $q(\lambda)$.  Let $\lambda_n\to\lambda$.  Every subsequence of $q(\lambda_n)$ has, by compactness of $I$, a further subsequence converging to some $q^*\in I$.  Joint continuity gives $H_\lambda(q^*)=\tau$, and uniqueness forces $q^*=q(\lambda)$.  Thus $q(\lambda_n)\to q(\lambda)$.  Finally, $\Lambda_S$ is compact and convex.  Its continuous image is compact and connected, hence a compact interval, and its endpoints are attained.
\end{proof}

\subsection{Welfare programs and simultaneous ranking}

Let the common finite support be $y_1<\cdots<y_K$. Denote the current
control probability vector by $p_{0,1}$ and place the historical treated-
control probability differences in the columns of the $K\times |S|$ matrix
$D_S$. For a vector $u=(u(y_1),\ldots,u(y_K))'$, define
\[
\Lambda=\{\lambda\in\mathbb R^{|S|}:\lambda\geq0,\ 
\mathbf 1'\lambda=1,\ p_{0,1}+D_S\lambda\geq0\}.
\]
If $\Lambda$ is nonempty, the exact upper and lower counterfactual welfare
levels solve
\[
\max_{\lambda\in\Lambda}\{u'p_{0,1}+u'D_S\lambda\},\qquad
\min_{\lambda\in\Lambda}\{u'p_{0,1}+u'D_S\lambda\}.
\]
They are attained because $\Lambda$ is a compact polytope. If $g$ is the
treated-post probability vector, the sharp interval for the welfare
contrast has endpoints $u'g$ minus the maximum and minimum above,
respectively. In particular, a strictly positive lower endpoint certifies
a positive contrast for every full-data law in the maintained model.

For first-order dominance on this support, it suffices to compare the CDFs
at the first $K-1$ support points. The assertion that the treated-post law
dominates every coherent counterfactual is equivalent to
\[
G(y_k)\leq
\min_{\lambda\in\Lambda}\left\{F_{0,1}(y_k)
+\sum_{s\in S}\lambda_s[F_{1,s}(y_k)-F_{0,s}(y_k)]\right\},
\quad k=1,\ldots,K-1.
\]
Each minimum is a linear program, but its optimizer may depend on $k$.
These pointwise minima test a universal ranking; they do not assert that
one weight vector realizes the entire lower envelope.

To account for sampling error, let $\mathfrak C_{T,N}$ cover the entire
treated-post CDF and let $\mathfrak C_{F,N}$ contain the entire sharp
counterfactual set on a joint event $E_N$ of probability at least
$1-\alpha$. The projected welfare region
\[
\mathfrak C_{W,N}(u)=\left\{\int u\,dG'-\int u\,dF':
G'\in\mathfrak C_{T,N},\ F'\in\mathfrak C_{F,N}\right\}
\]
contains $\mathcal W_{u,S}(P)$ on $E_N$. Thus it covers the complete
identified interval with probability at least $1-\alpha$ whenever that
containment event is measurable, and with inner probability at least
$1-\alpha$ otherwise. The same joint event validates every prespecified
bounded welfare index simultaneously; it does not require a separate
critical value for each index. A dominance conclusion is justified only
when the joint CDF region supports its universal inequalities.

\section{Relaxation and Breakdown}\label{S-sec:breakdown-supp}

This section proves the finite-grid programs behind Section~\ref{sec:breakdown}, gives the counterexample showing that the two obstructions in the breakdown radius must share one historical weight, and states the exclusion-restriction result used in Section~\ref{sec:mw}. On a grid with $J$ cells, write $Z$ for the $H\times J$ matrix of vertex rows, $u\in\{0,1\}^J$ for the indicator of the outcome event, and $d_{\TV}(q,z)=\tfrac12\|q-z\|_1$.

\[
\mathcal Q_\tau(Z)=\Bigl\{q\in\Delta_J:\ \exists\lambda\in\Delta_H,\ \tfrac12\|q-Z^{\top}\lambda\|_1\le\tau\Bigr\},\qquad
\Theta_\tau=\{g-u^{\top}q:\ q\in\mathcal Q_\tau(Z)\}.
\]
\begin{proposition}[Minimum-consistency and sign-breakdown programs]\label{S-prop:exact-min-repair-sign}
Define $\rho_*:=\inf\{\tau\ge0:\mathcal Q_\tau(Z)\ne\varnothing\}$ and $\tau_{\mathrm{sign}}:=\inf\{\tau\ge0:\exists q\in\mathcal Q_\tau(Z),\ g-u^{\top}q\ge0\}$. Then:
\begin{enumerate}
\item The minimum consistency relaxation is
\[\rho_*=\min_{\lambda\in\Delta_H}\sum_{j=1}^J\left(-[Z^{\top}\lambda]_j\right)_+ .\]
\item Whenever $\tau\ge\rho_*$, $\Theta_\tau$ is a nonempty compact interval with sharp endpoints obtained by minimizing and maximizing $g-u^{\top}q$ over $q\in\mathcal Q_\tau(Z)$. It is nested in $\tau$. The program is linear after introducing $t_j\ge0$ with $-t_j\le q_j-[Z^{\top}\lambda]_j\le t_j$ and $\sum_jt_j\le2\tau$.
\item The smallest relaxation allowing a nonnegative effect is given by the linear program
\[\tau_{\mathrm{sign}}=\min_{q\in\Delta_J,\,\lambda\in\Delta_H}\frac12\|q-Z^{\top}\lambda\|_1\quad\text{subject to }u^{\top}q\le g.\]
If $\Theta_0$ is nonempty and has a strictly negative upper endpoint, then $\tau_{\mathrm{sign}}>0$ and every member of $\Theta_\tau$ is strictly negative for $\tau<\tau_{\mathrm{sign}}$. The boundary is attained.
\end{enumerate}
\end{proposition}
\begin{proof}
For each fixed signed vector $z$ of total mass one, its negative mass $b(z)=\sum_j(-z_j)_+$ equals $\min_{q\in\Delta_J}\tfrac12\|q-z\|_1$: moving its negative masses to zero necessarily changes the negative and positive parts by at least $b(z)$ each, and a probability vector supported on the positive coordinates attains that amount. Minimization over the compact simplex $\Delta_H$ proves the first claim. The set of triples $(q,\lambda,t)$ obeying the displayed linear constraints and a finite radius is a compact projection after restricting the irrelevant slack to $0\le t_j\le2\tau$; its image under the affine map $q\mapsto g-u^{\top}q$ is a compact interval because the original feasible set is convex. The inequalities are equivalent to the stated $\ell_1$ bound and yield sharp extrema. Monotonicity follows by nesting. Finally, the event $g-u^{\top}q\ge0$ is the linear constraint $u^{\top}q\le g$. The constrained minimum of a continuous function over $\Delta_J\times\Delta_H$ exists; the epigraph of its objective is given by the same $t_j$ inequalities. If a negative upper endpoint were attained at $\tau=0$, compactness rules out $\tau_{\mathrm{sign}}=0$. At $\tau<\tau_{\mathrm{sign}}$, no feasible pair has nonnegative effect, proving the last assertion.
\end{proof}

\begin{proposition}[Event-face distance and exact sign-breakdown decomposition]\label{S-prop:exact-event-face-distance}
Let $z\in\mathbb R^J$ satisfy $\sum_jz_j=1$, and let $A$ be a nonempty proper subset of the $J$ cells. Write
\[
 b(z)=\sum_{j=1}^J(-z_j)_+,
 \qquad z^+(A)=\sum_{j\in A}(z_j)_+,
 \qquad K_g=\{q\in\Delta_J:q(A)\le g\},\quad g\in[0,1].
\]
Then
\begin{equation}\label{S-eq:exact-event-face-distance}
 \min_{q\in K_g}\frac12\|q-z\|_1
 =\max\big\{b(z),\ z^+(A)-g\big\}.
\end{equation}
Consequently, for the historical vertices $z_h$ of unit mass,
\begin{equation}\label{S-eq:reduced-sign-breakdown}
 \tau_{\mathrm{sign}}(Z,g)
 =\min_{\lambda\in\Delta_H}\max\big\{
 b(z_\lambda),\ z_\lambda^+(A)-g\big\},
 \qquad z_\lambda=\sum_{h=1}^H\lambda_hz_h.
\end{equation}
Both minimizations attain their values. The second is a finite linear program after introducing upper envelopes for the positive and negative coordinate parts. The condition that $A$ be a proper subset is essential: when $A$ is the entire outcome space, $q(A)=1$ for every probability law, so the sign event is impossible for $g<1$. The formula addresses one prespecified event $A$; joint certification across several nested events is a different optimization.
\end{proposition}
\begin{proof}
Write $p=z^+$ for the coordinatewise positive part, and let $P_A=p(A)$, $P_{A^c}=p(A^c)$; thus $P_A+P_{A^c}=1+b(z)$. For a probability vector $q$, the equal-mass identity gives
\[
 \frac12\|q-z\|_1
 =\sum_{j=1}^J(p_j-q_j)_+.
\]
Indeed the right side is the mass removed from positive coordinates, and the same amount is added to the negative or depleted coordinates to obtain a unit-mass probability. Every admissible $q$ must remove at least $b(z)$, since $p$ has total mass $1+b(z)$. It must also remove at least $P_A-g$ from positive coordinates in $A$, since $q(A)\le g$. This proves the lower bound in \eqref{S-eq:exact-event-face-distance}.

For attainment, set $r=\max\{b(z),P_A-g\}$ and remove $r$ units of mass from $p$, never removing more than $p_j$ from an individual coordinate. If $P_A-g\le b(z)$, we may remove the total $b(z)$ so that the remaining mass in $A$ is at most $g$: remove $\max\{0,P_A-g\}$ from $A$ and distribute any remaining removal between $A$ and $A^c$ without exceeding their positive masses. This is possible because the total positive mass is $1+b(z)$ and $g\ge0$. The remaining mass is one and is therefore itself an element of $K_g$ at distance $b(z)$.
If $P_A-g>b(z)$, remove exactly $P_A-g$ from the positive coordinates in $A$; the remaining positive vector has total mass $1+b(z)-(P_A-g)<1$ and mass exactly $g$ in $A$. Add the missing $P_A-g-b(z)$ units to any cell in $A^c$, which is nonempty. The resulting probability vector has event mass $g$ and total mass one. In either case the amount removed from positive coordinates equals $r$, which proves equality.
Finally apply \eqref{S-eq:exact-event-face-distance} to every $z_\lambda$ and minimize over the compact simplex. Piecewise linearity ensures attainment, and introducing auxiliary variables for $(z_{\lambda j})_+$ and $(-z_{\lambda j})_+$ gives the stated LP.
\end{proof}

\paragraph{Interpretation.}
The quantity $b(z_\lambda)$ measures the price of producing any probability law, whereas $z_\lambda^+(A)-g$ measures how much positive mass must be removed from the policy event to permit a nonnegative effect. Their maximum, not their sum, is the exact repair cost because one mass transfer can address both obstructions. The minimum over $\lambda$ allows the optimal historical mixture to depend on the proposed event, but it never permits different mixtures across individual outcome cells.

\begin{proposition}[Separate screens need not have a common witness]\label{S-prop:joint-obstruction-gap}
Let $A=\{1\}$, $g=3/10$, and let the three-cell signed historical vertices be
\[
 z_1=(4/5,1/5,0),\qquad z_2=(1/10,11/10,-1/5).
\]
Both vertices have total mass one. For $t\in[0,1]$ put
$z_t=t z_1+(1-t)z_2$. Then
\[
 \min_{t\in[0,1]}b(z_t)=0,
 \qquad \min_{t\in[0,1]}\big\{z_t^+(A)-g\big\}=-1/5,
\]
whereas the joint breakdown radius is
\begin{equation}\label{S-eq:joint-obstruction-gap}
 \min_{t\in[0,1]}\max\{b(z_t),z_t^+(A)-g\}=1/9.
\end{equation}
More generally, if $f(\lambda)=b(z_\lambda)$ and
$h(\lambda)=z_\lambda^+(A)-g$, then
\begin{equation}\label{S-eq:joint-obstruction-dual}
 \min_{\lambda\in\Delta_H}\max\{f(\lambda),h(\lambda)\}
 =\max_{\eta\in[0,1]}\min_{\lambda\in\Delta_H}
 \{\eta f(\lambda)+(1-\eta)h(\lambda)\}.
\end{equation}
This is an equivalent scalar minimax formulation, not an additional identifying assumption.
\end{proposition}
\begin{proof}
The mixture is
$z_t=(1/10+7t/10,\ 11/10-9t/10,\ -1/5+t/5)$.
The first two coordinates are positive throughout $[0,1]$, and the
third is nonpositive. Consequently
\[
 b(z_t)=(1-t)/5,\qquad
 z_t^+(A)-g=7t/10-1/5.
\]
The first function decreases in $t$ and the second increases. Their
maximum is minimized where they coincide, at $t=4/9$, with common
value $1/9$. Each obstruction nevertheless has its own favorable
weight: $b(z_1)=0$, while $z_2^+(A)-g=-1/5$. Thus separate
optimization gives only the uninformative lower bound zero, rather
than the joint minimum $1/9$.

For the general identity, observe that
$\max\{f(\lambda),h(\lambda)\}$ equals
$\max_{\eta\in[0,1]}\{\eta f(\lambda)+(1-\eta)h(\lambda)\}$.
Both $f$ and $h$ are convex and continuous in $\lambda$, since they
are sums of positive parts of affine coordinates, while the objective
is affine in $\eta$. The simplexes are compact and convex. The
finite-dimensional convex--concave minimax theorem therefore allows
interchanging the minimum over $\lambda$ and the maximum over $\eta$.
This establishes \eqref{S-eq:joint-obstruction-dual}. The identity does
not justify interchanging the two \emph{separate} minimizations with
the maximum: that operation is precisely what the example rules out.
\end{proof}

\paragraph{What the counterexample establishes.}
It tests a potential shortcut, not an actual empirical violation.
A user may solve the coherence screen and the event-share screen
separately and find both favorable, but those solutions may use
incompatible historical weights. Joint optimization is required by
the original distributional identifying restriction. An appropriately
optimized scalar minimax program is a valid computational substitute.

\subsection{Stability of minimum repair and sign-breakdown radii}\label{S-sec:breakdown-stability}
A bound for the entire optimization is useful because the minimizing historical weights may change discontinuously as estimated cell frequencies vary. The following stability result does not require uniqueness, differentiability, or separation between active constraints.

Let $Z,Z'\in\mathbb R^{H\times J}$ have unit-mass rows $z_h,z'_h$, and let $g,g'\in[0,1]$. Let $u\in\{0,1\}^J$ select an event containing at least one cell, whose complement also contains at least one cell. Write
\[
 d(z,z')=\tfrac12\|z-z'\|_1,\qquad
 \varepsilon=\max_{1\le h\le H}d(z_h,z'_h),\qquad \delta=|g-g'|.
\]
Both signed rows and the probability vectors below have total mass one. Let $\rho_*(Z)$ and $\tau_{\mathrm{sign}}(Z,g)$ be the optimal values defined in Proposition~\ref{S-prop:exact-min-repair-sign}.

\begin{proposition}[Uniform perturbation bounds for optimized sensitivity values]\label{S-prop:breakdown-stability}
With the preceding definitions,
\[
 |\rho_*(Z)-\rho_*(Z')|\le\varepsilon,
 \qquad
 |\tau_{\mathrm{sign}}(Z,g)-\tau_{\mathrm{sign}}(Z',g')|
 \le\varepsilon+\delta.
\]
In particular, on any simultaneous sampling event for which the estimated rows and event share obey these deterministic bounds, the population minimum repair and sign-breakdown values belong to the corresponding closed intervals around the plug-in solutions. If the population model is feasible at a relaxation $\tau$ and $\widehat\tau_{\mathrm{sign}}>\tau+\varepsilon+\delta$, every population effect admitted at that relaxation is strictly negative.
\end{proposition}
\begin{proof}
Set $z_\lambda=\sum_h\lambda_hz_h$ and $z'_\lambda=\sum_h\lambda_hz'_h$. For every $\lambda\in\Delta_H$, the triangle inequality yields
$ d(z_\lambda,z'_\lambda)\le\sum_h\lambda_hd(z_h,z'_h)\le\varepsilon$.
For any nonempty subset $K\subseteq\Delta_J$, the distance function
$D_K(z)=\min_{q\in K}d(q,z)$ is one-Lipschitz in $d$ by the triangle inequality; the reverse direction follows by exchanging $z$ and $z'$. Taking the minimum over $\lambda$ preserves the bound $\varepsilon$ when the same $K$ is used. Choosing $K=\Delta_J$ proves the first assertion.

For the second, put $K_g=\{q\in\Delta_J:u^\top q\le g\}$. For every $q\in K_g$, there is a $q'\in K_{g'}$ with $d(q,q')\le|g-g'|$. If $g'\ge g$, use $q'=q$. Otherwise, when $u^\top q>g'$, transfer mass $u^\top q-g'$ from the event cells to any fixed complementary cell; this is feasible because $u^\top q\le g$ and yields $d(q,q')=u^\top q-g'\le g-g'$. The same construction with the roles of $g$ and $g'$ exchanged proves that the Hausdorff distance (under $d$) between $K_g$ and $K_{g'}$ is at most $\delta$. Consequently, for any $z$, $|D_{K_g}(z)-D_{K_{g'}}(z)|\le\delta$. Combining this bound with $d(z_\lambda,z'_\lambda)\le\varepsilon$, and minimizing over the common weight simplex, proves the claimed bound for $\tau_{\mathrm{sign}}$. The last conclusion follows from the definition of the sign-breakdown radius and the strict inequality; population feasibility is stipulated separately, since an empty identified set would make a sign assertion vacuous.
\end{proof}

\subsection{Exclusion restrictions}\label{S-sec:anchoring}

Suppose the analyst knows a class $\mathcal U\subseteq\mathcal B$ of events on which the policy has no effect for the treated group, $\mu^1_{1,1}(B)=\mu^0_{1,1}(B)$ for $B\in\mathcal U$. Since $\mu^1_{1,1}=\mu_{1,1}$ is observed, this restricts the missing law.

\begin{proposition}[Sharp set under an exclusion restriction]\label{S-prop:anchoring}
Under Assumptions~\ref{ass:basic}--\ref{ass:hull} and the exclusion restriction, the sharp set of untreated laws is
\[
\{\mu_\lambda:\ \lambda\in\Lambda_S(P),\ \mu_\lambda(B)=\mu_{1,1}(B)\ \text{for all }B\in\mathcal U\}.
\]
On a finite grid with $\mathcal U$ generated by cells or unions of cells, the set of admissible weights is a polytope defined by the inequalities of $\Lambda_{S,\Pi}$ and the additional linear equalities, and bounds on linear contrasts are linear programs. If the set is empty, Assumption~\ref{ass:hull} and the exclusion restriction are jointly refuted at $P$.
\end{proposition}
\begin{proof}
Necessity is immediate. For sufficiency, the construction in the proof of Theorem~\ref{thm:sharp-id} (Section~\ref{S-sec:ident}) assigns the missing treated untreated outcome any coherent law in the hull, independently of observed data, and leaves the observed treated outcome unchanged; the exclusion restriction involves only these two marginals, so it holds for the constructed law whenever $\mu_\lambda$ satisfies the displayed equalities.
\end{proof}

In Section~\ref{sec:mw} we impose the single equality $\mu_\lambda(\{w\ge W\})=\mu_{1,1}(\{w\ge W\})$, which says that the policy does not change the population share employed at wages of at least $W$. Imposing equality cell by cell above $W$ is stronger and, with fine bins, would mostly test sampling noise.

\section{Inference: Proofs and Complements}\label{S-sec:supp-inference}

\subsection{Why plug-in coherence fails}

Let $\widehat F_{g,t}$ be within-group empirical CDFs and $\widehat H_\lambda=\sum_s\lambda_s(\widehat F_{0,1}+\widehat F_{1,s}-\widehat F_{0,s})$.

\begin{proposition}[Generic empirical incoherence]\label{S-prop:empirical-incoherence}
Suppose that, for every $s\in S$, $\widehat F_{0,s}$ jumps at a point where neither $\widehat F_{0,1}$ nor any $\widehat F_{1,u}$, $u\in S$, jumps. Then $\widehat H_\lambda$ is not a CDF for any $\lambda\in\Delta_S$. If outcomes are continuously distributed and every group--date sample is nonempty, the condition holds with probability one.
\end{proposition}
\begin{proof}
Some $\lambda_s>0$. At the specified jump of $\widehat F_{0,s}$, the term $-\lambda_s\widehat F_{0,s}$ produces a downward jump, the other negatively signed terms can only add to it, and no positively signed term jumps there. Hence $\widehat H_\lambda$ decreases at that point. With continuous outcomes, ties across distinct observations have probability zero.
\end{proof}

\subsection{Joint projection for quantile effects}

Let $\mathfrak C_{T,N}$ be a confidence region for the treated post-policy CDF $F_T$ and $\mathfrak C_N$ the coherence tube. Define $\mathfrak C_{\Theta,N}(\tau)=\{Q_{G}(\tau)-Q_F(\tau):\ G\in\mathfrak C_{T,N},\ F\in\mathfrak C_N\}$ and $\Theta_S(\tau)=\{Q_{F_T}(\tau)-Q_{H_\lambda}(\tau):\lambda\in\Lambda_S(P)\}$. Write $\Pp_*$ for inner probability.

\begin{corollary}[Quantile treatment effects]\label{S-cor:qte-projection}
If $\Pp_*\{F_T\in\mathfrak C_{T,N},\ \cF_S(P)\subseteq\mathfrak C_N\}\ge1-\alpha$, then $\Pp_*\{\Theta_S(\tau)\subseteq\mathfrak C_{\Theta,N}(\tau)\}\ge1-\alpha$. If the two events are measurable and calibrated separately at levels $\alpha_T$ and $\alpha_C$, the bound holds with $1-\alpha_T-\alpha_C$.
\end{corollary}
\begin{proof}
On the joint event, each $\theta\in\Theta_S(\tau)$ equals $Q_{F_T}(\tau)-Q_{H_\lambda}(\tau)$ with $F_T\in\mathfrak C_{T,N}$ and $H_\lambda\in\mathfrak C_N$, so $\theta\in\mathfrak C_{\Theta,N}(\tau)$. Monotonicity of inner probability gives the first claim and the union bound the second.
\end{proof}

Subtracting a plug-in treated quantile from a projected counterfactual quantile set is not a confidence procedure for $\Theta_S(\tau)$; the treated quantile is estimated too. Because the projection is applied to whole CDFs, it remains valid with atoms and flat stretches.

\subsection{The benchmark test}\label{S-sec:benchmark-supp}

Proposition~\ref{prop:benchmark-test} is deterministic given the simultaneous event. Two calibrations of that event are used in Section~\ref{sec:nsw}.

\emph{Finite-sample.} Suppose the benchmark sample, the treated sample and the comparison sample are independent, and within each the units are i.i.d.\ with complete histories. The $M+1=2|S|+2$ primitive CDFs are $F_{0,1}$, $F_{g,s}$ ($g=0,1$, $s\in S$), and $F_R$. Applying the two-sided DKW inequality with Massart's constant at level $\alpha/(M+1)$ to each, as in Proposition~\ref{S-prop:dkw-radius}, and the triangle inequality to $(\widehat F_R-F_R)-(\widehat H_s-H_s)$, gives the event with $c=r_R+\max_s(r_{0,1}+r_{1,s}+r_{0,s})$. Dependence across dates within a unit is unrestricted.

\emph{Bootstrap.} Let $\mathbb G_{N,s}=\sqrt N\{(\widehat F_R-\widehat H_s)-(F_R-H_s)\}$, $s\in S$. With the sample fractions bounded away from zero, the joint functional central limit theorem of Section~\ref{S-sec:emp}, applied separately to the independent samples, gives weak convergence of $(\mathbb G_{N,s})_{s\in S}$ in $\ell^\infty(\mathbb R)^{|S|}$ to a tight Gaussian process. The nonparametric bootstrap that resamples complete histories within each sample is consistent for this limit by the bootstrap empirical-process theorem for Donsker classes, and $\max_s\|\cdot\|_\infty$ is continuous. If the limit of $\max_s\|\mathbb G_{N,s}\|_\infty$ has a continuous distribution at its $1-\alpha$ quantile, the bootstrap critical value gives asymptotic coverage $1-\alpha$ for the event and hence asymptotic size $\alpha$ for the test and coverage for the lower confidence bound $\widehat D-c$. When comparison units are reweighted by an estimated propensity score, the weights must be re-estimated in each draw, and validity requires in addition that the weighted CDFs admit an asymptotically linear expansion; we use this version only as a robustness check.

The statistic $\widehat D$ is computed exactly by the linear program
\[
\min_{\lambda\in\Delta_S,\,t\ge0}\ t\quad\text{subject to}\quad
-t\le\widehat F_R(y_j)-\sum_s\lambda_s\widehat H_s(y_j)\le t,\quad j=1,\ldots,n,
\]
where $y_1<\cdots<y_n$ are the pooled sample points. Between consecutive pooled points all functions involved are constant, and below the smallest point all are zero, so the supremum over $\mathbb R$ equals the maximum over the pooled points.
\subsection{Finite grids: stability, sandwiches and certificates}\label{S-sec:finite-grid-proofs}
For a fixed partition with $K$ cells, let $S$ be the nonempty finite
set of historical dates, and let $q_s\in\mathbb R^K$ denote the
unit-mass signed cell vector transported from date $s\in S$.
Define the simplex $\Delta_S=\{\lambda:\lambda_s\ge0,\ \sum_{s\in S}\lambda_s=1\}$,
$q(\lambda)=\sum_s\lambda_sq_s$, and the coherent set
$\Lambda=\{\lambda\in\Delta_S:q_k(\lambda)\ge0\ \forall k\}$.
For corresponding estimates $\widehat q_s$, define $\widehat\Lambda$
by replacing $q_s$ with $\widehat q_s$ and set
$\varepsilon=\max_{s,k}|\widehat q_{s,k}-q_{s,k}|$.
Here $\widehat a_s$ estimates the prespecified linear-score
coefficient $a_s$ of vertex $s$; $d_{H,1}$ denotes Hausdorff
distance with respect to the simplex $\ell_1$ distance. For a
finite score vector $a=(a_s)_{s\in S}$, write
$\operatorname{osc}(a)=\max_s a_s-\min_s a_s$.
For deterministic $e\ge0$, define
$\widehat\Lambda_{\mathrm{in}}(e)=\{\lambda\in\Delta_S:
\sum_s\lambda_s\widehat q_{s,k}\ge e\ \forall k\}$ and
$\widehat\Lambda_{\mathrm{out}}(e)=\{\lambda\in\Delta_S:
\sum_s\lambda_s\widehat q_{s,k}\ge-e\ \forall k\}$.
The simultaneous vertex-error event is
$\mathcal E_e=\{\max_{s,k}|\widehat q_{s,k}-q_{s,k}|\le e\}$.
For the probability statements below, $e$ is a prespecified
nonnegative error allowance satisfying
$\Pr(\mathcal E_e)\ge1-\alpha$, where
$\alpha\in(0,1)$ is a specified error probability. The
``inner'' and ``outer'' sets refer to
$\widehat\Lambda_{\mathrm{in}}(e)$ and
$\widehat\Lambda_{\mathrm{out}}(e)$, respectively.

\begin{proposition}[Stability under strict finite-grid feasibility]
\label{S-prop:slater-stability}
Suppose $q_k(\lambda^\circ)\ge\kappa>0$ for every cell and some
$\lambda^\circ\in\Delta_S$, and $\varepsilon<\kappa$.
Then $\Lambda$ and $\widehat\Lambda$ are nonempty, and their
Hausdorff distance in simplex $\ell_1$ norm satisfies
\[
 d_{H,1}(\Lambda,\widehat\Lambda)\le 2\varepsilon/\kappa.
\]
For a linear score with vertex coefficients $a_s$ estimated by
$\widehat a_s$, let $\zeta=\max_s|a_s-\widehat a_s|$. The respective
minimum and maximum scores over the population and estimated feasible
sets differ by at most
$\zeta+(\varepsilon/\kappa)(\max_s\widehat a_s-\min_s\widehat a_s)$.
\end{proposition}

\begin{proposition}[Finite-grid feasibility sandwich]
\label{S-prop:grid-sandwich}
On $\mathcal E_e$, the inner set is contained in $\Lambda$
which is contained in the outer set. If the population $\Lambda$ is nonempty, rejecting its nonemptiness when the outer set is empty has probability at most $\alpha$. Conversely, nonrejection does not establish a coherent population model. If the joint event $\mathcal E_e\cap
\{\max_s|\widehat a_s-a_s|\le\zeta\}$ has probability
at least $1-\alpha$ for a prespecified $\zeta\ge0$, the whole identified range of the linear score lies in
\[
 \left[\min_{\lambda\in\widehat\Lambda_{\mathrm{out}}(e)}
        \sum_s\lambda_s\widehat a_s-\zeta,\quad
        \max_{\lambda\in\widehat\Lambda_{\mathrm{out}}(e)}
        \sum_s\lambda_s\widehat a_s+\zeta\right]
\]
with probability at least $1-\alpha$, provided $\Lambda$ is nonempty.
\end{proposition}

\begin{proposition}[CDF bands for data-dependent interval partitions]
\label{S-prop:random-interval-bands}
Let $F_{0,1}$ be the population CDF for untreated post-policy
comparison-group outcomes, and let $F_{1,s}$ and $F_{0,s}$
denote the population CDFs of the two groups at historical date
$s\in S$. Define their empirical counterparts with hats. Assume that,
on an event $\mathcal D$, the simultaneous bounds
\[
\|\widehat F_{0,1}-F_{0,1}\|_\infty\le\delta_{0,1},
\qquad
\|\widehat F_{g,s}-F_{g,s}\|_\infty\le\delta_{g,s}
\quad(g\in\{0,1\},\ s\in S)
\]
hold, where each $\delta_{g,t}\ge0$ is specified independently
of the realized errors. Here $\|H\|_\infty=\sup_{y\in\mathbb R}|H(y)|$.
Consider \emph{any} finite partition of the real line into
half-open intervals with endpoints that may be functions of the
same observations. Write $q_{s,k}$ and $\widehat q_{s,k}$ for the
population and empirical cell masses of the transported signed law
$F_{0,1}+F_{1,s}-F_{0,s}$, respectively, on its $k$th interval. Then,
for every such partition, simultaneously,
\[
\max_k|\widehat q_{s,k}-q_{s,k}|
\le 2(\delta_{0,1}+\delta_{1,s}+\delta_{0,s})
=:e_s,\qquad s\in S.
\]
Consequently the finite-grid no-gain certificate applies to the
realized, potentially data-dependent interval partition using
$e=\max_{s\in S}e_s$, whenever $\Pr(\mathcal D)\ge1-\alpha$.
The certified conclusion concerns the fixed population laws evaluated on the \emph{realized random partition}. The statistical target therefore varies with the selected cutoffs, even though the underlying laws do not. The result does not compare these masses with those of another, prespecified partition. In particular,
write $\mathcal N_e$ for the observable no-gain certificate on the realized
partition, and let $\mathcal I$ be the event that at least one population
historical candidate has a negative cell on that same partition. If
$\Pr(\mathcal D)\ge1-\alpha$, then
\[
\Pr(\mathcal N_e\cap\mathcal I)\le\alpha.
\]
The bound controls false no-gain declarations for the population law
restricted to the realized partition. It does not establish coherence of
the unpartitioned law, validate historical transport, or provide valid
inference for a policy score selected after examining the data without
a simultaneous score-error guarantee.
\end{proposition}
\begin{proof}
For any fixed or data-selected interval $(a,b]$, its population mass under a CDF $F$ equals $F(b)-F(a)$.
The two CDF errors on this interval are bounded in absolute value
by twice its uniform CDF error, independently of the locations of
$a$ and $b$. The same bound holds for the first and last unbounded
intervals (with an error of zero at infinite endpoints), and for
atoms because the half-open convention uses right-continuous CDFs.
Apply the triangle inequality to the three terms of each transported
signed cell mass. Since the CDF bounds hold at \emph{every} real
threshold on the single event $\mathcal D$, they hold also at every
random endpoint selected from the data. Thus the bound is
simultaneous over all interval partitions, with no further union bound
across cells or candidate cutoffs. On $\mathcal D$, the no-gain event
$\mathcal N_e$ rules out $\mathcal I$ by the cellwise error bound.
Consequently $\mathcal N_e\cap\mathcal I\subseteq\mathcal D^c$,
which proves the probability statement. The finite-partition
conclusion also follows from Corollary~\ref{S-cor:no-gain-certificate}
using the resulting coordinate-error bound.
\end{proof}
\paragraph{A finite-sample independent-unit benchmark.}
If, conditional on the realized sample sizes $n_{g,t}\ge1$, each primitive
empirical CDF is based on independent and identically distributed
observations from its corresponding group--date law (with those laws
unchanged by conditioning on sample size), the
Dvoretzky--Kiefer--Wolfowitz inequality and a union bound give
$\Pr(\mathcal D)\ge1-\alpha$ for
\[
\delta_{g,t}=\sqrt{\frac{\log(2m/\alpha)}{2n_{g,t}}},
\qquad m=1+2|S|,
\]
for the $m$ distinct primitive CDFs used above (the relevant
indices are $(0,1)$ and $(g,s)$). Cross-date dependence within a
complete-history unit is permitted; the observations used for each
primitive empirical CDF must still satisfy the stated independent-unit
condition. This benchmark does not supply a design-valid guarantee
for state-clustered data or complex-survey replicate weights.

\begin{corollary}[A simultaneous finite-grid certificate of no coherence gain]
\label{S-cor:no-gain-certificate}
Maintain the finite-grid notation of Proposition~\ref{S-prop:grid-sandwich}.
Suppose each population candidate $q_s$ has total cell mass one, and
$\Pr(\mathcal E_e)\ge 1-\alpha$. Define the observable event
\[
\mathcal N_e=\left\{\min_{s\in S,\,1\le k\le K}\widehat q_{s,k}\ge e\right\}.
\]
On $\mathcal E_e\cap\mathcal N_e$, every $q_s$ is a probability vector,
so $\Lambda=\Delta_S$. In particular, the population coherent and
unrestricted bounds agree for \emph{every} score vector on this
prespecified partition. For any fixed population array containing an
invalid historical vertex,
\[
\Pr(\mathcal N_e)\le\alpha.
\]
On the simultaneous error event, the equality of hulls holds for all
policy scores defined on the same partition; this deterministic
set-equality conclusion requires no multiplicity correction across scores.
It does not, by itself, provide simultaneous confidence bounds for the
values of data-selected policy scores. This is not a certificate of
full-outcome coherence or of the historical-mixture identifying assumption. It is sufficient, not necessary, for equality
of a particular policy endpoint: an invalid vertex can fail to attain
that endpoint. An estimated positive cell smaller than $e$ provides
no such conclusion. For intervals with data-dependent cutoffs, Proposition~\ref{S-prop:random-interval-bands} controls the realized, random partition when simultaneous CDF bands are available. Its error guarantee is then against falsely certifying no gain on that realized partition, not against deviations from a separately prespecified population partition. General sampling designs require their own valid bands.
\end{corollary}
\begin{proof}
On $\mathcal E_e$, $q_{s,k}\ge\widehat q_{s,k}-e$ for every
vertex and cell. Thus $\mathcal N_e$ implies $q_{s,k}\ge0$ for all
$s,k$. Together with unit total mass, each $q_s$ is a probability
vector. Convexity of the probability simplex yields
$\Lambda=\Delta_S$, so every score has identical coherent and
unrestricted extrema. If a population vertex has a negative cell,
$\mathcal E_e\cap\mathcal N_e$ is empty. Hence
$\Pr(\mathcal N_e)\le\Pr(\mathcal E_e^c)\le\alpha$.
\end{proof}

\begin{corollary}[Finite-sample certification of a policy-endpoint gain]
\label{S-cor:gain-certificate}
Maintain the notation and simultaneous-error event of
Proposition~\ref{S-prop:grid-sandwich}. Assume the population coherent
set $\Lambda$ is nonempty, and suppose
\[
\Pr\!\left(\mathcal E_e\cap
 \left\{\max_{s\in S}|\widehat a_s-a_s|\le\zeta\right\}\right)
\ge 1-\alpha.
\]
On the event in this display, the outer set is nonempty. Define the
population lower and upper \emph{coherence gains} by
\[
\Delta_L=\min_{\lambda\in\Lambda}a^\top\lambda-\min_{s\in S}a_s,
\qquad
\Delta_U=\max_{s\in S}a_s-\max_{\lambda\in\Lambda}a^\top\lambda.
\]
Both gains are nonnegative. When the estimated outer set is nonempty,
form the lower certificates
\[
\begin{split}
 B_L&=\min_{\lambda\in\widehat\Lambda_{\mathrm{out}}(e)}
           \widehat a^\top\lambda-\min_{s\in S}\widehat a_s-2\zeta,\\
 B_U&=\max_{s\in S}\widehat a_s-
       \max_{\lambda\in\widehat\Lambda_{\mathrm{out}}(e)}
           \widehat a^\top\lambda-2\zeta.
\end{split}
\]
Set $B_L=B_U=-\infty$ when the outer set is empty.
When the estimated inner set is nonempty, also define
\[
\begin{split}
 U_L&=\min_{\lambda\in\widehat\Lambda_{\mathrm{in}}(e)}
           \widehat a^\top\lambda-\min_{s\in S}\widehat a_s+2\zeta,\\
 U_U&=\max_{s\in S}\widehat a_s-
       \max_{\lambda\in\widehat\Lambda_{\mathrm{in}}(e)}
           \widehat a^\top\lambda+2\zeta.
\end{split}
\]
Set $U_L=U_U=+\infty$ when the inner set is empty. Then
\[
 \Pr\big\{\max(0,B_L)\le\Delta_L\le U_L,\quad
             \max(0,B_U)\le\Delta_U\le U_U\big\}
 \ge1-\alpha.
\]
Thus $B_L>0$ or $B_U>0$ certifies a strictly positive gain, with
the probability of any false declaration bounded by $\alpha$ under
the corresponding zero-gain null. A small finite upper certificate limits the size of a gain but does not establish exact equality of endpoints; a nonpositive lower certificate is inconclusive about the sign of the gain. At a probability boundary the
inner set may remain empty, leaving the upper certificate
uninformative. The two gains are covered jointly by the same event; there is no separate error budget for each endpoint. This is a confidence statement about gain magnitudes, not coverage of the historical-mixture identifying assumption itself. For several
prespecified policy scores, the statement extends if the assumed
score-error event bounds all associated vertex coefficients
simultaneously. Otherwise, separate confidence claims need a
multiplicity adjustment. The certificate itself does not require an independent pilot:
its guarantee follows from the stated joint error event, which must
be justified for the actual estimators and sampling design.
\end{corollary}
\begin{proof}
On the simultaneous event, Proposition~\ref{S-prop:grid-sandwich}
gives $\Lambda\subseteq\widehat\Lambda_{\mathrm{out}}(e)$.
Consequently,
\[
 \min_{\lambda\in\Lambda}a^\top\lambda
 \ge \min_{\lambda\in\widehat\Lambda_{\mathrm{out}}(e)}
       \widehat a^\top\lambda-\zeta,
 \qquad
 \min_{s\in S}a_s\le\min_{s\in S}\widehat a_s+\zeta.
\]
Subtracting yields $\Delta_L\ge B_L$. Similarly,
\[
 \max_{s\in S}a_s\ge\max_{s\in S}\widehat a_s-\zeta,
 \qquad
 \max_{\lambda\in\Lambda}a^\top\lambda
 \le\max_{\lambda\in\widehat\Lambda_{\mathrm{out}}(e)}
         \widehat a^\top\lambda+\zeta,
\]
which yields $\Delta_U\ge B_U$. If the inner set is
nonempty, Proposition~\ref{S-prop:grid-sandwich} gives
$\widehat\Lambda_{\mathrm{in}}(e)\subseteq\Lambda$.
Therefore
\[
\begin{split}
 \min_{\lambda\in\Lambda}a^\top\lambda
 &\le\min_{\lambda\in\widehat\Lambda_{\mathrm{in}}(e)}
           \widehat a^\top\lambda+\zeta,\\
 \max_{\lambda\in\Lambda}a^\top\lambda
 &\ge\max_{\lambda\in\widehat\Lambda_{\mathrm{in}}(e)}
           \widehat a^\top\lambda-\zeta.
\end{split}
\]
Combining these inequalities with
$\min_s a_s\ge\min_s\widehat a_s-\zeta$ and
$\max_s a_s\le\max_s\widehat a_s+\zeta$
gives $\Delta_L\le U_L$ and $\Delta_U\le U_U$.
When the inner set is empty, the upper bounds are vacuous; when
the outer set is empty, the lower bounds are vacuous. Finally,
$\Delta_L,\Delta_U\ge0$ deterministically. All inequalities
hold jointly on the assumed event, giving the claimed coverage and
false-declaration control.
\end{proof}

\paragraph{Proof of Proposition~\ref{S-prop:slater-stability}.}
Let $q(\lambda)=\sum_s\lambda_s q_s$ and
$\widehat q(\lambda)=\sum_s\lambda_s\widehat q_s$.
The vertex error implies
$\|\widehat q(\lambda)-q(\lambda)\|_\infty\le\varepsilon$
for every $\lambda\in\Delta_S$. First, for any $\widehat\lambda\in
\widehat\Lambda$, let $t=\varepsilon/(\kappa+\varepsilon)$ and
$\lambda'=(1-t)\widehat\lambda+t\lambda^\circ$.
Then $q_k(\widehat\lambda)\ge-\varepsilon$ and
$q_k(\lambda^\circ)\ge\kappa$, whence
$q_k(\lambda')\ge-(1-t)\varepsilon+t\kappa=0$.
Thus $\lambda'\in\Lambda$ and
$\|\lambda'-\widehat\lambda\|_1\le2t\le2\varepsilon/\kappa$.
Conversely, for any $\lambda\in\Lambda$, set
$u=\varepsilon/\kappa<1$ and
$\widehat\lambda'=(1-u)\lambda+u\lambda^\circ$.
The estimated masses obey $\widehat q_k(\lambda)\ge-\varepsilon$
and $\widehat q_k(\lambda^\circ)\ge\kappa-\varepsilon$, so
$\widehat q_k(\widehat\lambda')\ge
-(1-u)\varepsilon+u(\kappa-\varepsilon)=0$.
This also proves $\widehat\Lambda\ne\varnothing$ and gives
$\|\widehat\lambda'-\lambda\|_1\le2\varepsilon/\kappa$.
Together the two bounds establish the claimed Hausdorff inequality.

For a vector $a=(a_s)_s$ and probability weights $\lambda,\lambda'$,
$|a^\top(\lambda-\lambda')|\le
\tfrac12\operatorname{osc}(a)\|\lambda-\lambda'\|_1$.
For any population feasible weight, its nearby estimated feasible weight
therefore changes the estimated objective by at most
$\varepsilon\operatorname{osc}(\widehat a)/\kappa$.
The reverse feasible-weight approximation gives the same bound in the
opposite direction, and $|a^\top\lambda-\widehat a^\top\lambda|
\le\zeta$ for every simplex weight. Comparing infima and suprema
proves the stated extremum inequality. The case $\varepsilon=0$
is immediate.

\paragraph{Proof of Proposition~\ref{S-prop:grid-sandwich}.}
On the simultaneous event, every $\lambda\in\Delta_S$ satisfies
$|q_k(\lambda)-\widehat q_k(\lambda)|\le e$ for all $k$.
If $\widehat q_k(\lambda)\ge e$ for every $k$, then
$q_k(\lambda)\ge0$, proving inner containment.
If $q_k(\lambda)\ge0$ for every $k$, then
$\widehat q_k(\lambda)\ge-e$, proving outer containment.
Under the population coherence null, an empty outer set can therefore
occur only off the simultaneous event, so its rejection probability is
at most the event's specified failure probability. On the intersection
with the welfare-coefficient event, each feasible $\lambda$ satisfies
$|a(\lambda)-\widehat a(\lambda)|\le\zeta$.
Minimizing and maximizing the latter inequality over the containing
outer set gives the reported whole-identified-range interval. This is
an unconditional probability statement for the simultaneous event;
conditioning on a data-selected PASS decision is not justified.

\paragraph{An unbuffered endpoint-gain screen is not a test.}
Consider two outcome cells and two unit-mass historical candidates.
The population vectors are $p_1=(0,1)$ and $p_2=(1,0)$.
For the score equal to the first-cell probability, the unrestricted
and coherent population lower endpoints both equal zero: the
minimizing historical candidate $p_1$ is itself a probability vector.

To estimate these vectors, let each pilot unit contribute the joint
score $(Z_1,Z_2)$, where $Z_2=(1,0)$ deterministically and $Z_1$
takes $(-1,2)$ and $(1,0)$ with equal probability, independently
across units. These bounded scores are unbiased for $p_1,p_2$ and
sum to one for every realization. For an odd pilot sample size $n$,
write its mean as $\widetilde p_1=(X_n,1-X_n)$.
Here $X_n$ is the mean of $n$ independent Rademacher variables,
so $\Pr(X_n<0)=1/2$ and $\Pr(X_n=0)=0$.

When $X_n<0$, the unrestricted plug-in lower endpoint is $X_n$,
achieved only at the first historical vertex. The first vertex is
incoherent because its estimated first-cell mass is negative.
Nevertheless the plug-in coherent mixture is nonempty: along the
segment joining $\widetilde p_1$ and $p_2$, the first-cell
probability can equal zero. Its coherent lower endpoint is zero.
Thus the unbuffered procedure incorrectly reports a strict gain.
When $X_n>0$, both vertices are coherent and the plug-in lower
endpoint has no coherence gain. It follows that
\[
\Pr\{\text{plug-in strict gain despite zero population gain}\}
=\frac12\qquad\text{for every odd }n.
\]
The pilot estimates are consistent, but the strict-gain decision is
not uniformly reliable at this boundary. Simultaneous coordinate
bounds and an outer feasibility region, rather than an unbuffered
endpoint-face calculation, are needed for controlled inference.
This example concerns a sampling diagnostic, not an economically
estimated treatment effect.

\section{Empirical-Process Foundations}\label{S-sec:emp}

\subsection{Ratio expansion for group empirical CDFs}

Let $W_i=(G_i,\{Y_{it}\}_{t\in\mathcal T})$ be i.i.d. and suppose $\pi_g=P(G_i=g)>0$.  Define
\[
\widehat F_{g,t}(y)=\frac{\mathbb P_N 1\{G=g,Y_t\leq y\}}{\mathbb P_N1\{G=g\}}.
\]
On the event $\mathbb P_N1\{G=g\}=0$, define this ratio arbitrarily; that event has probability $(1-\pi_g)^N$ and hence does not affect the asserted stochastic expansion.

\begin{lemma}[Uniform influence expansion]\label{S-lem:ratio-expansion}
For a fixed finite date set,
\[
\sqrt N(\widehat F_{g,t}-F_{g,t})
=\mathbb G_N\psi_{g,t}+o_p(1)
\]
uniformly in $y$, where
\[
\psi_{g,t,y}(W)=\frac{1\{G=g\}}{\pi_g}\{1\{Y_t\leq y\}-F_{g,t}(y)\}.
\]
\end{lemma}

\begin{proof}
Let $A_y=1\{G=g,Y_t\leq y\}$ and $B=1\{G=g\}$.  Since $PA_y=\pi_gF_{g,t}(y)$,
\[
\widehat F_{g,t}(y)-F_{g,t}(y)
=\frac{(\mathbb P_N-P)(A_y-F_{g,t}(y)B)}{\mathbb P_NB}.
\]
The class $\{A_y-F_{g,t}(y)B:y\in\mathbb R\}$ is bounded VC type. Hence its empirical process is $O_p(1)$ in sup norm. On $E_N=\{\mathbb P_NB\geq\pi_g/2\}$, whose probability tends to one, the remainder from replacing the random denominator is exactly
\[
R_N(y)=\mathbb G_N\{A_y-F_{g,t}(y)B\}
\left\{\frac{1}{\mathbb P_NB}-\frac{1}{\pi_g}\right\},
\]
and therefore
\[
\sup_y|R_N(y)|
\leq \sup_y|\mathbb G_N\{A_y-F_{g,t}(y)B\}|
\frac{2|\mathbb P_NB-\pi_g|}{\pi_g^2}
=O_p(1)O_p(N^{-1/2})=o_p(1).
\]
Because the date set is finite, the same bound holds jointly over every entering $(g,t)$ coordinate.
\end{proof}

\subsection{Joint Donsker limit}

\begin{proposition}[Joint vertex FCLT]\label{S-prop:vertex-fclt}
For fixed finite $S$, the vector process
\[
\left\{\sqrt N(\widehat H_s-H_s):s\in S\right\}
\]
converges weakly in a separable version of the product sup-norm space to the centered Gaussian process generated by the vertex influence functions
\[
\psi_{s,y}=\psi_{0,1,y}+\psi_{1,s,y}-\psi_{0,s,y}.
\]
In particular, its covariance kernel is
\[
\operatorname{Cov}\{\mathbb Z_s(y),\mathbb Z_u(z)\}
=P[\psi_{s,y}\psi_{u,z}],
\qquad s,u\in S,\quad y,z\in\mathbb R.
\]
\end{proposition}

\begin{proof}
The relevant class is a finite union of bounded half-line indicator classes and their finite linear combinations. It is therefore pointwise measurable up to the standard rational-threshold separable modification, has a square-integrable envelope, and is $P$-Donsker. Apply the preceding joint expansion and the continuous linear map that forms the vertex combinations. The empirical-process covariance formula gives the displayed kernel, including every within-unit cross-date covariance.
\end{proof}

The observation in this proposition is the complete unit history $W_i$; consequently one multiplier $\xi_i$ must multiply all date coordinates of unit $i$. With independent repeated cross-sections, there is no common unit history and the influence process is instead assembled from the separate group--date samples with their own sampling fractions. The deterministic simplex reduction remains valid, but Proposition~\ref{S-prop:vertex-fclt} and the unit-level multiplier implementation must be replaced by the corresponding design-specific joint first-stage limit.

\subsection{Multiplier approximation}

\begin{assumption}[Multiplier moments]\label{S-ass:multiplier}
The multipliers are i.i.d., independent of the sample, have mean zero, variance one, and $E|\xi|^{2+\delta}<\infty$ for some $\delta>0$.
\end{assumption}

On $\{\widehat\pi_g>0\}$ define
\[
\widehat\psi_{g,t,y}(W_i)
=\frac{1\{G_i=g\}}{\widehat\pi_g}
\{1\{Y_{it}\leq y\}-\widehat F_{g,t}(y)\},
\qquad
\widehat\psi_{s,y}=\widehat\psi_{0,1,y}
+\widehat\psi_{1,s,y}-\widehat\psi_{0,s,y}.
\]
Then $\mathbb P_N\widehat\psi_{g,t,y}=0$ identically. On the vanishing-probability event that an entering group is empty, set the multiplier critical value to $+\infty$.

\begin{proposition}[Fixed-DGP multiplier validity]\label{S-prop:multiplier-valid}
Under the baseline i.i.d.-unit experiment, fixed finite $S$, group overlap, and Assumption~\ref{S-ass:multiplier}, define
\[
\widehat Z_{N,s}^*(y)=N^{-1/2}\sum_{i=1}^N\xi_i\widehat\psi_{s,y}(W_i),
\qquad s\in S.
\]
Let $\mathbb Z$ be the tight Gaussian limit of the empirical vertex process on a separable version $\mathbb B$ of the finite product sup-norm space, and let $d_{BL}$ be the bounded-Lipschitz metric on probability laws on $\mathbb B$. Then
\[
d_{BL}\!\left\{\mathcal L_\xi(\widehat Z_N^*\mid W_1,\ldots,W_N),
\mathcal L(\mathbb Z)\right\}\to_p0.
\]
Consequently, with $T_N^*=\max_{s\in S}\|\widehat Z_{N,s}^*\|_\infty$ and $T=\max_{s\in S}\|\mathbb Z_s\|_\infty$, the conditional CDF of $T_N^*$ converges in probability to the CDF of $T$ at every continuity point.
\end{proposition}

\begin{proof}
Let $\Psi$ be the finite union, over entering $(g,t)$ pairs, of the classes $\{\psi_{g,t,y}:y\in\mathbb R\}$. Group overlap gives a fixed bounded envelope. Half-line indicators are VC, and multiplication by a finite-valued group indicator and finite linear combination preserve VC type; hence $\Psi$ is $P$-Donsker and $P$-pre-Gaussian. The multiplier central limit theorem and $E|\xi|^{2+\delta}<\infty$ therefore give conditional weak convergence in outer probability of
\[
\left\{N^{-1/2}\sum_{i=1}^N\xi_i
[\psi_{g,t,y}(W_i)-\mathbb P_N\psi_{g,t,y}]:(g,t),y\right\}
\]
to the same tight Brownian bridge as the empirical process indexed by $\Psi$.

It remains to justify replacement of the true influence functions by their estimated counterparts.  We do this directly, avoiding an appeal to a maximal inequality for a sample-dependent class.  Put $I_{ig}=1\{G_i=g\}$.  On $\{\widehat\pi_g>0\}$, exact algebra gives
\begin{align*}
\widehat\psi_{g,t,y}(W_i)-\psi_{g,t,y}(W_i)
={}&(\widehat\pi_g^{-1}-\pi_g^{-1})I_{ig}
       \{1\{Y_{it}\le y\}-F_{g,t}(y)\}\\
&-\widehat\pi_g^{-1}I_{ig}
       \{\widehat F_{g,t}(y)-F_{g,t}(y)\}.
\end{align*}
The first multiplier remainder is bounded, uniformly in $y$, by
\[
|\widehat\pi_g^{-1}-\pi_g^{-1}|
\sup_y\left|N^{-1/2}\sum_{i=1}^N\xi_i I_{ig}
\{1\{Y_{it}\le y\}-F_{g,t}(y)\}\right|.
\]
The coefficient is $O_p(N^{-1/2})$.  The supremum is $O_p^*(1)$ in conditional outer probability by the multiplier maximal inequality for the fixed bounded VC class of half lines.  The second multiplier remainder is bounded by
\[
\widehat\pi_g^{-1}\|\widehat F_{g,t}-F_{g,t}\|_\infty
\left|N^{-1/2}\sum_{i=1}^N\xi_i I_{ig}\right|.
\]
The three factors are respectively $O_p(1)$, $O_p(N^{-1/2})$, and $O_p^*(1)$; the middle rate follows from the Donsker property of the same half-line class (equivalently, from its standard empirical-process maximal bound).  Because the number of entering $(g,t)$ blocks is fixed, both conclusions hold jointly.  Hence
\[
\max_{g,t}\sup_y\left|N^{-1/2}\sum_{i=1}^N\xi_i
[\widehat\psi_{g,t,y}(W_i)-\psi_{g,t,y}(W_i)]\right|=o_p^*(1).
\]
Here $o_p^*(1)$ means conditional convergence in outer probability: for every $\eta>0$, the outer probability that the conditional multiplier probability of exceeding $\eta$ is larger than $\eta$ tends to zero.

Finally, $\mathbb P_N\widehat\psi_{g,t,y}=0$ identically, whereas
$\sup_y|\mathbb P_N\psi_{g,t,y}|=O_p(N^{-1/2})$ and
$N^{-1/2}\sum_i\xi_i=O_p^*(1)$.  Thus empirical centering of the infeasible multiplier process changes it by $o_p^*(1)$.  The feasible process and the centered infeasible process are therefore asymptotically equivalent in the product sup norm.  Forming the finitely many vertex combinations is a continuous linear map, so conditional Slutsky gives the asserted joint limit.
\end{proof}

The map $z\mapsto\max_{s\in S}\|z_s\|_\infty$ is Lipschitz on $\mathbb B$, so the final assertion also follows directly from the conditional continuous-mapping theorem. If the limiting supremum statistic has a continuous distribution at its $(1-\alpha)$ quantile, conditional bootstrap quantiles make the simultaneous vertex event have asymptotic probability $1-\alpha$. The main deterministic theorem then gives whole-set coverage with lower limit at least $1-\alpha$, not necessarily equality: calibration also controls incoherent vertices that do not index the sharp set. Without continuity, the safe statement remains a lower-limit coverage result based on suitable upper quantiles. Proposition~\ref{S-prop:multiplier-valid} is neither a uniform-in-$P$ result nor a justification for treating survey replicate weights as multiplier draws.

\subsection{Closedness of the coherence tube}

Let \(\ell^\infty(\mathbb R)\) carry the sup norm and let \(\mathbb D\) denote its subset of CDFs. For finite \(S\), bounded functions \(h=(h_s:s\in S)\), and \(\varepsilon\geq0\), put \(h_\lambda=\sum_s\lambda_sh_s\) and
\[
\mathcal C(h,\varepsilon)=\left\{F\in\mathbb D:\inf_{\lambda\in\Delta_S}\|F-h_\lambda\|_\infty\leq\varepsilon\right\}.
\]

\begin{proposition}[Closed realized coherence tube]\label{S-prop:tube-closed}
For every realized \((h,\varepsilon)\), the infimum over \(\Delta_S\) is attained, \(\mathcal C(h,\varepsilon)\) is sup-norm closed, and
\[
\mathcal C(h,\varepsilon)=\bigcup_{\lambda\in\Delta_S}\{F\in\mathbb D:\|F-h_\lambda\|_\infty\leq\varepsilon\}.
\]
This deterministic statement does not by itself assert Effros measurability after estimated vertices and a random radius are substituted.
\end{proposition}

\begin{proof}
The map \(\lambda\mapsto h_\lambda\) is continuous because
\[
\|h_\lambda-h_{\lambda'}\|_\infty\leq\sum_{s\in S}|\lambda_s-\lambda_s'|\|h_s\|_\infty.
\]
For fixed $F$, continuity on the compact simplex gives a minimizer, which also proves the union identity. Now let \(F_n\in\mathcal C(h,\varepsilon)\) converge uniformly to \(F\), and choose a minimizing \(\lambda_n\) for each $n$. Compactness supplies a subsequence $\lambda_{n_k}\to\lambda$. Since $\|F_{n_k}-h_{\lambda_{n_k}}\|_\infty\leq\varepsilon$, continuity of $\lambda\mapsto h_\lambda$ and uniform convergence of $F_{n_k}$ imply
\[
\|F-h_\lambda\|_\infty
\leq \|F-F_{n_k}\|_\infty
    +\|F_{n_k}-h_{\lambda_{n_k}}\|_\infty
    +\|h_{\lambda_{n_k}}-h_\lambda\|_\infty
\leq\varepsilon+o(1).
\]
Letting $k\to\infty$ gives $\|F-h_\lambda\|_\infty\leq\varepsilon$. Uniform limits preserve monotonicity and right continuity. They preserve the CDF tails as well: approximate \(F\) uniformly by one \(F_n\), use that CDF's two tails, and then let the approximation error vanish. Hence \(\mathbb D\) is sup-norm closed, so \(F\in\mathbb D\) and therefore \(F\in\mathcal C(h,\varepsilon)\).
\end{proof}

\begin{proposition}[Measurable whole-set containment]\label{S-prop:tube-containment-measurable}
Let $K$ be a fixed nonempty compact subset of $\mathbb D$. Suppose the random
vertices $\widehat H_s$ have bounded right-continuous sample paths, with
$\widehat H_s(q)$ measurable for each rational $q$, and let $\varepsilon$
be a measurable $[0,+\infty]$-valued radius. Then
$\{K\subseteq\mathcal C(\widehat H,\varepsilon)\}$ is measurable. The
vertex event
$\{\max_s\|\widehat H_s-H_s\|_\infty\leq\varepsilon\}$ is
measurable whenever the fixed $H_s$ are bounded and right-continuous.
If $K$ is empty, the containment event is the whole sample space.
\end{proposition}

\begin{proof}
For any bounded right-continuous $f$ and $g$, their difference is
right-continuous, so
$\|f-g\|_\infty=\sup_{q\in\mathbb Q}|f(q)-g(q)|$.
Choose a countable dense subset $D$ of the compact metric space $K$ and a
countable dense subset $L$ of $\Delta_S$. For fixed $F\in K$, the map
$\lambda\mapsto\|F-\widehat H_\lambda\|_\infty$ is continuous, so
\[
 d(F,\widehat{\mathcal H})
 =\inf_{\lambda\in L}\sup_{q\in\mathbb Q}
 \left|F(q)-\sum_{s\in S}\lambda_s\widehat H_s(q)\right|
\]
is measurable. Distance to the realized hull is one-Lipschitz as a function
of $F$. It follows that
\[
 \{K\subseteq\mathcal C(\widehat H,\varepsilon)\}
 =\left\{\sup_{F\in D}d(F,\widehat{\mathcal H})\leq\varepsilon\right\},
\]
a measurable event. The same rational-threshold identity and finiteness of
$S$ establish measurability of the vertex event. No measurability claim
about the set-valued map $\mathcal C$ in the ambient nonseparable space is
needed.
\end{proof}

The sharp identified set is compact and supplies $K$ in Theorem~\ref{thm:coherent-inference}.

\subsection{Finite-sample DKW edge case}

The DKW inequality controls an empirical cumulative distribution function (CDF).
The result uses
i.i.d. observational units with complete observed histories; it does
not cover survey replicate-weight or dependent-state sampling.
Let $J=|S|$ and let
\[
\mathcal I=\{(0,1)\}\cup\{(1,s),(0,s):s\in S\},
\qquad M=|\mathcal I|=1+2J.
\]
For $N_g=\sum_{i=1}^N1\{G_i=g\}>0$, define
\[
r_{g,t}=\left\{\frac{\log(2M/\alpha)}{2N_g}\right\}^{1/2},
\qquad (g,t)\in\mathcal I.
\]
If an entering group count is zero, define its empirical CDF arbitrarily and set $r_{g,t}=+\infty$ for every corresponding primitive.

\begin{proposition}[Operational finite-sample DKW radius]\label{S-prop:dkw-radius}
Under the baseline i.i.d.-unit sampling experiment,
\[
P\!\left\{
\|\widehat F_{g,t}-F_{g,t}\|_\infty\le r_{g,t}
\ \text{for every }(g,t)\in\mathcal I
\right\}\ge1-\alpha.
\]
Consequently, with
\[
\varepsilon_N^{DKW}
=\max_{s\in S}\{r_{0,1}+r_{1,s}+r_{0,s}\},
\]
\[
P\!\left\{\max_{s\in S}\|\widehat H_s-H_s\|_\infty
\le\varepsilon_N^{DKW}\right\}\ge1-\alpha.
\]
\end{proposition}

\begin{proof}
Condition on the complete group-label vector. If both group counts are positive, the observations entering each fixed $(g,t)$ empirical CDF are i.i.d. draws from $F_{g,t}$. The two-sided DKW inequality gives
\[
P\!\left\{\|\widehat F_{g,t}-F_{g,t}\|_\infty>r_{g,t}
\mid G_1,\ldots,G_N\right\}
\le2e^{-2N_gr_{g,t}^2}=\frac{\alpha}{M}.
\]
A union bound over the primitive CDFs whose group count is positive proves conditional simultaneous coverage of at least $1-\alpha$, because there are at most $M$ such coordinates. A coordinate belonging to an empty group imposes no restriction: its radius is infinite, so its individual coverage event is certain. The other coordinates remain subject to their DKW bounds. This argument does not use independence across dates within an observational unit. Averaging the conditional bound over the group labels proves the first display unconditionally. Finally, for every $s\in S$, the triangle inequality gives
\[
\|\widehat H_s-H_s\|_\infty
\le\|\widehat F_{0,1}-F_{0,1}\|_\infty
+\|\widehat F_{1,s}-F_{1,s}\|_\infty
+\|\widehat F_{0,s}-F_{0,s}\|_\infty.
\]
Taking the maximum over $s$ proves the second claim.
\end{proof}

\subsection{Distance to the CDF class}\label{S-sec:cdfdist}

Let $h:\mathbb R\to\mathbb R$ be right-continuous with $h(-\infty)=0$ and $h(+\infty)=1$.  Let $\Dcal$ be the CDF class.

\begin{theorem}[Exact sup-norm distance to the CDF class]\label{S-thm:cdf-distance}
Define
\[
L=\sup_y[-h(y)]_+,\quad
U=\sup_y[h(y)-1]_+,\quad
V=\sup_{x<y}[h(x)-h(y)]_+.
\]
Then
\[
\inf_{F\in\Dcal}\|F-h\|_\infty=\max\{L,U,V/2\}.
\]
\end{theorem}

\begin{proof}
Let $\delta$ denote the right-hand side.  Any CDF approximation within radius $r$ requires $r\geq L$ and $r\geq U$.  Moreover, for $x<y$,
\[
h(x)-h(y)\leq |h(x)-F(x)|+F(x)-F(y)+|F(y)-h(y)|\leq2r,
\]
so $r\geq V/2$.  This gives the lower bound.

If $\delta=+\infty$, the lower bound implies that every CDF is at infinite sup-norm distance from $h$, so the asserted equality follows. If $\delta=0$, then $h$ is already a CDF. Henceforth suppose $0<\delta<+\infty$ and define
\[
G(y)=\max\left\{0,\sup_{x\leq y}[h(x)-\delta]\right\}.
\]
The range bounds implied by $L,U\leq\delta$ ensure $0\leq G\leq1$.  The function is nondecreasing.  For every $x\leq y$, the definition of $V$ gives $h(x)-\delta\leq h(y)+\delta$, so
\[
h(y)-\delta\leq G(y)\leq h(y)+\delta.
\]
To verify right continuity, take $y_n\downarrow y$. Monotonicity gives a limit $G(y_n)\downarrow\ell\geq G(y)$. For every $\epsilon>0$, right continuity of $h$ supplies $\eta>0$ such that $h(x)\leq h(y)+\epsilon$ for $y<x\leq y+\eta$. Hence, for all large $n$,
\[
\sup_{x\leq y_n}\{h(x)-\delta\}
\leq\max\left\{\sup_{x\leq y}\{h(x)-\delta\},h(y)-\delta+\epsilon\right\},
\]
whose positive part is at most $G(y)+\epsilon$. Thus $\ell=G(y)$. To prove the lower tail, fix $\epsilon>0$ and choose $M_-$ such that $|h(x)|<\epsilon$ for every $x\leq M_-$. For $y\leq M_-$, $0\leq G(y)\leq[\epsilon-\delta]_+\leq\epsilon$, so $G(y)\to0$ as $y\to-\infty$. Since $h(y)\to1$, choose $M$ such that $h(y)\geq1-\delta$ for all $y\geq M$. Define
\[
F(y)=G(y)\quad(y<M),\qquad F(y)=1\quad(y\geq M).
\]
The function $F$ is nondecreasing, right-continuous, and has limits zero and one. For $y<M$ its error is bounded by $\delta$ through the inequalities for $G$; for $y\geq M$, the two bounds $h(y)\geq1-\delta$ and $U\leq\delta$ imply $|1-h(y)|\leq\delta$. Hence $F$ is a CDF and $\|F-h\|_\infty\leq\delta$.
\end{proof}

The preceding formula also yields an exact finite representation, provided that the tail intervals are retained rather than silently discarded.

\begin{proposition}[Exact pooled-jump reduction]\label{S-prop:pooled-jump-reduction}
Suppose $h$ is right-continuous and constant on the intervals determined by distinct finite breakpoints $x_1<\cdots<x_K$, with $h(y)=0$ for $y<x_1$ and $h(y)=1$ for $y\geq x_K$. Write $h_j=h(x_j)$ for $j=1,\ldots,K$ and set $h_0=0$ for the left-tail value. Then
\[
L=\max_{1\leq j\leq K}[-h_j]_+,
\quad
U=\max_{1\leq j\leq K}[h_j-1]_+,
\quad
V=\max_{0\leq i<j\leq K}[h_i-h_j]_+.
\]
The index $i=0$ is essential: it retains comparisons between the zero left tail and a negative value at the first or a later finite breakpoint. The formula also covers $K=1$, in which case its only pair is $(0,1)$ and $h_1=1$.
Moreover,
\[
\inf_{F\in\Dcal}\|F-h\|_\infty
=\min_{0\leq z_1\leq\cdots\leq z_K=1}
\max_{1\leq j\leq K}|z_j-h_j|,
\]
and the minimum is attained.
\end{proposition}

\begin{proof}
For the range terms, the two tails contribute only zero and one, so every violation is attained at a finite breakpoint. For the downward-increment term, attach index $0$ to the left-tail interval and indices $1,\ldots,K$ to the intervals beginning at the finite breakpoints. If $x<y$ lie in different intervals, their values are $h_i$ and $h_j$ for some $0\leq i<j\leq K$; if they lie in the same interval, the difference is zero. Conversely, for every $i<j$, points can be chosen in the corresponding intervals with the stated values. This proves all three finite-maximum formulas, including the left-tail comparison.

For any feasible vector $z$, the right-continuous step function that equals zero below $x_1$, equals $z_j$ on $[x_j,x_{j+1})$, and equals one on $[x_K,\infty)$ is a CDF, and its uniform error is the displayed maximum. Conversely, take any CDF, record its values at $x_1,\ldots,x_{K-1}$ and set $z_K=1$; this vector is feasible, and since $h=1$ on $[x_K,\infty)$, its maximum error is at most the uniform error of the CDF. Taking infima proves equality. The feasible set is compact and the objective continuous, so the minimum is attained.
\end{proof}

For empirical CDFs, the breakpoints are the distinct pooled sample values and affine combinations preserve the common tail values zero and one. The added index $0$ records the left-tail value rather than an artificial data point. The proposition therefore gives an exact reduction without outcome discretization or unspecified one-sided values.

\section{Sample-Split Calibration on a Finite Grid}\label{S-sec:split-outer}

The procedure uses a prespecified outcome grid and independent pilot and estimation samples of bounded joint vertex scores with known support. Its guarantee is specific to that sampling experiment.

Fix integers $J\ge1$ and $K\ge2$.  On a $K$-cell partition, let
$p_j\in\mathbb R^K$ be historical signed probability vectors with
$\sum_{k=1}^Kp_{j,k}=1$.  Write $\Delta_J=\{\lambda\in\mathbb R^J:\lambda_j\ge0,\ \sum_j\lambda_j=1\}$
for the simplex of historical weights. Their coherent weight set is
\[
\Lambda=\{\lambda\in\Delta_J:
  p_{\lambda,k}:=\sum_{j=1}^J\lambda_jp_{j,k}\ge0
  \text{ for all }k\}.
\]
For the coverage theorem below we will assume
$\Lambda\ne\varnothing$; the subsequent specification test
also considers $\Lambda=\varnothing$.  Each observation is a vector
$Z=(Z_{j,k})_{j\le J,k\le K}$ with $E Z_{j,k}=p_{j,k}$ and
$Z\in\mathcal Z$ almost surely, where $\mathcal Z$ is a known,
nonempty, compact subset of $\mathbb R^{JK}$.  We use independent
pilot and estimation samples of sizes $n_1,n_2\ge1$, drawn i.i.d.
from this law.  Write $\widetilde p_j$ and $\widehat p_j$ for the
respective coordinatewise sample means.  Define the known coordinate
ranges
\[
D_{j,k}=\sup_{z\in\mathcal Z}z_{j,k}
          -\inf_{z\in\mathcal Z}z_{j,k},\qquad
D=\max_{j,k}D_{j,k},
\]
and for $0<\beta<1$ let
\[
\eta=D\sqrt{\frac{\log(2JK/\beta)}{2n_1}},\qquad
\widetilde\Lambda^+=\{\lambda\in\Delta_J:
\sum_j\lambda_j\widetilde p_{j,k}\ge-\eta
\ \forall k\}.
\]
If this polytope is empty, set $\widetilde\Lambda=\Delta_J$;
otherwise let $\widetilde\Lambda=\widetilde\Lambda^+$.
For $\lambda\in\Delta_J$, define the 
\emph{joint-support range}
\[
R_{\lambda,k}=
\sup_{z\in\mathcal Z}\sum_j\lambda_jz_{j,k}
-\inf_{z\in\mathcal Z}\sum_j\lambda_jz_{j,k}.
\]
Using the exact joint range retains cancellations that a sum of marginal ranges may conceal.

\begin{theorem}[Finite-sample outer-weight calibration]\label{S-thm:split-outer}
Suppose $\Lambda\ne\varnothing$. Let $V=\operatorname{ext}(\widetilde\Lambda)$ be its finite,
nonempty vertex set.  Given the pilot, define
\[
r_{n_2}=\max_{v\in V,\,1\le k\le K}
R_{v,k}\sqrt{\frac{\log(2K|V|/\alpha)}{2n_2}},
\qquad 0<\alpha<1.
\]
Define the random region of probability vectors
\[
\mathcal C_{n_1,n_2}
=\left\{q\in\Delta_K:\ \exists\lambda\in
\widetilde\Lambda:\
\left\|q-\sum_j\lambda_j\widehat p_j\right\|_\infty
\le r_{n_2}\right\}.
\]
Then, for every law satisfying the assumptions above,
\[
P\!\left(\{p_\lambda:\lambda\in\Lambda\}
\subseteq\mathcal C_{n_1,n_2}\right)\ge1-\alpha-\beta.
\]
The guarantee is uniform over this bounded-score class when the
specified support $\mathcal Z$ is a valid common bound.  No
interiority, uniqueness, or differentiability of the coherent set is
required.
\end{theorem}

\begin{proof}
For each pilot coordinate, Hoeffding's inequality gives
$P(|\widetilde p_{j,k}-p_{j,k}|>\eta)
\le2\exp(-2n_1\eta^2/D_{j,k}^2)\le\beta/(JK)$,
with zero-range coordinates treated as exact.  A union bound implies
that the event
$E_1=\{\max_{j,k}|\widetilde p_{j,k}-p_{j,k}|\le\eta\}$
has probability at least $1-\beta$.  For each
$\lambda\in\Lambda$ and $k$, the convex-weighted pilot error is
at most $\eta$, so on $E_1$ we have
$\Lambda\subseteq\widetilde\Lambda^+\subseteq
\widetilde\Lambda$.  In particular, the fallback case occurs only
outside $E_1$.

Conditional on the pilot, $V$ is fixed and the estimation sample is
independent.  For each $v\in V$ and $k$, the second-sample average of
$\sum_jv_jZ_{j,k}$ has expectation $p_{v,k}$ and known range
$R_{v,k}$.  Hoeffding's inequality and a union bound over the
$K|V|$ vertex-coordinate pairs show, with conditional probability at
least $1-\alpha$, that
\[
\max_{v\in V,k}\left|
\sum_jv_j(\widehat p_{j,k}-p_{j,k})\right|\le r_{n_2}.
\]
Any $\lambda\in\widetilde\Lambda$ is a convex combination of
points in $V$.  Linearity of each coordinate and the triangle
inequality extend the same error bound to every such $\lambda$.
On the intersection with $E_1$, every $\lambda\in\Lambda$ belongs
to $\widetilde\Lambda$ and the true probability vector
$p_\lambda$ lies in $\mathcal C_{n_1,n_2}$.  Independence and the
two conditional error guarantees yield at least
$(1-\beta)(1-\alpha)\ge1-\alpha-\beta$ coverage.
\end{proof}

\begin{proposition}[Finite-sample rejection and an infeasibility margin]\label{S-prop:split-falsification}
Retain the score-support, independent-pilot, and concentration
assumptions and the definitions of $J,K,D,\beta,\eta$ and
$\widetilde\Lambda^+$ preceding Theorem~\ref{S-thm:split-outer},
but allow the population coherent-weight set $\Lambda$ to be empty.
Define the population grid-infeasibility margin
\[
\gamma=\min_{\lambda\in\Delta_J}\max_{1\le k\le K}
  [-p_{\lambda,k}]_+,
\]
where $[x]_+=\max\{x,0\}$.  Thus $\gamma=0$ if and only if
$\Lambda\ne\varnothing$.  Use the pilot data and threshold $\eta$
from that theorem, and reject grid coherence when
$\widetilde\Lambda^+=\varnothing$, \emph{before} applying the
full-simplex fallback used to construct confidence regions.  Under
$\Lambda\ne\varnothing$ this rule has rejection probability at most
$\beta$.  If $\gamma>2\eta$, it rejects with probability at least
$1-\beta$.  In particular, when $D>0$ and $\gamma>0$, the latter
conclusion holds whenever
\[
 n_1>\frac{2D^2}{\gamma^2}\log\frac{2JK}{\beta}.
\]
These probabilities are uniform over laws obeying the stated joint
score support and independent-pilot assumptions; they do not test
whether the historical transport assumption identifies the true
untreated counterfactual.
\end{proposition}

\begin{proof}
By the pilot concentration argument in the proof of
Theorem~\ref{S-thm:split-outer} (which uses no coherence assumption),
with probability at least $1-\beta$ every coordinate satisfies
$|\widetilde p_{j,k}-p_{j,k}|\le\eta$.  On this event, every
population-coherent weight belongs to $\widetilde\Lambda^+$, proving
the size statement.  For any $\lambda\in\Delta_J$, convexity of the
weights gives
$|\sum_j\lambda_j(\widetilde p_{j,k}-p_{j,k})|\le\eta$
for each $k$.  If the pilot outer set contains $\lambda$, then
$\sum_j\lambda_j\widetilde p_{j,k}\ge-\eta$ for all $k$, hence
$p_{\lambda,k}\ge-2\eta$ for all $k$.  It follows that
$\gamma\le2\eta$.  Thus, when $\gamma>2\eta$, the pilot outer set
must be empty on the concentration event, proving the power bound.
Solving $\gamma>2D\sqrt{\log(2JK/\beta)/(2n_1)}$ for $n_1$
gives the last display.  Compactness of the simplex and continuity of
the finite-dimensional maximum ensure that $\gamma=0$ precisely
when a population-coherent weight exists.
\end{proof}

\section{Uniform Inference}\label{S-sec:uniform}

For probability laws on the real line, $d_{Pr}$ denotes the Prokhorov metric. For any such law $R$, define its generalized quantile by
\[
q_R(p)=\inf\{x\in\mathbb R:R(( -\infty,x])\ge p\},\qquad p\in(0,1).
\]
For a random conditional law $Q^*$ with jointly measurable CDF, write $q^*(p)$ for its generalized quantile. Under the tightness conditions below, the quantiles used for calibration are finite.

Let $\Pcal_0$ be a specified nonempty class of observed-data probability laws for the complete-history sampling experiment. For each $P\in\Pcal_0$, all historical vertex CDFs and the associated empirical process are defined under that law. This section states the additional approximation conditions required for uniform, rather than pointwise, validity over $\Pcal_0$. The historical index set $S$ is fixed and finite throughout, and the same norm and vertex statistic are used for every member of this class.

\begin{assumption}[U-EP]\label{S-ass:uep}
For each $P\in\Pcal_0$ and sample size $N$, let
\[
T_{N,P}=\max_{s\in S}\|\sqrt N(\widehat H_s-H_s)\|_\infty,
\qquad T_P=\max_{s\in S}\|Z_{P,s}\|_\infty,
\]
where $Z_P=(Z_{P,s})_{s\in S}$ is a tight, Borel-measurable, separable
Gaussian vertex process taking values in the finite product of sup-norm
spaces. Write
\[
T_{N,P}^*=\max_{s\in S}\|Z_{N,s}^*\|_\infty
\]
and let $Q_{N,P}^*$ denote a regular conditional law of $T_{N,P}^*$ given the data, chosen so that its CDF is jointly measurable in the realized data and its scalar argument. There exist deterministic sequences $\rho_N\downarrow0$ and $\eta_N\downarrow0$ such that
\[
\sup_{P\in\Pcal_0}d_{Pr}\{\mathcal L_P(T_{N,P}),\mathcal L(T_P)\}\le\rho_N
\]
and
\[
\sup_{P\in\Pcal_0}P\!\left[
d_{Pr}\{Q_{N,P}^*,\mathcal L(T_P)\}>\rho_N
\right]\le\eta_N.
\]
\end{assumption}

Assumption~\ref{S-ass:uep} is a high-level uniform approximation condition. Its first clause controls the measurable empirical statistic uniformly over $P\in\Pcal_0$; its second clause requires the measurable random conditional multiplier law to approximate the same Gaussian reference uniformly in probability.

Joint measurability of the conditional CDF makes the conditional quantile used
below measurable. Right-continuous empirical CDF paths and rational-threshold
reduction make $T_{N,P}$ measurable. A genuinely nonmeasurable process would
require an inner-probability coverage formulation instead.

\begin{lemma}[Buffered quantile transfer]\label{S-lem:buffered-quantile}
Let $X,Y$ be real random variables, let $Q$ be a fixed probability law on $\mathbb R$, and suppose
\[
d_{Pr}\{\mathcal L(X),\mathcal L(Y)\}\leq\rho,
\qquad d_{Pr}\{Q,\mathcal L(Y)\}\leq\rho.
\]
If $2\rho<\alpha$, then
\[
P\{X\leq q_Q(1-\alpha+2\rho)+2\rho\}\geq1-\alpha,
\]
where $q_Q$ is the generalized quantile under $Q$. If $Q^*$ is a random conditional law and the second distance bound holds on an event $A$, the corresponding unconditional statement is
\[
P\{X\leq q^*(1-\alpha+2\rho)+2\rho\}\geq1-\alpha-P(A^c).
\]
\end{lemma}

The generalized inverse is well defined at atoms. The condition $2\rho<\alpha$ keeps the quantile indices in $(0,1)$.

\begin{proof}
Write $F_R$ for the CDF of a law $R$.  The defining enlargement inequality
for the Prokhorov metric gives, for every $x\in\mathbb R$,
\[
F_Q(x)\leq F_Y(x+\rho)+\rho.
\]
Consequently, for every $p\in(\rho,1)$,
\[
q_Q(p)\geq q_Y(p-\rho)-\rho.
\]
This implication remains valid at atoms because both quantiles are generalized
inverses.  Set $p=1-\alpha+2\rho$.  The other Prokhorov bound gives
$F_X(x)\geq F_Y(x-\rho)-\rho$.  Evaluating it at
$x=q_Y(1-\alpha+\rho)+\rho$ yields
\[
P\{X\leq q_Y(1-\alpha+\rho)+\rho\}\geq1-\alpha.
\]
The quantile comparison shows that the threshold in the lemma is weakly larger than $q_Y(1-\alpha+\rho)+\rho$, proving the fixed-$Q$ claim. For random $Q^*$, let $B=\{X\le q_Y(1-\alpha+\rho)+\rho\}$. On $A$, the conditional-law distance yields $q^*(1-\alpha+2\rho)+2\rho\ge q_Y(1-\alpha+\rho)+\rho$ pointwise. Hence $A\cap B$ is contained in the event displayed in the lemma, and its probability is at least $1-\alpha-P(A^c)$.
\end{proof}

\begin{theorem}[Uniform buffered coverage]\label{S-thm:uniform-buffered}
Suppose Assumption~\ref{S-ass:uep} holds and $0<\alpha<1$.  For every sufficiently
large $N$ such that $2\rho_N<\alpha$, define
\[
\varepsilon_{N,P}^{\mathrm{buf}}
=N^{-1/2}\{q_{Q_{N,P}^*}(1-\alpha+2\rho_N)+2\rho_N\},
\]
where the conditional quantile is defined from the observed multiplier law
on every sample realization. The approximation event is unobservable and
is used only in the coverage proof. Then
\[
\inf_{P\in\Pcal_0}P\!\left\{
\Fcal_S(P)\subseteq\mathfrak C_{F,N}(\varepsilon_{N,P}^{\mathrm{buf}})
\right\}
\ge 1-\alpha-\eta_N.
\]
Consequently,
\[
\liminf_{N\to\infty}\inf_{P\in\Pcal_0}P\!\left\{
\Fcal_S(P)\subseteq\mathfrak C_{F,N}(\varepsilon_{N,P}^{\mathrm{buf}})
\right\}\geq1-\alpha.
\]
\end{theorem}

\begin{proof}
Fix $P\in\Pcal_0$ and let
\[
A_{N,P}=\left\{d_{Pr}\{Q_{N,P}^*,\mathcal L(T_P)\}\leq\rho_N\right\}.
\]
The second clause of U-EP gives
$\sup_{P\in\Pcal_0}P(A_{N,P}^c)\leq\eta_N$. Apply
Lemma~\ref{S-lem:buffered-quantile} with
$X=T_{N,P}$, $Y=T_P$, $Q^*=Q_{N,P}^*$, and $A=A_{N,P}$.  The first clause of
U-EP then yields
\[
P\!\left\{T_{N,P}\leq
q_{Q_{N,P}^*}(1-\alpha+2\rho_N)+2\rho_N\right\}
\geq1-\alpha-\eta_N.
\]
The event inside braces is exactly the simultaneous vertex event with radius
$\varepsilon_{N,P}^{\mathrm{buf}}$.  The deterministic coverage theorem in
the main paper maps this event into whole-set containment.  Since the bound is
independent of $P$, taking the infimum over $\Pcal_0$ proves the finite-$N$
claim.  The limiting statement follows from $\eta_N\to0$.
\end{proof}

The theorem does not use anti-concentration or continuity of the law of $T_P$.
The buffer is defined using an observable conditional law, but its numerical
value depends on a certified upper bound for $\rho_N$. A claim of uniform
validity cannot be based solely on the pointwise multiplier central limit
theorem or on survey replicate weights. No such bound is established for
the empirical applications here.

\subsection{Repeated cross-sections}\label{S-sec:rcs}

The deterministic coverage theorem does not require longitudinal observations.  Write
\[
H_s=F_{0,1}+B_s,
\]
where $F_{0,1}$ is the post-treatment comparison-group CDF and $B_s$ is the historical treated-control CDF difference.

\begin{proposition}[Block-radius transfer]\label{S-prop:block-radius}
Suppose random radii satisfy, on an event $E$,
\[
\|\widehat F_{0,1}-F_{0,1}\|_\infty\leq r_1,
\qquad
\|\widehat B_s-B_s\|_\infty\leq r_s\quad\forall s\in S.
\]
Then on $E$,
\[
\max_{s\in S}\|\widehat H_s-H_s\|_\infty
\leq r_1+\max_{s\in S}r_s.
\]
Hence any joint or Bonferroni construction for the block radii can be inserted into the main coherence-tube theorem.
\end{proposition}

No cross-year independence is required for the Bonferroni version.

\subsection{Cluster sampling}\label{S-sec:cluster}

Policy data are often sampled in clusters: states, localities or survey primary sampling units, with arbitrary dependence among the units, dates and outcome cells within each cluster. The coverage theorem in the main paper is deterministic once a simultaneous band for the vertex cell masses is available, so what changes under clustering is only how that band is calibrated. This subsection gives two calibrations on a fixed grid, one exact in finite samples and one asymptotic and uniform, and then bounds the power of the coherence test. Throughout, the grid $\Pi=\{A_1,\ldots,A_K\}$ and the set of historical dates $S$, with $J=|S|$, are fixed. The vertex cell masses form the $J\times K$ array $Z$ with entries $z_{s,k}=\mu_{0,1}(A_k)+\mu_{1,s}(A_k)-\mu_{0,s}(A_k)$, every row of $Z$ sums to one, and $Z_\lambda=\sum_s\lambda_sZ_{s\cdot}$. For an array $V$ write $\|V\|_\infty=\max_{s,k}|v_{s,k}|$. We use $C$ for the number of clusters; the group indicator remains $G$.

\paragraph{Design.} Treated clusters $c\in\mathcal T$ and comparison clusters $c\in\mathcal U$, with $C_1=|\mathcal T|$ and $C_0=|\mathcal U|$, are mutually independent. Each treated cluster reports its own estimated cell shares $\widehat\mu^{\,c}_{1,s}(A_k)$, $s\in S$, and each comparison cluster reports $\widehat\mu^{\,c}_{0,1}(A_k)-\widehat\mu^{\,c}_{0,s}(A_k)$, computed from any within-cluster sample. The estimator is the equal-weight average
\[
\widehat z_{s,k}=\frac1{C_0}\sum_{c\in\mathcal U}\bigl\{\widehat\mu^{\,c}_{0,1}(A_k)-\widehat\mu^{\,c}_{0,s}(A_k)\bigr\}+\frac1{C_1}\sum_{c\in\mathcal T}\widehat\mu^{\,c}_{1,s}(A_k),
\]
and the target $Z$ is its expectation: the cluster-average untreated laws, with clusters drawn from their respective populations. Nothing is assumed about the joint distribution of a cluster's reports across dates, groups or cells. Weighting units equally, as when records from all clusters are pooled, turns $\widehat Z$ into a ratio of cluster totals. Each cell is then a smooth function of the means of $C$ independent bounded vectors (the cluster's counts by group, date and cell, and its group sizes). If cluster sizes are bounded and their mean is bounded away from zero, a first-order expansion gives clause~(i) of Assumption~\ref{S-ass:cluster} below with $\psi_c$ the cluster's linearized contribution and $R_C=O_P(C^{-1})$, and the Gaussian calibration applies; the nonparametric cluster bootstrap is its resampling counterpart. The finite-sample radius does not apply.

In many applications both groups are observed in the same clusters: mothers and childless women live in the same states. In this \emph{shared-cluster design} each of $C$ independent clusters reports the whole array $V_c$ with entries $v_{c,s,k}=\widehat\mu^{\,c}_{0,1}(A_k)+\widehat\mu^{\,c}_{1,s}(A_k)-\widehat\mu^{\,c}_{0,s}(A_k)$, computed from its own residents, and $\widehat Z=C^{-1}\sum_cV_c$. The two groups' estimation errors within a cluster may then be correlated in any way, so the groups must be resampled together, cluster by cluster.

\begin{proposition}[Finite-sample cluster radius]\label{S-prop:cluster-hoeffding}
For $0<\alpha<1$, $\Pp(\|\widehat Z-Z\|_\infty\le t_C)\ge1-\alpha$, where
\[
t_C=\begin{cases}\Bigl\{\dfrac{\log(2JK/\alpha)}{2}\Bigl(\dfrac1{C_1}+\dfrac4{C_0}\Bigr)\Bigr\}^{1/2} & \text{separate clusters},\\[2ex] \Bigl\{\dfrac{9\log(2JK/\alpha)}{2C}\Bigr\}^{1/2} & \text{shared clusters}.\end{cases}
\]
Consequently the confidence region of Theorem~\ref{thm:coherent-inference} on the grid, with radius $t_C$, contains the coherent set of cell vectors with probability at least $1-\alpha$, and the test that rejects coherence when $d(\widehat Z)>t_C$, where $d(V)=\min_{\lambda\in\Delta_S}\max_k[-(V_\lambda)_k]_+$, has size at most $\alpha$.
\end{proposition}

\begin{proof}
Fix $(s,k)$. The deviation $\widehat z_{s,k}-z_{s,k}$ is a sum of independent mean-zero terms, one per cluster. A treated cluster's term is $C_1^{-1}\widehat\mu^{\,c}_{1,s}(A_k)$ minus its mean, and $\widehat\mu^{\,c}_{1,s}(A_k)\in[0,1]$, so the term ranges over an interval of length $1/C_1$; a comparison cluster's term ranges over an interval of length $2/C_0$, since the difference of two shares lies in $[-1,1]$. Hoeffding's inequality gives $\Pp(|\widehat z_{s,k}-z_{s,k}|>t)\le2\exp\{-2t^2/(C_1\cdot C_1^{-2}+C_0\cdot4C_0^{-2})\}$, which equals $\alpha/(JK)$ at $t=t_C$. In the shared-cluster design each $v_{c,s,k}$ lies in $[-1,2]$, so each cluster's term ranges over an interval of length $3/C$ and the bound is $2\exp(-2Ct^2/9)$. A union bound over the $JK$ coordinates proves the first claim. On that event, every coherent $Z_\lambda$ satisfies $\|\widehat Z_\lambda-Z_\lambda\|_\infty\le\|\widehat Z-Z\|_\infty\le t_C$ by Lemma~\ref{lem:vertex}, which gives the region. If $\lambda_0$ is coherent, $(Z_{\lambda_0})_k\ge0$, so $[-(\widehat Z_{\lambda_0})_k]_+\le|(\widehat Z_{\lambda_0}-Z_{\lambda_0})_k|\le t_C$ for every $k$ and $d(\widehat Z)\le t_C$.
\end{proof}

The radius depends on the numbers of clusters only, not on the number of units in each, and it is dominated by the smaller group. For a design of this form with the five treated states, four historical dates and 118 cells of the minimum-wage application, $t_C\ge0.99$ at $\alpha=0.05$, the scale of the cells themselves (the separate-cluster formula is used here because treated and comparison states are distinct); event windows that overlap in calendar time add dependence across events on top of this. Cluster sampling therefore cannot make that application inferential, which is why we report it descriptively.

\paragraph{Gaussian calibration.} Sharper critical values come from the cluster central limit theorem. Write $\widehat Z-Z=C^{-1}\sum_{c=1}^C\psi_c+R_C$ with $C=C_0+C_1$, where the $\psi_c$ are independent, mean-zero $J\times K$ arrays; in the design above $\psi_c$ is $C/C_1$ or $C/C_0$ times the centred report of cluster $c$ and $R_C=0$, and the reports are bounded, so clause (i) below holds when $C_1/C$ and $C_0/C$ stay bounded away from zero.

\begin{assumption}[Clusters]\label{S-ass:cluster}
For a class $\Pcal_C$ of data-generating processes and constants $B<\infty$, $\sigma_{\min}>0$, uniformly over $P\in\Pcal_C$: (i) $J$ and $K$ are fixed and the $\psi_c$ are independent with $\|\psi_c\|_\infty\le B$; (ii) the average covariance $\Omega_C=C^{-1}\sum_c\mathrm{Var}\{\mathrm{vec}(\psi_c)\}$ has every diagonal entry at least $\sigma_{\min}^2$, except on coordinates that are almost surely constant, which are removed from the band and imposed exactly; (iii) $\sqrt C\|R_C\|_\infty\le r_C$ with deterministic $r_C\to0$; (iv) an estimator $\widehat\Omega$ satisfies $\Pp(\max_{i,j}|\widehat\Omega_{ij}-\Omega_{C,ij}|>\Delta_C)\le\eta_C$ for deterministic $\Delta_C,\eta_C\to0$.
\end{assumption}

The shared-cluster design satisfies clause (i) with $\psi_c=V_c-\E V_c$, bounded by 3, and $R_C=0$. Clause (iv) holds in either design with $\widehat\Omega=C^{-1}\sum_c\widehat\psi_c\widehat\psi_c^{\top}$, where $\widehat\psi_c$ is the cluster's report minus the average report of its group (or of all clusters, in the shared design), when the clusters within each group are identically distributed: then Hoeffding's inequality for the bounded products gives $\Delta_C$ of order $(\log C/C)^{1/2}$. When clusters differ in their means, $\widehat\Omega$ estimates $\Omega_C$ plus a positive semidefinite term; by Anderson's theorem \citep{Anderson1955} the symmetric band below is then conservative, but the one-sided band need not be, and its validity then requires an estimator of $\Omega_C$ that is consistent under heterogeneous means, for instance one that removes a modelled cluster-specific mean.

\begin{theorem}[Uniform cluster calibration]\label{S-thm:cluster-gauss}
Under Assumption~\ref{S-ass:cluster}, let $\widehat q_{\rm sym}$ and $\widehat q_{\rm one}$ be the $1-\alpha$ quantiles, conditional on the data, of $\|\widehat W\|_\infty$ and $\max_{s,k}[-\widehat W_{s,k}]_+$ for $\mathrm{vec}(\widehat W)\sim N(0,\widehat\Omega)$. Then
\[
\begin{aligned}
&\liminf_{C\to\infty}\inf_{P\in\Pcal_C}\Pp\bigl(\sqrt C\|\widehat Z-Z\|_\infty\le\widehat q_{\rm sym}\bigr)\ge1-\alpha,\\
&\limsup_{C\to\infty}\sup_{P\in\Pcal_C:\,\Lambda_{S,\Pi}(P)\ne\varnothing}\Pp\bigl(\sqrt C\,d(\widehat Z)>\widehat q_{\rm one}\bigr)\le\alpha .
\end{aligned}
\]
\end{theorem}

\begin{proof}
Both events have the form $\{\sqrt C(\widehat Z-Z)\in\mathcal R\}$ for a hyperrectangle $\mathcal R$ in $\mathbb R^{JK}$ (up to the remainder), because $\{v:\|v\|_\infty\le q\}$ and $\{v:\min_{s,k}v_{s,k}\ge-q\}$ are rectangles; for the test we use the bound of step (ii) of Proposition~\ref{S-prop:cluster-power} below, $d(\widehat Z)\le\max_{s,k}[-(\widehat Z-Z)_{s,k}]_+$ under the null. Let $W\sim N(0,\Omega_C)$. (a) By the central limit theorem over hyperrectangles for independent, not necessarily identically distributed, bounded vectors in \citet{CCK2017}, $\sup_{\mathcal R}|\Pp\{C^{-1/2}\sum_c\psi_c\in\mathcal R\}-\Pp(W\in\mathcal R)|\le\kappa_{1,C}$, where $\kappa_{1,C}$ is of order $\{B^2\log^7(JKC)/C\}^{1/6}$ with a constant depending only on $\sigma_{\min}$. (b) The remainder shifts the statistic by at most $r_C$; by the anti-concentration inequality for Gaussian maxima in \citet{CCK2015}, which needs only the variance lower bound in (ii), this changes probabilities by at most $\kappa_{2,C}=r_C\{(2\log(2JK))^{1/2}+2\}/\sigma_{\min}$. (c) On the event in clause (iv), the comparison inequality for Gaussian maxima in \citet{CCK2015} bounds the difference between the distribution functions of the two maxima under $N(0,\widehat\Omega)$ and $N(0,\Omega_C)$ by $\kappa_{3,C}$, of order $\Delta_C^{1/3}\{1\vee\log(JK/\Delta_C)\}^{2/3}$, with a constant depending only on $\sigma_{\min}$ and $B$. Steps (a)--(c) bound the Kolmogorov distance between the law of the statistic and its conditional Gaussian approximation by $\kappa_C=\kappa_{1,C}+\kappa_{2,C}+\kappa_{3,C}$ outside an event of probability $\eta_C$, uniformly over $\Pcal_C$, because every constant depends on $P$ only through $(B,\sigma_{\min},J,K)$. The standard quantile transfer then gives coverage at least $1-\alpha-2\kappa_C-\eta_C$ for the band and size at most $\alpha+2\kappa_C+\eta_C$ for the test; both error terms vanish.
\end{proof}

A studentized band, dividing each coordinate by $\widehat\sigma_{s,k}\vee f$ for a floor $f>0$ fixed in advance, is covered by the same argument after a rescaling of the coordinates when $f\le\sigma_{\min}$ and the $\widehat\sigma_{s,k}$ are uniformly consistent; with $f>\sigma_{\min}$ it remains valid but is conservative on the affected cells. If a cell's variance is not bounded below, Assumption~\ref{S-ass:cluster}(ii) fails. A cell may be imposed exactly only when the design makes it constant, for instance a cell above a top-code that no report can reach; a sample variance of zero is not evidence of this, and such a cell should instead be placed in a separate Hoeffding band with part of $\alpha$. If the benchmark of Proposition~\ref{prop:benchmark-test} shares clusters with the comparison group, its reports must be resampled jointly with theirs, not independently.

\paragraph{Power.} The test statistic is the deficit $d$. The following properties hold for any estimator and need no distributional assumption.

\begin{proposition}[Deficit properties and power bounds]\label{S-prop:cluster-power}
(i) $|d(V)-d(V')|\le\|V-V'\|_\infty$ for all $J\times K$ arrays $V,V'$. (ii) If $\Lambda_{S,\Pi}(P)\ne\varnothing$, then $d(\widehat Z)\le\max_{s,k}[-(\widehat Z-Z)_{s,k}]_+\le\|\widehat Z-Z\|_\infty$. (iii) For any threshold $t$, the test that rejects when $d(\widehat Z)>t$ satisfies
\[
\Pp\bigl(\|\widehat Z-Z\|_\infty<d(Z)-t\bigr)\le\Pp\bigl(d(\widehat Z)>t\bigr)\le\Pp\bigl(\|\widehat Z-Z\|_\infty>t-d(Z)\bigr).
\]
(iv) Along local alternatives $Z_C$ with $\sqrt C\,d(Z_C)\to\gamma$, thresholds with $\sqrt Ct_C\to q$, and $\sqrt C(\widehat Z-Z_C)\Rightarrow W$ with $W$ Gaussian and nondegenerate, the limiting rejection probability, at continuity points, lies between $\Pp(\|W\|_\infty<\gamma-q)$ and $\Pp(\|W\|_\infty>q-\gamma)$. In particular a test whose threshold is the $1-\alpha$ quantile of $\|W\|_\infty$ has asymptotic power at most $\alpha$ against alternatives with $\gamma=0$ and power tending to one as $\gamma\to\infty$.
\end{proposition}

\begin{proof}
(i) For each $\lambda$, $\max_k[-(V_\lambda)_k]_+\le\max_k[-(V'_\lambda)_k]_++\|V_\lambda-V'_\lambda\|_\infty$ because $x\mapsto[-x]_+$ is 1-Lipschitz, and $\|V_\lambda-V'_\lambda\|_\infty\le\|V-V'\|_\infty$ because $\lambda$ is a probability vector. Minimizing over $\lambda$ on both sides and exchanging $V$ and $V'$ gives (i). (ii) Let $\lambda_0$ be coherent. For each $k$, $(\widehat Z_{\lambda_0})_k\ge(\widehat Z_{\lambda_0}-Z_{\lambda_0})_k$, so $[-(\widehat Z_{\lambda_0})_k]_+\le[-(\widehat Z_{\lambda_0}-Z_{\lambda_0})_k]_+\le\sum_s\lambda_{0,s}[-(\widehat Z-Z)_{s,k}]_+\le\max_{s}[-(\widehat Z-Z)_{s,k}]_+$ by convexity of $x\mapsto[-x]_+$, and $d(\widehat Z)$ is at most the maximum over $k$. (iii) By (i), $d(\widehat Z)\ge d(Z)-\|\widehat Z-Z\|_\infty$ and $d(\widehat Z)\le d(Z)+\|\widehat Z-Z\|_\infty$; the two inclusions of events follow. (iv) Multiply (iii) by $\sqrt C$, with $Z_C$ in place of $Z$, and pass to the limit with the continuous mapping theorem at continuity points. With $\gamma=0$ the upper bound is $\Pp(\|W\|_\infty>q)=\alpha$; as $\gamma\to\infty$ the lower bound tends to one.
\end{proof}

Part (ii) shows that the one-sided critical value $\widehat q_{\rm one}$, which is never larger than $\widehat q_{\rm sym}$, is valid; part (iv) shows that power requires a deficit of order $C^{-1/2}$. A smaller critical value for the same statistic gives weakly higher power pathwise; it says nothing about power relative to other statistics. The bounds in (iii) and (iv) are not the power function itself. The next result gives it.

\paragraph{The exact local limit.} Fix $Z_0$ with $d(Z_0)=0$ and let $\Lambda^*=\{\lambda\in\Delta_S:(Z_{0,\lambda})_k\ge0\text{ for all }k\}$ be its nonempty coherent set, where $Z_{0,\lambda}=\sum_s\lambda_sZ_{0,s\cdot}$. For $\lambda\in\Lambda^*$ let $K^*(\lambda)=\{k:(Z_{0,\lambda})_k=0\}$ be its binding cells and $T(\lambda)=\{v\in\mathbb R^J:\sum_sv_s=0,\ v_s\ge0\text{ whenever }\lambda_s=0\}$ the cone of feasible directions of $\Delta_S$ at $\lambda$. For a $J\times K$ array $H$ write $H_\lambda=\sum_s\lambda_sH_{s\cdot}$, and define
\[
D(H)=\inf_{\lambda\in\Lambda^*}\ \inf_{v\in T(\lambda)}\ \max\Bigl\{0,\ \max_{k\in K^*(\lambda)}\bigl[-(Z_{0,v})_k-(H_\lambda)_k\bigr]\Bigr\},
\]
with the inner maximum over an empty set read as $-\infty$.

\begin{theorem}[Local limit of the coherence statistic]\label{S-thm:local-limit}
(i) The deficit $d$ is Hadamard directionally differentiable at $Z_0$ with derivative $D$: for $t\downarrow0$ and $H_t\to H$, $d(Z_0+tH_t)/t\to D(H)$. The map $D$ is positively homogeneous and $|D(H)-D(H')|\le\|H-H'\|_\infty$. (ii) If $Z_C=Z_0+C^{-1/2}H+o(C^{-1/2})$ and $\sqrt C(\widehat Z-Z_C)\Rightarrow W$, then $\sqrt C\,d(\widehat Z)\Rightarrow D(H+W)$, and a test with thresholds $\sqrt Ct_C\to q$ has limiting rejection probability $\Pp\{D(H+W)>q\}$ at continuity points. (iii) If some $\lambda\in\Lambda^*$ has $K^*(\lambda)=\varnothing$, then $D\equiv0$, $Z_C$ is coherent for all large $C$, and the limiting rejection probability is zero for every $q>0$ and every $H$. (iv) If $Z_C$ is coherent for every $C$, the law of $D(H+W)$ is stochastically dominated by that of $\max_{s,k}[-W_{s,k}]_+$.
\end{theorem}

\begin{proof}
(i) Write $\phi(\lambda,V)=\max\{0,\max_k[-(V_\lambda)_k]\}$, so that $d(V)=\min_{\lambda\in\Delta_S}\phi(\lambda,V)$, and let $V_t=Z_0+tH_t$. \emph{Upper bound.} Fix $\lambda\in\Lambda^*$ and $v\in T(\lambda)$. Because $\Delta_S$ is a polytope, $\lambda+tv\in\Delta_S$ for small $t$. For $k\notin K^*(\lambda)$ we have $(Z_{0,\lambda})_k>0$, so the $k$-th term of $\phi(\lambda+tv,V_t)$ is negative for small $t$; for $k\in K^*(\lambda)$ it equals $-t(Z_{0,v})_k-t(H_{t,\lambda})_k-t^2(H_{t,v})_k$. Hence $\limsup_t d(V_t)/t\le\limsup_t\phi(\lambda+tv,V_t)/t=\max\{0,\max_{k\in K^*(\lambda)}[-(Z_{0,v})_k-(H_\lambda)_k]\}$, and taking the infimum over $(\lambda,v)$ gives $\limsup_td(V_t)/t\le D(H)$. \emph{Lower bound.} Let $\lambda_t$ minimize $\phi(\cdot,V_t)$. For any $\lambda_0\in\Lambda^*$, $d(V_t)\le\phi(\lambda_0,V_t)\le t\|H_t\|_\infty$, so $\phi(\lambda_t,Z_0)\le\phi(\lambda_t,V_t)+t\|H_t\|_\infty\le2t\|H_t\|_\infty$. The set $\Lambda^*$ is a polytope defined by linear inequalities, and $\phi(\lambda,Z_0)$ is the largest violation of those inequalities that are not already imposed by $\lambda\in\Delta_S$; by Hoffman's error bound \citep{Hoffman1952} there is $\kappa<\infty$, depending only on $Z_0$, and a point $\mu_t\in\Lambda^*$ with $\|\lambda_t-\mu_t\|_1\le\kappa\phi(\lambda_t,Z_0)\le2\kappa t\|H_t\|_\infty$. Put $v_t=(\lambda_t-\mu_t)/t$; then $\|v_t\|_1\le2\kappa\|H_t\|_\infty$ and $v_t\in T(\mu_t)$ because $\mu_t+tv_t\in\Delta_S$. Since $\lambda_t=\mu_t+tv_t$, the $k$-th cell of $V_{t,\lambda_t}$ equals $t\{(Z_{0,v_t})_k+(H_{t,\mu_t})_k\}+t^2(H_{t,v_t})_k$ when $k\in K^*(\mu_t)$. Keeping only these cells, and using $|(H_{t,v_t})_k|\le\|H_t\|_\infty\|v_t\|_1$ and the fact that $(\mu_t,v_t)$ is admissible in the definition of $D(H_t)$,
\[
\begin{aligned}\frac{d(V_t)}t=\frac{\phi(\lambda_t,V_t)}t&\ge\max\Bigl\{0,\max_{k\in K^*(\mu_t)}\bigl[-(Z_{0,v_t})_k-(H_{t,\mu_t})_k\bigr]\Bigr\}-t\|H_t\|_\infty\|v_t\|_1\\&\ge D(H_t)-2\kappa t\|H_t\|_\infty^2 .\end{aligned}
\]
Since $|(H_{t,\lambda})_k-(H_\lambda)_k|\le\|H_t-H\|_\infty$ for every probability vector $\lambda$, the expression inside $D$ changes by at most $\|H_t-H\|_\infty$, which proves the Lipschitz property and $D(H_t)\to D(H)$. Thus $\liminf_td(V_t)/t\ge D(H)$. Positive homogeneity holds because $T(\lambda)$ is a cone. (ii) Write $\sqrt C\,d(\widehat Z)=\sqrt C\{d(Z_0+C^{-1/2}h_C)-d(Z_0)\}$ with $h_C=\sqrt C(\widehat Z-Z_C)+\sqrt C(Z_C-Z_0)\Rightarrow W+H$; the delta method for Hadamard directionally differentiable maps \citep{Shapiro1991,FangSantos2019} gives the limit, and the rejection probability follows at continuity points. (iii) The inner maximum is over an empty set for such $\lambda$, so the bracket equals zero. Coherence of $Z_C$ for large $C$ follows because every cell of $Z_{0,\lambda}$ is strictly positive and $\|Z_C-Z_0\|_\infty\to0$. (iv) By Proposition~\ref{S-prop:cluster-power}(ii), $\sqrt C\,d(\widehat Z)\le\max_{s,k}[-\sqrt C(\widehat Z-Z_C)_{s,k}]_+$ for every $C$; weak limits preserve this ordering of distribution functions.
\end{proof}

Three consequences. First, the limit law depends on the configuration only through $\Lambda^*$ and the binding cells. With one binding cell at a unique weight on the boundary of the simplex, $\lambda_0=(1,0)$ in the design of Table~\ref{S-tab:cluster-sims}, the cone is $\{(-a,a):a\ge0\}$, moving along it only lowers the binding cell, and $D(H)=[-H_{1,1}]_+$. With two binding cells at the unique interior weight $\lambda_0=(1/2,1/2)$, the weight can move in either direction; in that design moving it raises one binding cell and lowers the other at equal rates, and minimizing over it gives $D(H)=[-\{(H_{\lambda_0})_1+(H_{\lambda_0})_2\}/2]_+$: the statistic averages the two binding cells instead of taking the worse one. A numerical check reproduces both formulas draw by draw: exactly in the first case, and in the second with a mean absolute discrepancy of 0.002 at $C=50$ that falls to 0.0007 at $C=800$. Second, part (iv) explains the conservativeness in the simulations: the vertex critical value is the quantile of $\max_{s,k}[-W_{s,k}]_+$ over all $JK$ coordinates, while the null limit $D(W)$ uses a few averaged binding cells. Third, part (iii) describes the interior of the null: when the coherent set contains a weight with no binding cell, every local perturbation of $Z_0$ is still coherent, so these sequences are null sequences, and the test's rejection probability along them tends to zero rather than to $\alpha$. Nontrivial local power therefore requires a configuration in which every coherent weight has a binding cell, and a direction $H$ that pushes all of them out. Critical values estimated from $D(\cdot)$, with $\Lambda^*$ and $K^*$ estimated using a tuning sequence, give tests that are valid pointwise \citep{FangSantos2019} and more powerful; their uniform validity requires conditions that we have not established, and the contact-set rule in Table~\ref{S-tab:cluster-sims}, a crude version, over-rejects.

\paragraph{Simulations.} Table~\ref{S-tab:cluster-sims} reports a Monte Carlo study with $J=2$ historical dates and $K=5$ cells. Each of $C\in\{20,40,80,160\}$ clusters contributes a bounded array whose two rows are correlated ($\rho\in\{0,0.9\}$) and whose scale differs across clusters by a factor of one or $\sqrt{10}$. Four null configurations are used: every cell interior; a unique coherent weight with one binding cell; a unique weight with two binding cells; and a set of coherent weights with binding cells at its edges. Local alternatives move mass of order $\delta C^{-1/2}$ out of the binding cells. Size is the maximum over the 16 combinations of $(C,\rho,\text{scale})$ with 2,000 replications each (Monte Carlo standard error at most 0.5 points at 5 percent); power uses 1,000 replications with $\rho=0.5$; Gaussian critical values use 1,000 draws.

\begin{table}[h]\centering\small
\caption{Coherence tests under cluster sampling: worst-case size and local power}\label{S-tab:cluster-sims}
\setlength{\tabcolsep}{4pt}
\footnotesize\begin{tabular}{@{}lrrrrrr@{}}\toprule
 & $\sqrt C\,d$ & Symm. & One-sided & Student. & Hoeffding & Contact\\\midrule
\multicolumn{7}{@{}l}{\emph{Size: worst case over $C$, $\rho$ and cluster scales}}\\
Interior & 0 & 0.000 & 0.000 & 0.000 & 0.000 & 0.000\\
Unique weight, one binding cell & 0 & 0.007 & 0.011 & 0.013 & 0.000 & 0.067\\
Unique weight, two binding cells & 0 & 0.000 & 0.000 & 0.000 & 0.000 & 0.003\\
Weight set, binding at edges & 0 & 0.000 & 0.000 & 0.000 & 0.000 & 0.000\\
\multicolumn{7}{@{}l}{\emph{Power, $C=160$, alternatives of size $\delta C^{-1/2}$}}\\
\quad One binding cell, $\delta=0.5$ & 0.15 & 0.033 & 0.058 & 0.065 & 0.000 & 0.243\\
\quad One binding cell, $\delta=1$ & 0.30 & 0.211 & 0.282 & 0.285 & 0.000 & 0.629\\
\quad One binding cell, $\delta=2$ & 0.60 & 0.861 & 0.910 & 0.918 & 0.000 & 0.989\\
\quad One binding cell, $\delta=4$ & 1.20 & 1.000 & 1.000 & 1.000 & 0.000 & 1.000\\
\quad Two binding cells, $\delta=0.5$ & 0.08 & 0.000 & 0.000 & 0.000 & 0.000 & 0.009\\
\quad Two binding cells, $\delta=1$ & 0.15 & 0.002 & 0.004 & 0.005 & 0.000 & 0.094\\
\quad Two binding cells, $\delta=2$ & 0.30 & 0.061 & 0.141 & 0.146 & 0.000 & 0.686\\
\quad Two binding cells, $\delta=4$ & 0.60 & 0.985 & 0.996 & 0.995 & 0.000 & 1.000\\
\bottomrule\end{tabular}

\begin{minipage}{0.97\linewidth}\footnotesize\emph{Notes:} Nominal level 5 percent. Symmetric and one-sided: Gaussian calibration of Theorem~\ref{S-thm:cluster-gauss}. Studentized: one-sided with coordinates divided by $\widehat\sigma_{s,k}\vee0.02$. Hoeffding: Proposition~\ref{S-prop:cluster-hoeffding}. Contact set: critical value from the cells within $(\log C)^{1/2}$ standard errors of zero at the estimated weight; no uniform validity result. $\sqrt C\,d$: scaled population deficit of the alternative.\end{minipage}
\end{table}

Three findings stand out. All the vertex-band tests control size in every configuration, with worst-case rejection rates of at most 1.3 percent: they are conservative because the deficit is a minimum over weights while the band covers every vertex. The one-sided calibration is uniformly more powerful than the symmetric one, by 4 to 8 points at intermediate alternatives, and studentizing adds little in these designs. The finite-sample Hoeffding test is valid but has no power at these cluster counts. The contact-set rule is much more powerful near the boundary, but its worst-case size is 6.7 percent in the one-binding-cell configuration, above the nominal level by more than the simulation error, which is the danger the uniformity literature warns about. When the coherent set has slack, local perturbations of the binding cells leave the deficit at zero and no test has power, as part (iii) of Theorem~\ref{S-thm:local-limit} predicts. We recommend the one-sided Gaussian calibration for clustered data, and the Hoeffding radius when a finite-sample guarantee is required and clusters number in the hundreds.

The high-dimensional central limit theorem behind Theorem~\ref{S-thm:cluster-gauss} has an error of order $\{\log^7(JKC)/C\}^{1/6}$, which is not small when $K=118$ and $C=50$. Two further experiments therefore split the last three cells into 20 and 116 cells, with $C\in\{15,25,50\}$ and cluster scales differing by a factor of $\sqrt{10}$. The Gaussian vertex bands themselves cover at rates between 0.94 and 0.98 across all $(K,C)$, slightly above nominal for many cells and few clusters, and the coherence tests built on them reject at most 1.8 percent of the time under the null, including with $K=118$ and $C=15$. The contact-set rule over-rejects with five cells (up to 7.5 percent at $C=25$) and becomes conservative with many. A cell whose cluster reports barely vary, with standard deviation 0.003 against a floor of 0.02, leaves all three Gaussian tests valid. These designs have bounded, light-tailed reports; with heavy-tailed cluster reports the approximation can be worse, and the Hoeffding radius is then the safe choice.
\section{What Pre-Treatment Data Cannot Identify}\label{S-sec:adapt}

For $S\subset S'$, let $\mathfrak M_S\subseteq\mathfrak M_{S'}$ be the
corresponding structural model classes and let $\theta(Q)=F^0_{1,1,Q}$.
Write $O_N$ for the observed sample. In the testing and confidence-set claims
below, assume its law under a full-data law $Q$ is
$\mathsf K_N\{\mathcal O(Q)\}$ for a sampling kernel $\mathsf K_N$ common
to the compared laws. Independent sampling of complete panel histories or
independent repeated cross-sections with a common fixed cell-sampling rule
satisfy this condition. The claim is conditional on the sampling kernel:
equal marginal observed-data laws alone need not determine a survey sample's
joint design distribution when the design itself differs. A randomized procedure must use the same auxiliary
randomization law under the compared models.

\begin{theorem}[Exact observational equivalence]\label{S-thm:obs-equiv}
If for an observed law $P$,
\[
\Fcal_S(P)\subsetneq\Fcal_{S'}(P),
\]
then there exist $Q_a\in\mathfrak M_S$ and $Q_b\in\mathfrak M_{S'}\setminus\mathfrak M_S$ such that
\[
\mathcal O(Q_a)=\mathcal O(Q_b)=P.
\]
Under the common sampling-kernel condition, every observed-data test has
the same rejection probability under the two laws at every sample size.
\end{theorem}

\begin{proof}
Choose $F^a\in\Fcal_S(P)$ and $F^b\in\Fcal_{S'}(P)\setminus\Fcal_S(P)$.
Sharpness under $S$ supplies $Q_a\in\mathfrak M_S$ with
$\mathcal O(Q_a)=P$ and $\theta(Q_a)=F^a$. Sharpness under $S'$ supplies
$Q_b\in\mathfrak M_{S'}$ with $\mathcal O(Q_b)=P$ and
$\theta(Q_b)=F^b$. If $Q_b$ belonged to $\mathfrak M_S$, sharpness under
$S$ would put $F^b$ in $\Fcal_S(P)$, a contradiction. The common sampling
kernel gives
$\mathcal L_{Q_a}(O_N)=\mathsf K_N(P)=\mathcal L_{Q_b}(O_N)$
for every $N$. This equality also covers a randomized test when its
auxiliary randomization has the same conditional law under both models.
\end{proof}

\begin{theorem}[No honest oracle adaptation]\label{S-thm:no-adapt}
Fix $0<\alpha<1$ and retain the common sampling-kernel condition. Suppose
$C_N$ is an observed-data-measurable random nonempty closed subset of the CDF
space, with the Borel sigma-field induced by Hausdorff distance under the
CDF sup norm, and is uniformly honest for the structural target over
$\mathfrak M_{S'}$ in the sense that
\[
\liminf_{N\to\infty}\inf_{Q\in\mathfrak M_{S'}}
Q\{\theta(Q)\in C_N\}\ge 1-\alpha.
\]
Suppose also that, at an observationally equivalent $Q_a\in\mathfrak M_S$,
\[
d_H(C_N,\Fcal_S(P))\to_p0.
\]
If $\Fcal_S(P)\subsetneq\Fcal_{S'}(P)$, these requirements are incompatible.
\end{theorem}

\begin{proof}
Choose $F^b\in\Fcal_{S'}(P)\setminus\Fcal_S(P)$.  Compactness of $\Fcal_S(P)$ implies
\[
\delta=d_\infty(F^b,\Fcal_S(P))>0.
\]
Construct an observationally equivalent $Q_b$ with $\theta(Q_b)=F^b$.
Let
\[
E_N=\{d_H(C_N,\Fcal_S(P))<\delta/2\}.
\]
If $E_N$ holds and
$F^b\in C_N$, then
\[
\delta=d_\infty(F^b,\Fcal_S(P))
\le \sup_{F\in C_N}d_\infty(F,\Fcal_S(P))<\delta/2,
\]
a contradiction. By measurability and the common sampling kernel,
$Q_b(E_N^c)=Q_a(E_N^c)\to0$. Consequently,
\[
Q_b\{\theta(Q_b)\in C_N\}\leq Q_b(E_N^c)\longrightarrow0.
\]
Since $Q_b\in\mathfrak M_{S'}$, the honesty infimum is at most this
quantity at every $N$ and cannot have lower limit at least $1-\alpha>0$.
\end{proof}

\section{Simulations}\label{S-sec:sims}

\paragraph{Designs.} All designs have six ordered cells, the first being ``no earnings'', and group--date samples drawn independently from multinomial distributions. In Design~1 the post-policy comparison law is $c=(.10,.20,.25,.20,.15,.10)$, and three historical dates have treated laws $(.55,.25,.12,.05,.02,.01)$, $(.45,.28,.15,.07,.03,.02)$, $(.40,.30,.16,.08,.04,.02)$ and comparison laws $(.10,.20,.25,.20,.10,.15)$, $c$, and $(.12,.22,.25,.19,.14,.08)$. The first vertex has mass $-0.04$ in the top cell. The true untreated law is the coherent mixture with weights $(0.2,0.5,0.3)$, and the benchmark sample is drawn from it. The target is the counterfactual share with no earnings, whose sharp interval is $[0.380,0.483]$ against $[0.380,0.550]$ for the signed hull. Design~2 shifts the benchmark law by $\gamma(0,-1,0,0,1,0)$ with $\gamma\in\{0.05,0.10,0.15\}$, which places it at the CDF distances from the hull reported in Table~\ref{S-tab:sims}. In Design~3 two vertices lie on the line $q+t\cdot0.1(e_1-e_6)$, with $q=(.40,.20,.15,.13,.10,.02)$ and $e_k$ the $k$th unit vector, generated from valid treated and comparison laws; the coherent set is a single point when the nearer vertex is at $t=0.2$, and empty with population repair $d$ when it is at $t=0.2+10d$.

\paragraph{Procedures.} The unstudentized coherence test uses the bootstrap quantile of $\max_{s,k}|\widehat z^*_{s,k}-\widehat z_{s,k}|$; the DKW test uses the finite-sample cell band $2\varepsilon^{\rm DKW}_N$. The studentized test and the confidence region for the zero-earnings share are described in Section~\ref{sec:inference}: the region optimizes the share over weights satisfying the studentized cell constraints and adds the studentized band for the vertex shares, with one critical value for all cells and shares. The signed region in the table adds an unstudentized band to the signed-hull interval. The benchmark tests are those of Proposition~\ref{prop:benchmark-test}, with distribution functions evaluated at the cell boundaries.

\begin{table}[h]\centering\footnotesize
\caption{Monte Carlo results (1,000 replications, 299 bootstrap draws)}\label{S-tab:sims}
\setlength{\tabcolsep}{3pt}
\begin{tabular}{@{}lccc@{}}\toprule
$(N_1,N_0,N_R)$ & $(200,2000,300)$ & $(1000,5000,1000)$ & $(5000,20000,5000)$\\\midrule
\multicolumn{4}{@{}l}{\emph{Design 1: transport holds; population bound $[0.380,0.550]$ signed,}}\\
\multicolumn{4}{@{}l}{\emph{\quad $[0.380,0.483]$ coherent}}\\
\quad Rejection: studentized coherence test & 0.000 & 0.000 & 0.000\\
\quad Rejection: benchmark test, bootstrap / DKW & 0.000 / 0.000 & 0.000 / 0.000 & 0.000 / 0.000\\
\quad Coverage of identified interval & 0.999 & 1.000 & 1.000\\
\quad Median width: coherent / signed region & 0.375 / 0.345 & 0.239 / 0.250 & 0.166 / 0.206\\[2pt]
\multicolumn{4}{@{}l}{\emph{Design 2: benchmark at CDF distance $D$ from the hull; rejection rate, bootstrap / DKW}}\\
\quad $D=0.031$ & 0.002 / 0.000 & 0.010 / 0.000 & 0.852 / 0.000\\
\quad $D=0.068$ & 0.098 / 0.000 & 0.845 / 0.000 & 1.000 / 0.342\\
\quad $D=0.118$ & 0.622 / 0.000 & 1.000 / 0.024 & 1.000 / 1.000\\
\multicolumn{4}{@{}l}{\emph{Design 3: population repair $d$; rejection rate,}}\\
\multicolumn{4}{@{}l}{\emph{\quad plug-in rule / unstudentized / studentized}}\\
\quad $d=0.000$ & 0.50 / 0.00 / 0.01 & 0.48 / 0.00 / 0.01 & 0.50 / 0.00 / 0.01\\
\quad $d=0.005$ & 0.64 / 0.00 / 0.03 & 0.79 / 0.00 / 0.04 & 0.96 / 0.00 / 0.17\\
\quad $d=0.010$ & 0.78 / 0.00 / 0.06 & 0.96 / 0.00 / 0.17 & 1.00 / 0.00 / 0.81\\
\quad $d=0.020$ & 0.95 / 0.00 / 0.19 & 1.00 / 0.00 / 0.69 & 1.00 / 0.55 / 1.00\\
\quad $d=0.040$ & 0.99 / 0.00 / 0.64 & 1.00 / 0.37 / 1.00 & 1.00 / 1.00 / 1.00\\
\bottomrule\end{tabular}
\begin{minipage}{0.97\linewidth}\footnotesize\emph{Notes:} $N_1$ and $N_0$ are the sizes of each treated and comparison group--date sample and $N_R$ the size of the benchmark sample. Nominal level 5\%. The plug-in rule rejects when the estimated coherent set is empty. The DKW coherence test never rejected in these designs and is omitted. In Design 3 with $d=0$ the population coherent set is a single point.\end{minipage}
\end{table}
\paragraph{Findings.} All procedures control size. The naive plug-in rule does not: when the coherent set is a single point it is empty in about half the samples. The coherence tests are conservative at the boundary, as expected from a test whose null includes the whole coherent region. Studentization is what gives the coherence test power, because infeasibility arises in rare cells, whose cell masses are estimated precisely; the unstudentized band is dominated by the most variable cells. With samples of the NSW size the coherent confidence region is slightly wider than the signed one, because the joint critical value for many studentized cells exceeds the one for a single score; coherence sharpens the region once samples are larger. The benchmark test has good power against departures of 5 to 10 points in CDF distance with a thousand treated units. The finite-sample DKW calibration is valid but far more conservative.

\paragraph{Validity of the studentized band.} Let $\sigma_{s,k}>0$ and suppose that, with probability at least $1-\alpha$, $|\widehat z_{s,k}-z_{s,k}|\le c\,\sigma_{s,k}$ for all $s,k$. If $\lambda\in\Lambda_{S,\Pi}$, then on this event $\widehat z_{\lambda,k}\ge z_{\lambda,k}-\sum_s\lambda_s|\widehat z_{s,k}-z_{s,k}|\ge-c\sum_s\lambda_s\sigma_{s,k}$ for every $k$. Hence the set $\{\lambda:\widehat z_{\lambda,k}+c\sum_s\lambda_s\sigma_{s,k}\ge0\ \forall k\}$ contains $\Lambda_{S,\Pi}$ on the event, and the test that rejects when it is empty has size at most $\alpha$. With estimated $\sigma_{s,k}$ and a bootstrap critical value for the maximal studentized error, validity is asymptotic and requires the cell variances to be bounded away from zero, or floored as in our implementation.

\subsection{Pooling under heterogeneous weights}\label{S-sec:pooling-sim}

This experiment is at the population level: no sampling. Each of three events, with equal weights, has six outcome cells and two historical dates. For event $e$ we draw a true weight $\lambda_e$ from the uniform distribution on $\Delta_2$ and a true untreated law $\mu_e$ from the uniform distribution on $\Delta_6$, and set the two vertices to $\mu_e+0.15\,d_{e,s}/\max|d_e|$, where the $d_{e,s}$ are Gaussian directions centred to have zero total mass and then shifted so that $\sum_s\lambda_{e,s}d_{e,s}=0$. Each event's true counterfactual is therefore the $\lambda_e$-mixture of its vertices and is a probability law, while individual vertices can be signed: the event-specific model holds and the common-weight models generally do not. The target is the average over events of the counterfactual probability of the first two cells. Over 2,000 designs:

\begin{center}\small
\begin{tabular}{@{}lccc@{}}\toprule
Model & Contains truth & Empty & Excludes truth\\\midrule
Event-specific weights & 1.00 & 0.00 & 0.00\\
Stacked & 0.71 & 0.00 & 0.29\\
Eventwise, common weight & 0.28 & 0.25 & 0.47\\\bottomrule
\end{tabular}
\end{center}

\noindent The median widths of the nonempty bounds are 0.077 (event-specific), 0.059 (stacked) and 0.015 (eventwise). Stacking buys a narrower bound than the correctly specified model by imposing a common weight on the pooled history, and in this design that restriction is false 29 percent of the time in a way that moves the bound off the truth.

\section{Application I: Data and Additional Results}\label{S-sec:nsw-supp}

\subsection{Samples}

The samples are those distributed with the replication archive of \citet{ImbensXu2025}: the Dehejia--Wahba subsample of the NSW men (185 participants, 260 controls), the CPS-1 comparison sample (15,992 men) and the PSID-1 comparison sample (2,490 men). They contain the same records as the files distributed by Rajeev Dehejia. Earnings are in 1982 dollars, and CPS earnings are top-coded at \$25,862. Table~\ref{S-tab:nsw-counts} gives the cell counts on the seven-cell grid, which suffice to reproduce every seven-cell calculation without the microdata.

\begin{table}[h]\centering\small
\caption{Counts by earnings cell, NSW and comparison samples}\label{S-tab:nsw-counts}
\setlength{\tabcolsep}{3.5pt}
\begin{tabular}{@{}lrrrrrrrrrrrr@{}}\toprule
 & \multicolumn{3}{c}{Participants} & \multicolumn{3}{c}{Exp.\ controls} & \multicolumn{3}{c}{CPS-1} & \multicolumn{3}{c}{PSID-1}\\
Earnings (\$000) & 74 & 75 & 78 & 74 & 75 & 78 & 74 & 75 & 78 & 74 & 75 & 78\\\midrule
0 & 131 & 111 & 45 & 195 & 178 & 92 & 1913 & 1748 & 2172 & 215 & 249 & 286\\
$(0,5]$ & 25 & 54 & 56 & 28 & 57 & 70 & 2300 & 2483 & 1649 & 132 & 125 & 97\\
$(5,10]$ & 15 & 15 & 43 & 21 & 17 & 56 & 1682 & 1778 & 1606 & 246 & 252 & 179\\
$(10,15]$ & 10 & 3 & 23 & 5 & 5 & 29 & 2054 & 2209 & 1938 & 362 & 362 & 300\\
$(15,20]$ & 1 & 1 & 9 & 4 & 2 & 9 & 2323 & 2378 & 2096 & 447 & 424 & 323\\
$(20,30]$ & 2 & 1 & 6 & 4 & 1 & 3 & 5720 & 5396 & 6531 & 660 & 654 & 730\\
$>30$ & 1 & 0 & 3 & 3 & 0 & 1 & 0 & 0 & 0 & 428 & 424 & 575\\\bottomrule
\end{tabular}
\end{table}

Mean characteristics differ sharply between participants and the survey samples. Participants average 25.8 years of age and 10.4 years of schooling; 84 percent are black, 19 percent married, and 71 percent lack a high-school degree. The corresponding figures are 33.2, 12.0, 7, 71 and 30 percent in CPS-1 and 34.9, 12.1, 25, 87 and 31 percent in PSID-1. The experimental controls resemble participants (25.1, 10.1, 83, 15 and 83 percent).

\subsection{Grids, outcomes and reweighting}

Table~\ref{S-tab:nsw-grids} reports the plug-in repair and the distance between the benchmark and the signed hull on three grids: the seven-cell grid of the main text, a 12-cell grid with \$2,500 steps to \$20,000 and cells $(20,25]$, $(25,30]$ and above 30, and a 23-cell grid with \$1,000 steps to \$20,000 and cells $(20,30]$ and above 30. The 12- and 23-cell grids are not nested, so Proposition~\ref{prop:partition-refinement} does not order them. For the outcome ``earnings above \$5,000'' the experimental effect is 0.077; the signed-hull intervals are $[0.27,0.32]$ (CPS-1) and $[0.31,0.35]$ (PSID-1), and with reweighting $[0.20,0.21]$ and $[0.22,0.32]$, with coherent intervals $[0.200,0.204]$ and $[0.302,0.304]$ on the seven-cell grid.

\begin{table}[h]\centering\small
\caption{NSW: repair and benchmark distance by grid}\label{S-tab:nsw-grids}
\begin{tabular}{@{}lcccccc@{}}\toprule
 & \multicolumn{3}{c}{Minimum repair (pp)} & \multicolumn{3}{c}{TV distance to benchmark}\\
Cells & 7 & 12 & 23 & 7 & 12 & 23\\\midrule
CPS-1 & 0.88 & 1.27 & 1.16 & 0.338 & 0.371 & 0.395\\
CPS-1 reweighted & 0 & 0.93 & 3.86 & 0.287 & 0.293 & 0.356\\
PSID-1 & 3.73 & 4.33 & 4.64 & 0.351 & 0.407 & 0.432\\
PSID-1 reweighted & 0 & 4.55 & 8.20 & 0.319 & 0.433 & 0.447\\\bottomrule
\end{tabular}
\end{table}

The reweighted samples weight each comparison man by $\widehat p(x)/\{1-\widehat p(x)\}$, normalized to mean one, where $\widehat p$ is a logit of participation on age, age squared, education, education squared, and indicators for black, Hispanic, married and no degree, estimated on participants and the comparison sample. These weights target the distribution of characteristics among participants. Pre-program earnings are deliberately excluded: conditioning on them would mechanically shrink the historical biases that the method is meant to use. The effective sample sizes after weighting are 431 (CPS-1) and 145 (PSID-1).

\subsection{Inference details}

The DKW critical values use $\alpha=0.05$, the six primitive CDFs listed in Section~\ref{S-sec:benchmark-supp}, and the actual sample sizes. The bootstrap resamples individuals with replacement within the participant, experimental-control and comparison samples separately, keeping each person's 1974, 1975 and 1978 earnings together; it uses 2,000 draws (999 when weights are re-estimated in every draw) and seed 20261121. The coherence screen and the outer region for the employment effect use a joint band formed by a Bonferroni split of $\alpha$ between the maximal cell error over vertices and cells (bootstrap quantile 0.087 for CPS-1, 0.087 for PSID-1) and the error in the treated employment share (0.070). Because the vertex cell errors sum to zero across cells, the error in a vertex's employment share is at most the cell band. The experimental confidence interval for the employment effect uses the normal approximation for a difference in two independent proportions.

\subsection{A grid-free confidence region}\label{S-sec:gridfree}

Let $x_1<\cdots<x_n$ be the pooled distinct earnings values, $\widehat H_s$ the estimated vertex distribution functions, and $c$ the bootstrap 95 percent quantile of $\max\{\max_s\|\widehat H^*_s-\widehat H_s\|_\infty,\ |\widehat g^*-\widehat g|\}$, resampling complete histories within each sample. On the event that the band covers, every $\lambda\in\Lambda_S(P)$ has a nondecreasing $H_\lambda$ with values in $[0,1]$, so $\widehat H_\lambda(w)-\widehat H_\lambda(u)\ge-2c$ for all $u<w$ and $-c\le\widehat H_\lambda\le1+c$; that is, $\lambda\in\widehat\Lambda_\infty(c)$. Each interval-cell constraint of Proposition~\ref{prop:grid-choice} is one of these inequalities, so $\widehat\Lambda_\infty(c)\subseteq\widehat\Lambda_\Pi(c)$ for every interval partition $\Pi$.

\paragraph{Exact computation.} $\widehat H_\lambda$ is constant between consecutive $x_i$ and its left limits are its values at the preceding points, so it suffices to impose the conditions at $x_1,\ldots,x_n$. Introduce $m_1,\ldots,m_n$ and the constraints
\[
m_{i-1}\le m_i,\qquad \widehat H_\lambda(x_i)\le m_i,\qquad m_i-\widehat H_\lambda(x_i)\le2c,\qquad -c\le\widehat H_\lambda(x_i)\le1+c .
\]
If $\lambda\in\widehat\Lambda_\infty(c)$, the running maximum $m_i=\max_{j\le i}\widehat H_\lambda(x_j)$ satisfies them. Conversely, any feasible $m$ dominates the running maximum, so $\max_{j\le i}\widehat H_\lambda(x_j)-\widehat H_\lambda(x_i)\le m_i-\widehat H_\lambda(x_i)\le2c$. The projection of this polyhedron onto $\lambda$ is therefore exactly $\widehat\Lambda_\infty(c)$, and the endpoints of the region for any linear contrast are two linear programs with $|S|+n$ variables and $5n-1$ inequalities. We solve them with every right-hand side enlarged by $10^{-9}$, so the computed set contains the exact one.

\paragraph{Results.} The region for the employment effect is $[\widehat g-\max\widehat e_\lambda-2c,\ \widehat g-\min\widehat e_\lambda+2c]$ over $\lambda\in\widehat\Lambda_\infty(c)$, where $\widehat e_\lambda=1-\widehat H_\lambda(0)$; it covers the identified interval with asymptotic probability at least 0.95. For CPS-1 ($n=22{,}339$), $c=0.092$, the admissible weights on 1974 form $[0.108,0.485]$, and the region is $[0.209,0.615]$. Weights near one are excluded by the range condition: the 1974 vertex exceeds one by more than $c$ near the CPS top-code. On the 7-, 12- and 23-cell grids with the same band the region is $[0.199,0.666]$. For PSID-1, $c=0.090$, all weights are admissible, and the region is $[0.191,0.674]$ on every grid and without one. The outer regions in Table~\ref{tab:nsw} differ because their bands are calibrated on the seven cell masses rather than on whole distribution functions. All exclude the experimental estimate of 0.111.

\subsection{Positive control: transport by design}\label{S-sec:positive-control}

To check that the tests do not reject when Assumption~\ref{ass:hull} holds, we split the 260 experimental controls at random into two halves of 130, use one half as the comparison group and the other as the benchmark, and keep the 185 participants as the treated group. Randomization makes the participants' and controls' untreated earnings histories identically distributed, so every historical bias is zero in the population and transport holds exactly. Across 200 random splits, with 199 bootstrap draws each, the benchmark test of Proposition~\ref{prop:benchmark-test} rejects at the 5 percent level in 1 split (0.5 percent), the studentized coherence test rejects in none, and the plug-in coherent set on the seven-cell grid is nonempty in 94 percent of splits. The median CDF distance between the benchmark half and the nearest transported distribution is 0.10, against 0.32 and 0.35 with the survey comparison groups. The rejections in Section~\ref{sec:nsw} are therefore not produced by the sparse cells of these small samples. As in the simulations, both tests are conservative.

\subsection{Sensitivity of the studentized test}

The studentized test divides each vertex cell error by its bootstrap standard error, with a floor to avoid division by zero. For CPS-1 one cell, earnings above \$30{,}000, has zero bootstrap variance: no CPS-1 earnings exceed the top-code, so the 1975 vertex has exactly zero mass there in every draw. For PSID-1 the smallest bootstrap standard error is 0.007. Varying the floor from $10^{-6}$ to $5\times10^{-3}$ leaves the $p$-values unchanged in both samples. With a different bootstrap seed the PSID-1 $p$-value is 0.008 instead of 0.013; the CPS-1 $p$-value is 0.84 in both.

\subsection{Wasserstein breakdown}\label{S-sec:w1}

On the seven-cell grid let $x_1<\cdots<x_7$ be cell representatives (0, 2,500, 7,500, 12,500, 17,500, 25,000 and, for the open top cell, 35,000 dollars). For a probability vector $q$ and a unit-mass signed vector $z$ define $W_1(q,z)=\sum_{k=1}^{6}|Q_k-Z_k|(x_{k+1}-x_k)$, where $Q_k$ and $Z_k$ are cumulative sums; for two probability vectors this is the Wasserstein distance between the corresponding discrete laws. The minimum repair, the sign breakdown radius and the realized error are the linear programs of Section~\ref{sec:breakdown} with $W_1$ in place of total variation, with auxiliary variables for the absolute values. For CPS-1 the minimum repair is \$44, the sign of the employment effect is lost at \$1{,}053, and the experimental control distribution lies \$2{,}630 from the historical hull. For PSID-1 the three numbers are \$260, \$1{,}236 and \$3{,}793. In total variation the corresponding radii are 0.38 and 0.37 against realized errors of 0.34 and 0.35. The two distances therefore disagree about whether the sign survives the departure the experiment reveals: moving mass between zero and high earnings is cheap in total variation and expensive in $W_1$, and the NSW departure consists largely of such moves. The $W_1$ numbers depend on the representative chosen for the open top cell, which enters only through the last term of the sum.

\subsection{Other estimators that use pre-program gaps}

Distributional difference-in-differences with a single base year $s$ uses the vertex $H_s$, so its employment effect is an endpoint of the signed interval: 0.481 (1974) and 0.383 (1975) for CPS-1, and 0.493 and 0.372 for PSID-1. For changes-in-changes we apply the bounds of \citet[Theorem~4.1]{AtheyImbens2006} for discrete outcomes to the counterfactual probability of zero earnings, with the comparison group as the control and the base year as the first period. Because the quantile of the comparison group's base-year distribution at the relevant level falls where that distribution is continuous, the lower and upper bounds coincide. The implied employment effects are 0.465 (base 1974) and 0.405 (base 1975) for CPS-1, and 0.568 and 0.508 for PSID-1.

\subsection{The studentized coherence test}

The test uses the seven-cell grid. For each vertex and cell, $\sigma_{s,k}$ is the bootstrap standard deviation of $\widehat z_{s,k}$ (floored at $10^{-4}$), computed from 2,000 draws that resample complete histories within the participant and comparison samples. The statistic is the smallest $c$ for which some $\lambda$ satisfies $\widehat z_{\lambda,k}+c\sum_s\lambda_s\sigma_{s,k}\ge0$ in every cell; it equals 1.35 for CPS-1 and 3.37 for PSID-1, against bootstrap 95 percent critical values of 3.00 and 2.93, giving the $p$-values in Table~\ref{tab:nsw}.

\subsection{The AFDC women}

The women's sample of \citet{CalonicoSmith2017} contains 600 participants and 585 experimental controls from the AFDC target group, with earnings in 1975 and 1979, and 648 PSID women as the comparison group. With $S=\{1975\}$ the hull is a single vertex, so historical transport reduces to distributional difference-in-differences. On the seven-cell grid the vertex has negative mass 0.006; its implied employment effect is 0.269 against an experimental 0.012; the total-variation distance to the experimental control law is 0.29; and the CDF distance is $\widehat D=0.282$, against bootstrap and DKW critical values of 0.081 and 0.256. The 95 percent lower confidence bounds on the transport error are 0.20 (bootstrap) and 0.03 (DKW). As with the men, the mechanism is visible before the program: 73 percent of participants but 28 percent of PSID women had no 1975 earnings, against a 1979 gap of 0.20 between experimental controls and PSID women. Selection on long-term welfare receipt might have been expected to favor transport, since it is a more persistent state than current joblessness; it did not.

\subsection{Breakdown against a known error}

For CPS-1 the relaxed model is nonempty for $\tau\ge0.009$, admits a zero employment effect at $\tau=0.383$, and has realized transport error (total variation from the experimental control law to the signed hull) of $0.338$. At $\tau=0.338$ the employment-effect interval is $[0.046,0.757]$. For PSID-1 the three numbers are $0.037$, $0.372$ and $0.351$, and the interval at the realized error is $[0.020,0.757]$. The upper endpoint, the treated employment share, is reached because the relaxation allows every participant to have been nonemployed absent the program. A bootstrap 95 percent confidence interval for the CPS-1 radius, from Proposition~\ref{prop:breakdown} with the joint bound $\max_s d_{\TV}(\widehat z^*_s,\widehat z_s)+|\widehat g^*-\widehat g|$, is $[0.26,0.51]$; for PSID-1 it is $[0.24,0.51]$.
\section{Application II: The EITC Expansion, Details}\label{S-sec:eitc-supp}

\subsection{Data and prespecification}

The data are a public extract of March Current Population Survey records for tax years 1991--1996, widely distributed for teaching the difference-in-differences comparison of \citet{EissaLiebman1996}. It contains 13,746 women aged 20 to 54 with eleven or fewer years of schooling, described by its distributors as unmarried; we cannot check that description against the original files, and the extract carries no survey weights, household identifiers or design variables. Variables used: tax year, number of children, annual earnings and, in Section~\ref{S-sec:eitc-state-supp}, state of residence. Earnings are in 1997 dollars: dividing each recorded amount by the ratio of the 1997 to the tax-year consumer price index (CPI-U annual averages) recovers whole-dollar nominal amounts. States are recorded with the Census state codes of the CPS (11 for Maine to 95 for Hawaii); we map them to states and confirm the mapping against the state unemployment rate carried in the extract (for instance 9.3 percent for California in 1992). The indicator of work in the extract does not coincide with positive earnings (1,351 women with positive earnings are coded as not working), so we use earnings throughout. The file's SHA-256 hash is in the replication package.

Before computing any outcome by group and year we recorded the design in a dated plan, which the replication package reproduces: groups, historical years, outcome years, the five-cell grid, a 21-cell secondary grid (zero, then 2,500-dollar bins to 47,500, and above), the three contrasts, the diagnostics, the comparisons and the success criteria. Three additions were made after the first estimates and are reported as such: the comparison-group environment check of Section~\ref{sec:content}, which the main text's protocol calls for; the bootstrap outer regions based on the vertex reduction, for every year; and the studentized tests on the 21-cell grid for 1994 and 1995. None of them changes a bound.

\subsection{Mothers of two or more children}

\begin{center}\small
\begin{tabular}{@{}lccc@{}}\toprule
 & 1994 & 1995 & 1996\\\midrule
Employment effect, bound & $[-0.025,0.012]$ & $[0.049,0.086]$ & $[0.134,0.171]$\\
\quad 95\% outer region & $[-0.09,0.08]$ & $[-0.03,0.16]$ & $[0.06,0.24]$\\
\quad Grid-free 95\% region & $[-0.18,0.17]$ & $[-0.11,0.24]$ & $[-0.02,0.33]$\\
Phase-out share effect & $[-0.019,-0.003]$ & $[-0.004,0.011]$ & $[0.062,0.077]$\\
Phase-in and plateau share effect & $[-0.014,0.001]$ & $[0.031,0.046]$ & $[0.045,0.059]$\\
Sign breakdown radius, employment & 0 & 0.049 & 0.134\\
Comparison-law check, $p$-value & 0.78 & 0.39 & 0.01\\\bottomrule
\end{tabular}
\end{center}

\noindent The studentized coherence test does not reject on either grid in any year ($p=1.00$), and the 1993 placebo benchmark gives a CDF distance of 0.036 against a bootstrap critical value of 0.089 ($p=0.94$). For all mothers, the phase-in and plateau share effect is $[-0.005,0.012]$, $[0.018,0.035]$ and $[0.014,0.031]$ in 1994, 1995 and 1996.

\subsection{Sensitivity and comparisons}

Adding 1993 as a third historical year leaves the 1996 employment bounds unchanged ($[0.115,0.139]$ for all mothers and $[0.134,0.171]$ for mothers of two or more), widens the phase-in and plateau share effect to $[0.014,0.043]$ and $[0.045,0.093]$, and leaves the coherence test at $p=1.00$. The 1996 single-base distributional difference-in-differences effects on employment are 0.115 (base 1992) and 0.139 (base 1991) for all mothers, the endpoints of the transport bound; changes-in-changes with base 1992 gives $[0.153,0.164]$, and $[0.174,0.182]$ for mothers of two or more.

On the 21-cell grid the plug-in repair is 0.003, 0.001 and 0 for all mothers in 1994--1996 and 0.005, 0.003 and 0.003 for mothers of two or more; the studentized test does not reject in any of these cases, and a common weight across the three outcome years is feasible on the five-cell grid but not, in the plug-in, on the 21-cell grid.

The bootstrap resamples women within group and year, treating the records as independent draws. With independent group--year samples each vertex is a sum of independent empirical distribution functions, and the within-sample bootstrap is consistent by the argument of Section~\ref{S-sec:emp} applied sample by sample. CPS households are sampled in clusters within states; Section~\ref{S-sec:eitc-state-supp} resamples whole states instead, with almost identical results.

\subsection{States, clustering and welfare waivers}\label{S-sec:eitc-state-supp}

Table~\ref{S-tab:eitc-states} extends Table~\ref{tab:eitc-waiver} to 1994 and to mothers of two or more children, and compares regions that resample whole states with regions that resample women. The state-level analysis was added in revision, after the results of Table~\ref{tab:eitc} were known; the waiver classification was taken from the published table before any outcome was computed by state. Bounds in each design are computed from the pooled records of the retained states, so the target is the population of those states, weighted by their sample sizes. The state bootstrap draws states with replacement and keeps every record of a drawn state, so any dependence within a state, across years, groups and households, is preserved. Its justification is the cluster theory of Section~\ref{S-sec:cluster} applied to ratio estimators (see the paragraph before Proposition~\ref{S-prop:cluster-hoeffding}); with 15 states it is approximate.

\begin{table}[p]\centering\footnotesize
\caption{The EITC application by welfare-waiver design}\label{S-tab:eitc-states}
\setlength{\tabcolsep}{4pt}\renewcommand{\arraystretch}{0.95}
\begin{tabular}{@{}lccc@{}}\toprule
 & 1994 & 1995 & 1996\\\midrule
\multicolumn{4}{@{}l}{\textbf{All mothers}}\\
\multicolumn{4}{@{}l}{\emph{All states (51)}}\\
Employment effect, bound & $[-0.010,0.014]$ & $[0.047,0.070]$ & $[0.115,0.139]$\\
\quad 95\% outer region, states resampled & $[-0.08,0.08]$ & $[-0.01,0.13]$ & $[0.05,0.20]$\\
\quad 95\% outer region, women resampled & $[-0.07,0.08]$ & $[-0.01,0.13]$ & $[0.05,0.21]$\\
Phase-out share effect, bound & $[-0.031,-0.008]$ & $[-0.009,0.014]$ & $[0.053,0.076]$\\
Comparison-law check, $p$-value & 0.70 & 0.38 & 0.02\\
Placebo benchmark at 1993, $p$-value & \multicolumn{3}{c}{0.92}\\
\multicolumn{4}{@{}l}{\emph{No waiver approved by 1995 (24)}}\\
Employment effect, bound & $[0.056,0.103]$ & $[0.055,0.102]$ & $[0.084,0.118]$\\
\quad 95\% outer region, states resampled & $[-0.04,0.19]$ & $[-0.04,0.20]$ & $[0.01,0.20]$\\
\quad 95\% outer region, women resampled & $[-0.04,0.19]$ & $[-0.04,0.20]$ & $[-0.01,0.22]$\\
Phase-out share effect, bound & $[0.023,0.029]$ & $[0.009,0.014]$ & $[0.044,0.048]$\\
Comparison-law check, $p$-value & 0.48 & 0.61 & 0.20\\
Placebo benchmark at 1993, $p$-value & \multicolumn{3}{c}{0.37}\\
\multicolumn{4}{@{}l}{\emph{No waiver approved by 1996 (15)}}\\
Employment effect, bound & $[0.000,0.089]$ & $[-0.010,0.079]$ & $[0.031,0.120]$\\
\quad 95\% outer region, states resampled & $[-0.15,0.23]$ & $[-0.16,0.22]$ & $[-0.06,0.21]$\\
\quad 95\% outer region, women resampled & $[-0.12,0.21]$ & $[-0.13,0.20]$ & $[-0.08,0.23]$\\
Phase-out share effect, bound & $[-0.004,0.015]$ & $[-0.038,-0.019]$ & $[0.023,0.041]$\\
Comparison-law check, $p$-value & 0.79 & 0.93 & 0.56\\
Placebo benchmark at 1993, $p$-value & \multicolumn{3}{c}{0.50}\\\midrule
\multicolumn{4}{@{}l}{\textbf{Mothers of two or more children}}\\
\multicolumn{4}{@{}l}{\emph{All states (51)}}\\
Employment effect, bound & $[-0.025,0.012]$ & $[0.049,0.086]$ & $[0.134,0.171]$\\
\quad 95\% outer region, states resampled & $[-0.10,0.09]$ & $[-0.02,0.15]$ & $[0.06,0.25]$\\
\quad 95\% outer region, women resampled & $[-0.09,0.08]$ & $[-0.02,0.16]$ & $[0.06,0.24]$\\
Phase-out share effect, bound & $[-0.019,-0.003]$ & $[-0.004,0.011]$ & $[0.062,0.077]$\\
Comparison-law check, $p$-value & 0.71 & 0.39 & 0.03\\
Placebo benchmark at 1993, $p$-value & \multicolumn{3}{c}{0.90}\\
\multicolumn{4}{@{}l}{\emph{No waiver approved by 1995 (24)}}\\
Employment effect, bound & $[0.058,0.108]$ & $[0.058,0.108]$ & $[0.098,0.125]$\\
\quad 95\% outer region, states resampled & $[-0.05,0.22]$ & $[-0.04,0.21]$ & $[0.00,0.24]$\\
\quad 95\% outer region, women resampled & $[-0.04,0.21]$ & $[-0.05,0.21]$ & $[-0.01,0.26]$\\
Phase-out share effect, bound & $[0.035,0.047]$ & $[0.015,0.027]$ & $[0.048,0.054]$\\
Comparison-law check, $p$-value & 0.46 & 0.58 & 0.20\\
Placebo benchmark at 1993, $p$-value & \multicolumn{3}{c}{0.59}\\
\multicolumn{4}{@{}l}{\emph{No waiver approved by 1996 (15)}}\\
Employment effect, bound & $[-0.001,0.089]$ & $[-0.015,0.049]$ & $[0.045,0.135]$\\
\quad 95\% outer region, states resampled & $[-0.16,0.25]$ & $[-0.16,0.22]$ & $[-0.09,0.27]$\\
\quad 95\% outer region, women resampled & $[-0.13,0.22]$ & $[-0.15,0.21]$ & $[-0.09,0.27]$\\
Phase-out share effect, bound & $[-0.002,0.006]$ & $[-0.041,-0.035]$ & $[0.016,0.025]$\\
Comparison-law check, $p$-value & 0.78 & 0.92 & 0.58\\
Placebo benchmark at 1993, $p$-value & \multicolumn{3}{c}{0.62}\\\bottomrule
\end{tabular}
\begin{minipage}{0.97\linewidth}\footnotesize\emph{Notes:} Designs drop the states whose first statewide AFDC waiver was approved by the year shown \citep[Table A3]{SchanzenbachStrain2021}; the number of states retained is in parentheses. Bounds are on the five-cell grid with historical years 1991 and 1992; the bounds impose coherence, and they coincide with the signed bounds except for all mothers in 1996 and mothers of two or more in 1996 under the 24-state design, and mothers of two or more in 1995 under the 15-state design (see text). Outer regions use the vertex reduction with a joint bootstrap band for the two vertex effects, resampling whole states (all years and both groups) or women within group and year. Comparison-law checks and placebo benchmarks resample states. The studentized coherence test does not reject in any design, group or year ($p=1.00$).\end{minipage}
\end{table}

Three features stand out. First, resampling states rather than women moves no endpoint of an outer region by more than 0.035 in any design, group or year, so dependence within states is not a material concern in this extract. Second, coherence binds after waiver states are dropped. In the 24-state design the signed bound on the 1996 employment effect is $[0.084,0.131]$ for all mothers and $[0.098,0.148]$ for mothers of two or more; coherence cuts the upper ends to 0.118 and 0.125. In the 15-state design it cuts the upper end of the 1995 bound for mothers of two or more from 0.075 to 0.049. In each case some weights in the historical hull put negative mass on a cell and are removed. Third, in the 24-state design the 1994 bound is also positive, $[0.056,0.103]$, unlike in the full sample, so in these states the employment gain appears in the first year of the expansion; its outer region, $[-0.04,0.19]$, includes zero.

The finite-sample shared-cluster radius of Proposition~\ref{S-prop:cluster-hoeffding} is of no use here: with $J=2$, $K=5$ and 15 states it equals 1.34, more than any deficit can be. An equal-weight, state-average design would also change the target and rely on state--year--group cells with as few as one woman (18 of the 612 cells have one or two). We therefore keep the pooled target and the state bootstrap, and treat the waiver designs as sensitivity analyses: approval dates are not implementation dates, several states also had their own credits, and dropping states changes the population.

\section{Application III: Data, Robustness and Audit Details}\label{S-sec:mw-supp}

\subsection{Data and timing}

The panel has $51\times36\times118=216{,}648$ state--quarter--cell records for 2007--2015. Within each state--quarter the 117 positive-wage cells count workers per capita in quarter-dollar bins of hourly wages in 2016 dollars (bin labels are lower edges), and the residual cell equals one minus the total. Employment is benchmarked to QCEW job counts and divided by resident population, so in the District of Columbia the residual is negative in every quarter (between $-0.45$ and $-0.33$); the District is excluded from all comparison pools. Rebuilding the panel from the original CPS processing inputs in the \citet{CengizEtAl2019} deposit reproduces every released cell to within $5.5\times10^{-8}$; all calculations use the released panel.

\begin{table}[h]\centering\small
\caption{The five 2014 events}\label{S-tab:mw-events}
\begin{tabular}{@{}llllc@{}}\toprule
State & Enacted & Effective & Outcome quarter & Pool size\\\midrule
New Jersey & Nov.\ 5, 2013 (ballot) & Jan.\ 1, 2014 & 2014Q1 & 34\\
Delaware & Jan.\ 30, 2014 & June 1, 2014 & 2014Q3 & 30\\
California & Sept.\ 25, 2013 & July 1, 2014 & 2014Q3 & 30\\
Minnesota & Apr.\ 14, 2014 & Aug.\ 1, 2014 & 2014Q4 & 30\\
Michigan & May 27, 2014 & Sept.\ 1, 2014 & 2014Q4 & 30\\\bottomrule
\end{tabular}
\begin{minipage}{0.95\linewidth}\footnotesize\emph{Notes:} The outcome quarter is the first calendar quarter fully covered by the new minimum. Pool size is the number of comparison states with no increase from 2010Q1 to the outcome quarter, excluding the District of Columbia. Minnesota's law set different minima for large and small employers.\end{minipage}
\end{table}

For event $e$ with outcome quarter $q_e$, comparison pool $\mathcal C_e$ and treated state $j_e$, the vertex at offset $h\in\{-16,-12,-8,-4\}$ is
\[
z_{eh}=\overline C_e(q_e)+T_{j_e}(q_e+h)-\overline C_e(q_e+h),
\]
where $T_j(t)$ is state $j$'s 118-cell vector and $\overline C_e(t)$ is the unweighted average over $\mathcal C_e$. The stacked vertex is $z_h=\frac15\sum_ez_{eh}$ and the stacked treated vector $g=\frac15\sum_eT_{j_e}(q_e)$.

\subsection{From seven to 118 cells}

The seven-cell grid consists of the residual and the intervals cut at \$7.25, \$9.25, \$11.25, \$14.25 and \$18.25. Table~\ref{S-tab:mw-refine} refines it by repeatedly splitting the cell with the most quarter-dollar bins (ties to the lowest wage) at its midpoint, keeping the five cutoffs as boundaries. The intermediate rows depend on the splitting rule; the first and last rows do not.

\begin{table}[h]\centering\small
\caption{Refinement path for the below-\$9.25 bound}\label{S-tab:mw-refine}
\begin{tabular}{@{}rccc@{}}\toprule
Cells & Invalid vertices & Coherent counterfactual share (pp) & Effect (pp)\\\midrule
7 & 0 & $[7.363,7.750]$ & $[-0.614,-0.228]$\\
15 & 0 & $[7.363,7.750]$ & $[-0.614,-0.228]$\\
30 & 1 & $[7.363,7.750]$ & $[-0.614,-0.228]$\\
60 & 3 & $[7.453,7.719]$ & $[-0.584,-0.317]$\\
118 & 4 & $[7.459,7.653]$ & $[-0.517,-0.324]$\\\bottomrule
\end{tabular}
\end{table}

\subsection{Eventwise coherence, matched windows and relaxation}

On the full grid the stacked hull is coherent, but no common weight makes all five event vectors $\sum_h\lambda_hz_{eh}$ nonnegative; the smallest average negative mass over events is 0.00303. The test of Proposition~\ref{S-prop:eventwise-test} returns negative values $m_{ej}$ for 131 of the $5\times118$ event--cell pairs, each supplying a stacked-coherent weight that violates eventwise coherence; the most negative is $-0.0065$, for California's \$7.75--\$8.00 bin. With the three offsets $(-12,-8,-4)$ instead of four, the policy-date hull is again coherent and the sign breakdown radius is unchanged at $0.00288$; the one-year-earlier placebo, computed with the comparison pools frozen at their 2014 definitions, has minimum repair $1.19\times10^{-5}$ with three offsets and $5.93\times10^{-6}$ with four. These repairs are tiny; the informative placebo statistic is the total-variation distance from the observed treated law to the signed hull, 0.0492 for 2013 and 0.0473 for 2012. The relaxed bound on the below-\$9.25 effect is $[-0.70,-0.19]$ at $\tau=0.001$ and $[-0.85,-0.04]$ at $\tau=0.0025$.

\subsection{Audit details}

\begin{figure}[h]\centering
\includegraphics[width=0.72\linewidth]{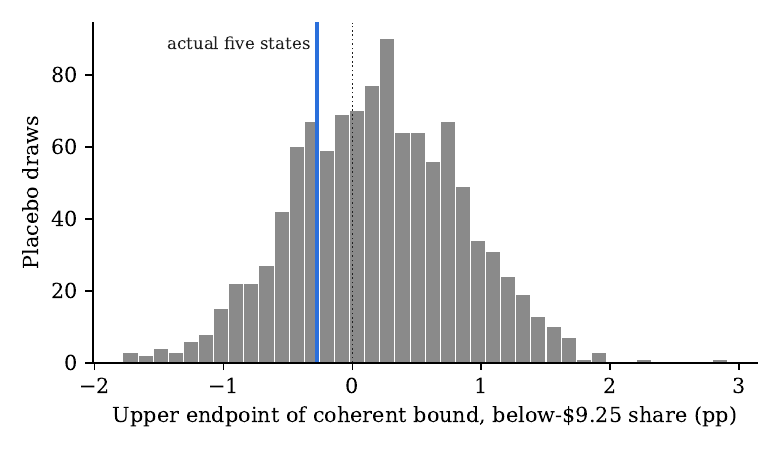}
\caption{Placebo assignments: upper endpoints of the coherent bound}\label{S-fig:placebo}
\begin{minipage}{0.97\linewidth}\footnotesize\emph{Notes:} The 1,090 of 2,000 placebo draws with a nonempty coherent set. Each draw gives the five actual event dates to five randomly chosen never-treated states and uses the remaining 18 as comparisons. Vertical line: the actual states, with the never-treated comparison pool.\end{minipage}
\end{figure}

Table~\ref{S-tab:mw-audit} collects the audit results summarized in Section~\ref{sec:mw} of the paper, and Figure~\ref{S-fig:placebo} shows the placebo distribution.

\begin{table}[h]\centering\small
\caption{Audit of the below-\$9.25 bound (percentage points)}\label{S-tab:mw-audit}
\setlength{\tabcolsep}{4pt}
\begin{tabular}{@{}lccccc@{}}\toprule
 & Signed & Coherent & Eventwise & $\tau^{\rm sign}$ & Placebo \\
 & hull & hull & repair & & distance\\\midrule
Baseline & $[-0.61,-0.23]$ & $[-0.52,-0.32]$ & 0.30 & 0.29 & 4.92\\[2pt]
\multicolumn{6}{@{}l}{\emph{Historical dates before enactment}}\\
\quad Drop $-4$ & $[-0.57,-0.23]$ & empty & 0.41 & -- & \\
\quad Drop $-4$ and $-8$ & $[-0.57,-0.23]$ & empty & 0.59 & -- & \\[2pt]
\multicolumn{6}{@{}l}{\emph{Aggregation}}\\
\quad Population-weighted events & $[-0.56,\ \ 0.20]$ & $[-0.01,\ \ 0.08]$ & & & \\[2pt]
\multicolumn{6}{@{}l}{\emph{Comparison states}}\\
\quad Never treated, 2010--2015 & $[-0.54,-0.18]$ & $[-0.42,-0.27]$ & 0.34 & 0.25 & 4.91\\
\quad Not bordering a treated state & $[-0.72,-0.39]$ & $[-0.65,-0.54]$ & 0.30 & 0.46 & 5.03\\
\quad Both restrictions & $[-0.56,-0.25]$ & $[-0.46,-0.39]$ & 0.35 & 0.30 & 5.05\\[2pt]
\multicolumn{6}{@{}l}{\emph{No effect on the share employed at or above $W$}}\\
\quad $W=\$18.25$ & & $[-0.51,-0.37]$ & & & \\
\quad $W=\$15.25$ & & $[-0.45,-0.39]$ & & & \\\bottomrule
\end{tabular}
\begin{minipage}{0.97\linewidth}\footnotesize\emph{Notes:} ``Drop $-4$'' removes the historical quarter one year before the outcome; California's increase was signed in it (September 25, 2013), and New Jersey's amendment had passed the legislature. Population weights are the treated states' populations in the outcome quarter (California carries 60 percent). ``Never treated'' excludes states with any increase in 2010--2015, which removes California, Delaware, Michigan and Minnesota from New Jersey's pool; 23 comparison states remain. ``Not bordering'' removes states adjacent to a treated state. The last rows impose $\mu_\lambda(\{w\ge W\})=\mu_{1,1}(\{w\ge W\})$.\end{minipage}
\end{table}

\emph{Bracketing.} Averaging over events, the comparison-state nonemployment rate is 44.41, 44.04, 44.08 and 43.83 percent at offsets $-16$, $-12$, $-8$ and $-4$, and 43.42 percent at the outcome quarter; the treated-minus-comparison gaps are 0.61, 0.60, 0.64 and 0.30 points. The comparison-state share employed below \$9.25 is 8.02, 8.59, 8.57 and 8.93 percent historically and 8.82 percent at the outcome quarter, so on this dimension the policy date is bracketed. The total-variation distance from the 2014 comparison-state law to the convex hull of its own historical laws is 0.120 on the 118 cells, and 0.127 one year earlier; at this resolution even the comparison states' distribution moves outside its historical span, partly because nominal wages drift across real quarter-dollar bins.

\emph{Population growth.} Over the four years before the outcome quarter, population grew by 1.0 percent in Michigan, 3.2 percent in New Jersey, 3.8 percent in Minnesota, 5.6 percent in Delaware and 5.9 percent in California, against 4.4 percent on average in the comparison pools. The outcome is per capita, so differential population growth enters only through composition, which the public panel cannot measure.

\emph{Upper tail.} At the policy date the observed treated share employed at or above \$18.25 is 28.67 percent; the coherent hull's range is $[28.14,28.98]$ and the signed hull's $[28.00,29.34]$. At or above \$15.25 the corresponding numbers are 33.94, $[33.59,34.25]$ and $[33.29,34.49]$. In the 2013 placebo the observed shares, 27.97 and 33.61 percent, lie below the signed-hull ranges $[28.48,29.38]$ and $[33.97,34.81]$.

\emph{Comparison pools.} The never-treated pool contains the 23 jurisdictions with no increase in 2010--2015. Removing states that border any treated state (Oregon, Nevada, Arizona, Maryland, Pennsylvania, New York, Ohio, Indiana, Wisconsin, Iowa, South Dakota and North Dakota, together with the treated states themselves) leaves 23 states under the baseline rule and 18 when combined with the never-treated rule.

\emph{Uncertainty in the bracketing check.} Resampling comparison states with replacement within each event's pool (2,000 draws), the difference between policy-date nonemployment and its historical minimum is $-0.41$ points with 95 percent interval $[-0.47,-0.35]$.

\emph{Population weights.} Equal event weights answer a question about the average state event; population weights answer one about the average resident of a treated state. Neither is privileged by the data, which is why we report both. With weights proportional to the treated states' populations in the outcome quarter (New Jersey 0.14, California 0.60, Delaware 0.01, Michigan 0.16, Minnesota 0.09), the stacked hull is still coherent, the signed bound is $[-0.56,0.20]$ and the coherent bound $[-0.007,0.078]$.

\emph{Placebo assignments.} Each of 2,000 draws (seed 7) selects five of the 23 never-treated states without replacement and assigns them, in random order, the outcome quarters of New Jersey, California, Delaware, Michigan and Minnesota; the other 18 never-treated states are the comparison pool. The coherent set is nonempty in 1,090 draws (54.5 percent); conditioning on a nonempty set is a selection, so we also report unconditional shares. The quantiles of the placebo upper endpoint (in pp) at 5, 10, 50 and 90 percent are $-0.89$, $-0.62$, $0.19$ and $1.04$. The actual upper endpoint with the never-treated pool is $-0.27$. Among the coherent draws, 24.6 percent have an upper endpoint at or below it; unconditionally, 13.4 percent of all 2,000 draws produce a coherent set with such an endpoint. In a second exercise the five pseudo-treated states are drawn from the 11 largest never-treated states (by population in 2014Q1), with all other never-treated states as comparisons; the coherent set is nonempty in 77 percent of 2,000 draws, and among these 15 percent have an upper endpoint at or below $-0.27$ and 29 percent are entirely negative.

\subsection{What a convincing minimum-wage analysis would add}

The audit points to four requirements. (i) Timing: date histories from enactment or credible announcement, not from effective dates, and check that the policy date's labor-market conditions lie within the historical range on the dimensions that drive treated--control differences (for example cyclical employment and the share of workers near the new minimum). (ii) Composition: use worker-level CPS records to reweight treated and comparison states to a fixed demographic composition before forming the 118-cell vectors, and test the upper-tail exclusion cell by cell with valid simultaneous bands. (iii) Comparison groups: use never-treated, non-contiguous states, or explicitly model spillovers, and measure policy bite directly. (iv) Sampling: obtain a design-based joint covariance for the state--quarter cells, or replace the CPS with administrative wage records, so that the breakdown radius can be compared with a sampling error of known coverage. Until then the bounds are descriptive.

\subsection{Where the negative mass lies}

For each state, take the weight that minimizes the negative mass of its own counterfactual (Proposition~\ref{S-prop:event-specific}). The negative cells are wage bins; the residual cell, one minus the employment-to-population ratio, is positive in every case. Across the five states there are 40 such bins: 6 for New Jersey, 13 for California, 13 for Delaware, 2 for Michigan and 6 for Minnesota. Twenty-six lie below \$8.00 (2016 dollars), below or near the old minimum wages, where cells are sparse; the others are scattered between \$10.75 and \$29.75. No negative count exceeds 0.13 percent of the population; the largest are California's at \$14.75 and \$19.50. A negative count of jobs in a wage bin is impossible whether or not the cells describe individual outcomes, so the infeasibility of the state counterfactuals does not rest on the individual interpretation of the panel.

\end{document}